\documentstyle[12pt,epsf,a4]{article}
\begin{document}

\begin{titlepage}
\begin{flushright}
UTHEP-327\\
UTCCP-P-8\\
December 1995 \\
\end{flushright}

\begin{centering}
\vfill
{\bf \LARGE Hadron Masses and Decay Constants \\
with Wilson Quarks at $\beta=5.85$ and 6.0} 

{\baselineskip=18pt
\vspace{1cm}   
{\it QCDPAX Collaboration}\\[0.5cm]
Y. Iwasaki$^{a,b}$, K. Kanaya$^{a,b}$, T. Yoshi{\'e}$^{a,b}$, \\
T. Hoshino$^c$, T. Shirakawa$^c$, Y. Oyanagi$^d$, \\
S. Ichii$^e$ and T. Kawai$^f$ \\[0.5cm]

\vspace{0.5cm}   
$^a$Institute of Physics, University of Tsukuba,
Ibaraki 305, Japan \\
$^b$Center for Computational Physics, University of Tsukuba,
Ibaraki 305, Japan \\
$^c$Institute of Engineering Mechanics, University of Tsukuba,
Ibaraki 305, Japan \\
$^d$Department of Information Science, University of Tokyo,
Tokyo 113, Japan \\
$^e$Computer Centre, University of Tokyo,
Tokyo 113, Japan \\
$^f$Department of Physics, Keio University,
Yokohama 223, Japan \\
}

\vspace{2.5cm}
{\bf Abstract} 
\end{centering}

\vspace{0.3cm}\noindent
We present results of a high statistics calculation of 
hadron masses and meson decay constants in the  
quenched approximation to lattice QCD with Wilson quarks
at $\beta=$ 5.85 and 6.0 on $24^3 \times 54$ lattices.
We analyze the data paying attention in particular
to the systematic errors
due to the choice of fitting range and
due to the contamination from excited states. 
We find that the systematic errors 
for the hadron masses with quarks
lighter than the strange quark amount to 
1 --- 2 times the statistical errors.
When the lattice scale is fixed from the $\rho$ meson mass,
the masses of the $\Omega^{-}$ baryon and the $\phi$ meson 
at two $\beta$'s
agree with experiment within about one standard deviation.
On the other hand, the central value of
the nucleon mass at $\beta=6.0$ (5.85)
is larger than its experimental value 
by about 15\% (20\%) and that of
the $\Delta$ mass by about 15\% (4\%):
Even when the systematic errors are included,
the baryon masses at $\beta=6.0$ do not agree with experiment.
Vector meson decay constants at two values of $\beta$ 
agree well with each other and are consistent with experiment
for a wide range of the quark mass,
when we use current renormalization constants determined
nonperturbatively by numerical simulations.
The pion decay constant agrees with experiment
albeit with large errors.
Results for the masses of excited states of
the $\rho$ meson and the nucleon are also presented.
\vfill
\end{titlepage}

\section{Introduction}\label{sec:Intro}
Although there have been many efforts 
to calculate hadron masses
in lattice QCD by numerical simulations,
it has turned out that derivation of convincing results is
much harder than thought at the beginning,
even in the quenched approximation.
For example, before 1988, there was large discrepancy among
the results for the mass ratio $m_{N}/m_{\rho}$ obtained 
for $\beta=6/g^2= 5.7$ --- 6.0 and in the quark mass region
corresponding to $m_{\pi}/m_{\rho} \ge 0.5$.
The discrepancy was caused by systematic errors due to 
contamination from excited 
states~\cite{STD1648B585,STDB585}
and effects of finite lattice spacing~\cite{APE8889}
and finite lattice volume.
Recent high statistics simulations employ
lattices with large temporal 
extent~\cite{QCDPAX92,QCDPAX93,APE18}
and/or extended quark sources
\cite{QCDPAX93,APE18,APE24,APE63,HEMCGC,GF11,LANL}
to reduce fluctuations as well as the contamination
from excited states.
However, a long plateau in an effective mass is rarely seen 
and data of effective masses
frequently show large fluctuations at large time separations. 
The uncertainty in the choice of fitting range is therefore
another source of systematic errors.
In order to obtain reliable values for the spectrum, 
it is essential to make a quantitative study
on these systematic errors.

In this paper we report results of a high statistics 
calculation of the quenched QCD spectrum
with the Wilson quark action
at $\beta=5.85$ and 6.0 on $24^3 \times 54$ lattices.
Our major objection is  to calculate light hadron masses
as well as meson decay constants
paying attention in particular to the systematic errors
due to the choice of fitting range and
due to the contamination from excited states. 
In order to estimate the magnitude of these systematic errors,
we perform correlated one-mass fits to hadron propagators 
systematically varying fitting 
ranges~\cite{QCDPAX93,UKQCD62-60}.
Assuming the ground state dominance at large time separations,
we estimate systematic errors in hadron masses
which cannot be properly taken into account
by the standard least mean square fit 
when the fitting range is fixed.
It is shown that,
for the hadron masses with quarks
lighter than the strange quark,
the systematic errors amount to 
1 --- 2 times the statistical errors.
We then perform correlated two-mass fits,
again varying fitting ranges.
We find that the ground state mass 
is consistent with that obtained from the one-mass fit
within the statistical and systematic errors.
Finally, we extrapolate the results of hadron masses
at finite quark mass to the chiral limit,
taking account of systematic errors 
both due to the choice of extrapolation
function and due to the fitting range.
We also study meson decay constants in a similar way.

We use the point source in this study.
Historically there was a report that
numerical results for hadron masses
appear to depend on the type of the source 
adopted~\cite{LANLsmear},
although it has afterward reported in some works 
that masses are independent within the 
statistical errors~\cite{QCDPAX93,APE18}.
Note in this connection that
there is no proof that the value of a hadron mass 
is independent on the type of sources
in the case of the quenched approximation 
due to the lack of the transfer matrix and
that there is the so-called Gribov problem for gauge fixing
which is necessary for almost all smeared sources.
Under these circumstances
it may be worthwhile to present the
details of the results and the analyses
with the point source as a reference.
The method of analyses of the systematic errors
in this work can be applied
to the cases of smeared sources too.

Numerical simulations are performed 
with the QCDPAX~\cite{PAXmachine},
a MIMD parallel computer constructed 
at the University of Tsukuba.
For the calculations performed in this work,
we use $24\times 18$ processing units interconnected
in a toroidal two-dimensional mesh 
with a peak speed of 12.4 GFLOPS.
(The maximum number of nodes is $24 \times 20 $ 
with a peak speed of 14.0 GFLOPS.)
The sustained speed for the Wilson quark matrix 
multiplication is approximately 5 GFLOPS.
The calculations described here 
took about six months on the QCDPAX.

We start by giving in Sec.~\ref{sec:Numer} 
some details about our numerical simulations.
Then we derive hadron masses at finite quark mass
in Sec.~\ref{sec:Mass}
and perform two-mass fits to estimate the masses 
of excited states of the $\rho$ meson and the nucleon 
in Sec.~\ref{sec:Excited}.
We extrapolate the results to the chiral limit 
in Sec.~\ref{sec:Chiral}.
Sec.~\ref{sec:Decay} is devoted to the evaluation of 
meson decay constants.
In Sec.~\ref{sec:conclusions}, we give conclusions and 
discussion on the results.

\section{Numerical Calculation}\label{sec:Numer}

We use the standard one-plaquette gauge action
\begin{equation}
S_g = {2\over g^2 } \sum_{P} {\rm Re} \ {\rm Tr} (U_P)
\end{equation}
and the Wilson quark action 
\begin{equation}
S_q = - \sum_{n,m} \bar\psi(n)D(K,n,m)\psi(m),
\end{equation}
\begin{equation}
D(K,n,m) = \delta_{n,m} - K \sum_{\mu} {
\{ (I-\gamma_{\mu})U_{n,\mu}\delta_{n+\mu,m} +
(I+\gamma_{\mu})U^{\dagger}_{m,\mu}\delta_{m+\mu,n}    \} },
\end{equation}
where $g$ is the bare coupling constant and
$K$ is the hopping parameter.

Simulations are done on $24^3 \times 54$ lattices 
at $\beta=6/g^2 = 5.85\ $ and 6.0 
for the five values of hopping parameter 
listed in table \ref{tab:K}.
The mass ratio $m_{\pi}/m_{\rho}$ takes a value 
from 0.97 to 0.52
and roughly agrees with each other at two $\beta$'s 
for the five cases of hopping parameter.
We choose the values of the third largest hopping parameter 
in such a way that they approximately 
correspond to the strange quark.

We generate 100 (200) configurations 
with periodic boundary conditions at $\beta= 5.85\ (6.0)$
by a Cabibbo-Marinari-Okawa algorithm
with 8 hit pseudo heat bath algorithm
for three $SU(2)$ subgroups.
The acceptance rate is about 0.95 for both $\beta$'s.
Each configuration is separated by 1000 sweeps
after a thermalization of 6000 (22000) sweeps
at $\beta=5.85 \ (6.0)$.

The quark propagator $G$ on a configuration given by
\begin{equation} 
\sum_{m} D(K,n,m)G(m) = B(n) \label{qprop}
\end{equation}
is constructed using a red/black minimal residual algorithm, 
taking periodic boundary conditions in all directions.
We employ the point source
at the origin $B(n) = \delta_{n,0}$.

The convergence criterion we take
for the quark matrix inversion
is that both of the following two conditions be satisfied:
\begin{equation}
\sqrt{|R|^2 / (3\times 4\times V)} < 10^{-9},
\end{equation}
\begin{equation}
{\rm max}_{n,c,s} \{|R_{c,s}(n)/G_{c,s}(n)|\} < 0.03,
\end{equation} 
where $|R|$ is the norm of the residual vector 
$R = B - D(K)G$, $V =L^3 \times T$ is the lattice volume 
($L=24$ is the lattice size in the spatial directions and 
 $T=54$ is that in the temporal direction),
and $c$ and $s$ are color and spin indices.
The average number of iterations needed for 
the convergence is given in table \ref{tab:K}.

Selecting several configurations,
we have solved exactly eq.~\ref{qprop} 
within single precision
to construct an exact hadron propagator
and compared it with that obtained 
with the stopping conditions above.
We find that the difference in a hadron propagator
(for any particle at any time slice) is at most one percent 
of the statistical error estimated using all (100 or 200) 
configurations.
Therefore the error due to truncation of iterations
is small enough and does not affect 
the following analyses and results.

We use ${\bar u} \Gamma d$ for meson operators
with $\Gamma=\gamma_5$ for $\pi$, 
$i\gamma_0 \gamma_5$ for $\pi$ ($\tilde\pi$), and
$\gamma_i$ for $\rho$.
For baryons, we use non-relativistic operators 
\begin{equation}
N_{l} = \epsilon^{abc} \sum_{i,j}^2 u_i^a 
\tau_3^{ij} d_j^b u_l^c \ \ \ l=1,2
\label{nucleon}
\end{equation}
\begin{equation}
\Delta_{l} = \epsilon^{abc} \sum_{i,j,k}^2 S_l^{ijk} 
u_i^a u_j^b u_k^c \ \ \ 
l=\pm 3/2, \pm 1/2, \label{delta}
\end{equation}
where $\tau_3$ is the third component of Pauli matrices and 
$S_l$ is the projection operator to $J=3/2, J_z=l$ state.
We also use anti-baryon operators obtained by replacing
the upper components of the Dirac spinor 
in eqs.~\ref{nucleon} and 
\ref{delta} with the lower components.

We average zero momentum hadron propagators
over all states with the same quantum numbers;
three polarization states for the $\rho$ meson and 
two (four) spin states for the nucleon ($\Delta$). 
Then we average the propagators 
for particle and anti-particle:
For mesons we average the propagator 
at $t$ and that at $T-t$,
for baryons we average the propagator 
for particle at $t$
and that of anti-particle at $T-t$.
In this work
we only calculate the masses of hadrons 
composed of degenerate mass quarks.

\begin{figure}[t]
\begin{center}
\leavevmode
\epsfysize=9.0cm
  \epsfbox{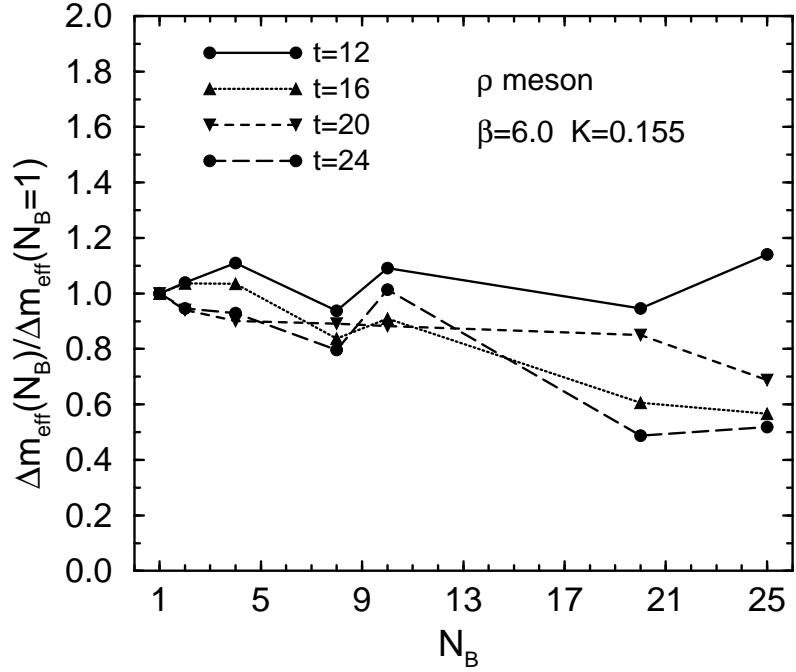}
\end{center}
\vskip -5mm
\caption{
Statistical errors in effective masses 
for the $\rho$ meson at $\beta=6.0$, $K=0.155$ 
versus the bin size $N_B$.
The errors are normalized by those for $N_B=1$.} 
\label{fig:Block}
\end{figure}

\begin{figure}[b]
\begin{center}
\leavevmode
\epsfysize=9.0cm
  \epsfbox{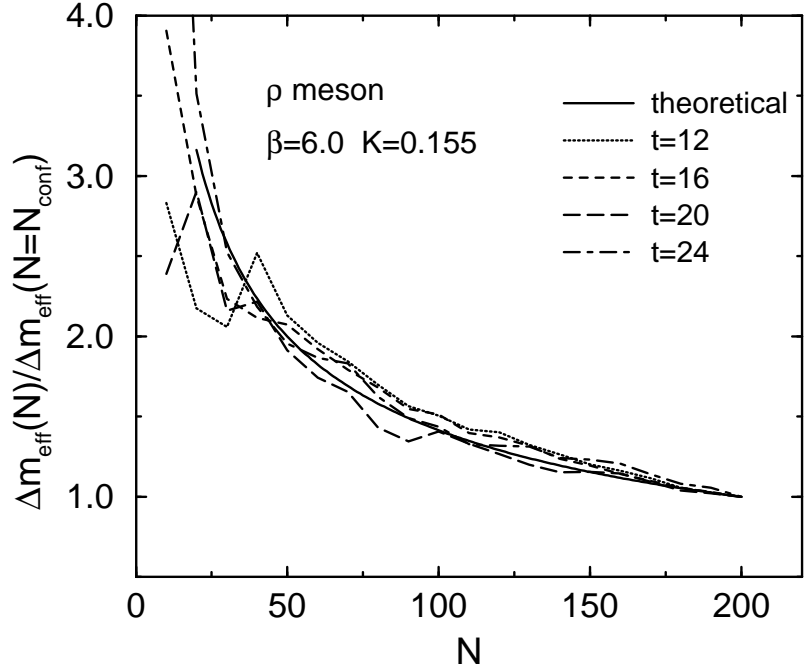}
\end{center}
\vskip -5mm
\caption{
Statistical errors in effective masses
for the $\rho$ meson at $\beta=6.0$, $K=0.155$ 
versus the number of configurations $N$.
The errors are normalized by those for $N=N_{conf}=200$.}
\label{fig:N-dep}
\end{figure}

Statistical independence of hadron propagators 
on each configuration is investigated 
by the following two methods.
1) We divide the total propagators 
into bins of $N_B$ successive
ones and apply the single elimination jack-knife method
to $N_{conf}/N_B$ block-averaged propagators.
We find that the errors in various quantities 
do not change significantly even if we change the bin size.
Fig.\ \ref{fig:Block} shows typical results 
for the bin size dependence
of the error in effective masses.
2) If configurations are independent,
we expect that the error obtained
for the set of $N$ configurations $\Delta(N)$ behaves as
\begin{equation}
\Delta (N) \sim 1/ \sqrt{N}.
\end{equation}
We check that this behavior is approximately satisfied 
using the propagators on the first $N$ configurations.
Fig.\ \ref{fig:N-dep} shows 
typical results for the $N$ dependence 
of the error in effective masses.

\section{Hadron Masses}\label{sec:Mass}

\subsection{Fitting procedure}   
Ground state masses of hadrons 
are extracted by fitting hadron propagators $G(t)$
to their asymptotic forms: 
\begin{equation}
G_0(t) = A \{ \exp(-m t) + \exp(-m (T-t) ) \}
\end{equation}
for mesons and
\begin{equation}
G_0(t) = A \exp(-m t) 
\end{equation}
for baryons.
(We will discuss the masses of excited states later.)
We perform least mean square fits
taking account of time correlations
minimizing $\chi^2$ defined by
\begin{equation}
\chi^2 = \sum_{t,t'=t_{min}}^{t_{max}}
   \{ G(t) - G_0(t)\} \, C^{-1}(t,t') \, \{ G(t') - G_0(t') \}
 \label{lms}
\end{equation}
where $C^{-1}(t,t')$ is the inverse of correlation matrix
$C(t,t')\ \ ( t_{min} \le t, t' \le t_{max} )$.
Errors are estimated by two methods.
One is the single elimination jack-knife method
taking account of the correlations 
among the propagators at different time separations.
Another estimate of the error is obtained from 
the least mean square fit itself.
Linear approximation to the fitting function 
around the minimum of $\chi^2$ gives a linear relation 
between the variance of fit parameters
and the variance of propagator $G(t)$ 
for the fitting range $t=t_{min}$ --- $t_{max}$. 
The relation leads to the error propagation rule
which relates the correlation matrix $C(t,t')$ 
to the error (and the correlation) on the fit parameters.
We find that the errors obtained by the two methods
are of the same order and that 
the error obtained by the jack-knife method is 
slightly (0\% to at most 40\%) larger 
than that by the least mean square fit.
Hereafter we quote the former error for the sake of safety,
unless otherwise stated.

\subsection{Fitting ranges and systematic error analyses}
\begin{figure}[htbp]
\begin{center}
\leavevmode
\epsfysize=21.0cm
  \epsfbox{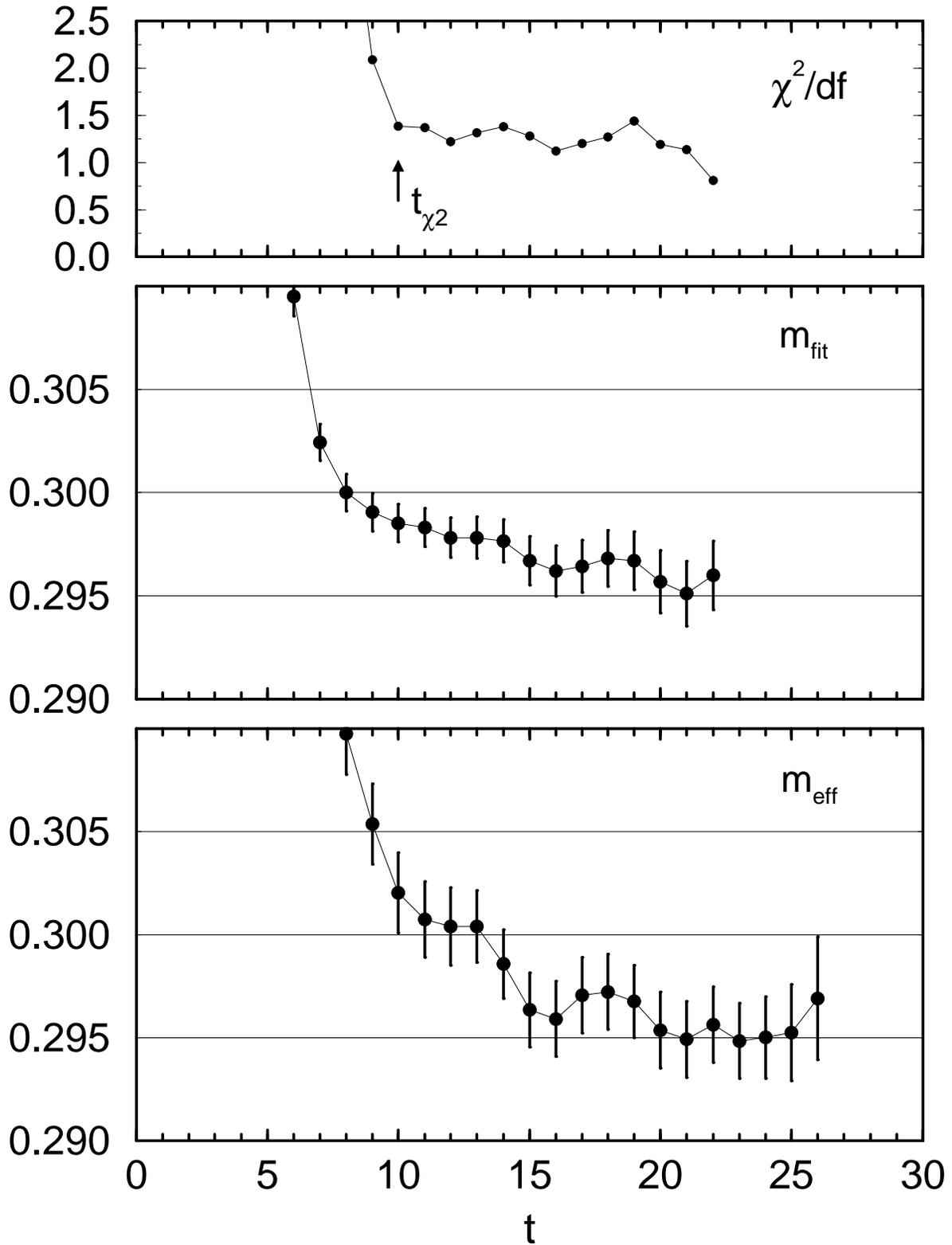}
\end{center}
\vskip -5mm
\caption{
Fitted mass $m_{fit}$ for the pion at 
$\beta=6.0$, $K=0.155$, obtained from one-mass fit 
to a range $t$ --- $T/2$ and the value of $\chi^2/df$
of the fit versus $t$.
The error bars for $m_{fit}$ are statistical uncertainties
estimated by the least mean square fit.
Effective masses $m_{\it eff}$ with errors estimated by
the jack-knife method are also given.}
\label{fig:Chi2-pi}
\end{figure}

\begin{figure}[htbp]
\begin{center}
\leavevmode
\epsfysize=21.0cm
  \epsfbox{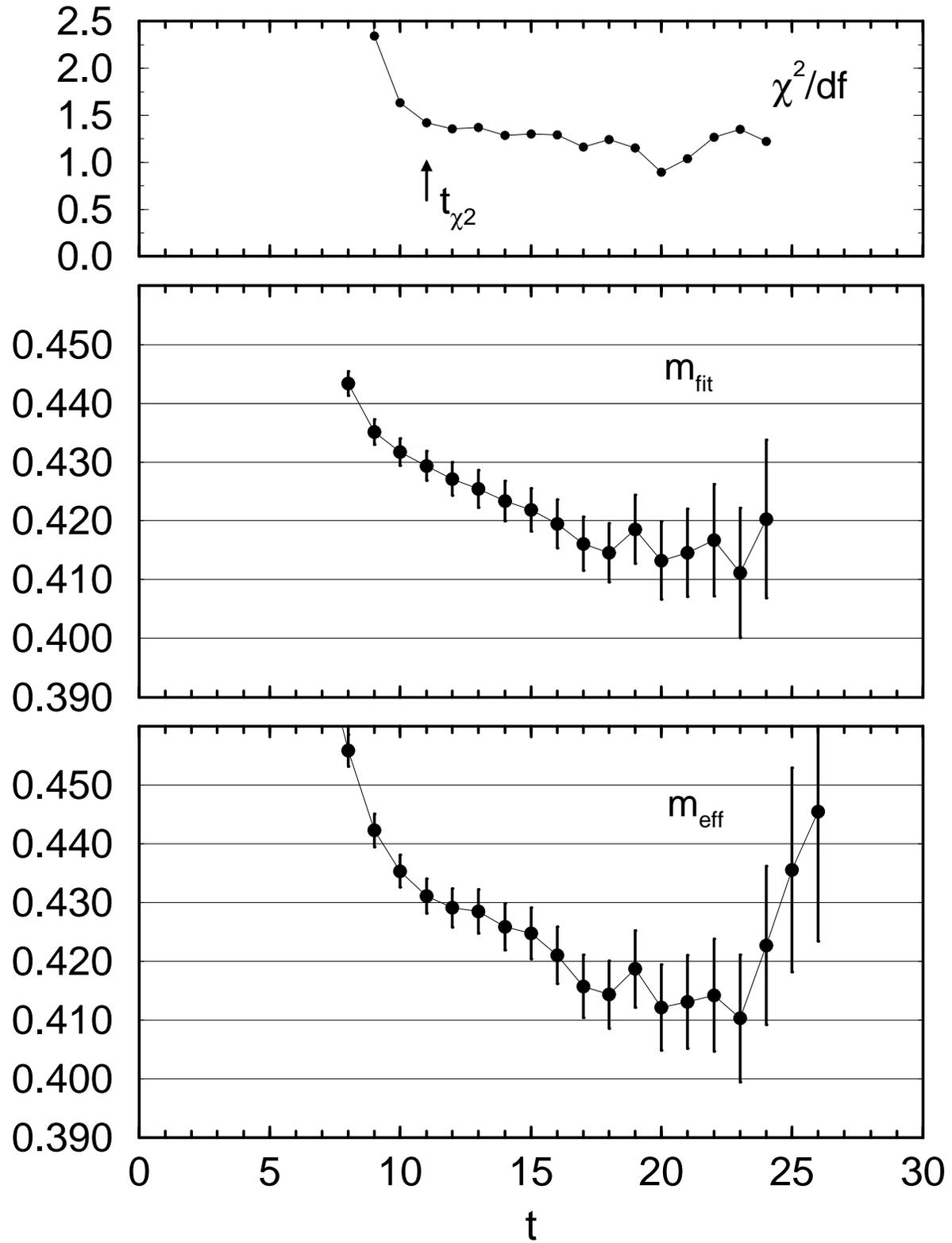}
\end{center}
\vskip -5mm
\caption{
The same as Fig.\ \protect\ref{fig:Chi2-pi}
for the $\rho$ meson.}
\label{fig:Chi2-rho}
\end{figure}

\begin{figure}[htbp]
\begin{center}
\leavevmode
\epsfysize=21.0cm
  \epsfbox{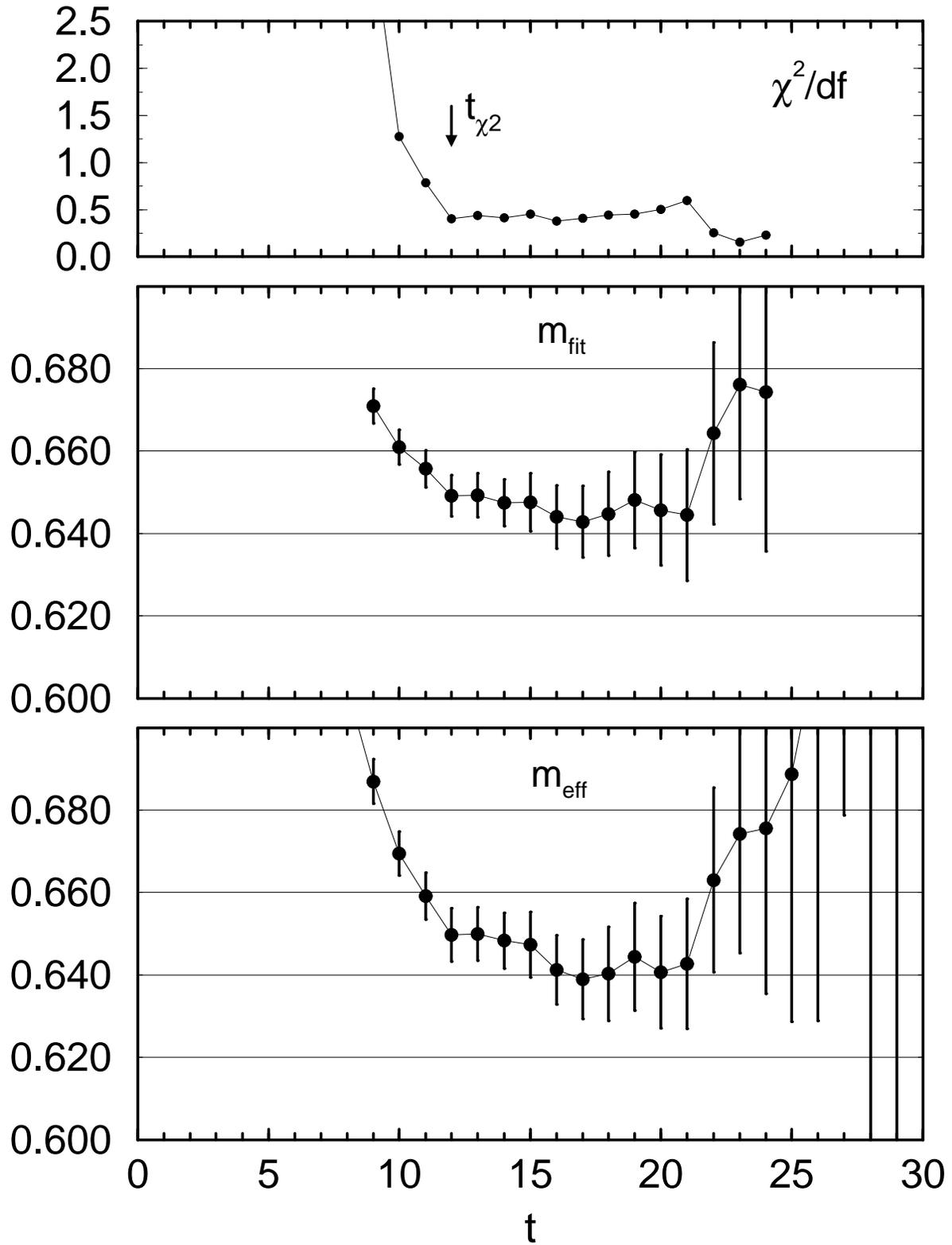}
\end{center}
\vskip -5mm
\caption{
The same as Fig.\ \protect\ref{fig:Chi2-pi}
for the nucleon.}
\label{fig:Chi2-pro}
\end{figure}

In order to obtain a ground state mass,
we have to choose carefully the fitting range 
$t_{min}$ --- $t_{max}$ in such a way that
the contamination from excited states is negligibly small.
We fix $t_{max} = T/2$ in order to take into account the data 
at as large distances as possible.
For the purpose of fixing $t_{min}$,
we make fits to a range $t_0$ --- $T/2$
varying $t_0$ which is a candidate for $t_{min}$.
Then we investigate the $t_0$ dependence 
of the fitted mass $m_{fit}$ and
$\chi^2/df$, $df$ being the number of degrees of freedom,
together with the $t$ dependence of 
the effective mass $m_{\it eff}$ defined by
\begin{equation}
G(t)/G(t+1) = G_0(t,m_{\it eff}(t))/G_0(t+1,m_{\it eff}(t)).
\end{equation}
We plot in Figs.\ \ref{fig:Chi2-pi},
\ref{fig:Chi2-rho} and \ref{fig:Chi2-pro},
as examples, the results for $\chi^2/df$,
$m_{fit}$ and $m_{\it eff}$ at $\beta=6.0$, $K=0.155$
for the pion, the $\rho$ meson and the nucleon, respectively.
Common features of the time slice dependences of $\chi^2/df$,
$m_{fit}$ and $m_{\it eff}$ for all cases
including the other cases which are not shown here
can be summarized as follows.
(Discussion on each particle together with 
a complete set of figures for effective masses
will be given below.)

1) 
When we increase $t_0$ starting from a small value
such as $t_0=4$,
$\chi^2/df$ decreases rapidly from a large value
down to a value around 2.0 --- 0.5 and stabilizes.
We denote $t_0$ where the stabilization starts as $t_{\chi^2}$.
The stabilized value of $\chi^2/df$
depends on particle, $\beta$ and $K$.
In table \ref{tab:chi2} we give $t_{\chi^2}$ and $\chi^2/df$
at $t_{\chi^2}$.
We note that $t_{\chi^2}$ for lighter quarks are smaller 
than those for heavier quarks.
From a point of view of the least mean square fit,
$t_{\chi^2}$ as well as any value of $t > t_{\chi^2}$ are 
candidates for $t_{min}$. 

2) 
Although $m_{\it eff}(t)$ and $m_{fit}(t)$ almost stabilize
around $t \sim t_{\chi^2}$,
a clear long plateau in $m_{\it eff}$ is rarely seen
and the data of $m_{\it eff}$
frequently show large and slowly varying fluctuations 
at large time separations, as shown in the figures. 
If the fitting range is fixed case by case based on a short 
plateau of $m_{\it eff}$,
this may lead to a sizable underestimate 
of statistical errors. 

3)
The value of $m_{\it eff}$ in many cases is
still decreasing at $t \sim t_{\chi^2}$.
Similar phenomena are reported by the UKQCD
collaboration~\cite{UKQCD62-60}.
Although probably the large statistical fluctuations
mentioned above is a partial cause of this phenomenon, 
the possibility that excited states still 
contribute at $t \sim t_{\chi^2}$ 
can not be excluded.
It is difficult to clearly separate out the effects of 
excited states from the statistical fluctuations.

From these considerations, we do not simply 
take $t_{\chi^2}$ as $t_{min}$. 
In order to remove the contamination from exited states 
as much as possible, we proceed in the following way: 
We take $t_{min}$ common to all $K$'s 
for the mesons and for the baryons, respectively,
at each $\beta$, 
in order to avoid a subjective choice case by case.
Therefore, we require $t_{min} \ge t_{\chi^2}$ for all $K$'s. 
We further require that 
$t_{min}$ always lies in a plateau when a clear plateau is seen 
in the effective mass plot. 
In cases where two plateaus are seen
(e.g.\ see Figs.\ \ref{fig:Chi2-pi} --- \ref{fig:Chi2-pro}),
we require that $t_{min}$ is larger than the beginning point 
of the first plateau. 
We also pay attention to the consistency 
between the choices of $t_{min}$ at two $\beta$'s 
in such a way that the ratio of the values of $t_{min}$
is approximately equal to 
that of the lattice spacings at two $\beta$'s.
Thus we have chosen $t_{min}=12$ (15) for mesons
and $t_{min}=13$ (16) for baryons at $\beta=5.85$ (6.0),
respectively.
The ratio of $t_{min}$ at $\beta=5.85$ to
that at $\beta=6.0$ is approximately equal to 
the ratio of the lattice spacings 
$a(\beta=5.85)/a(\beta=6.0) \sim 1.2$.

In addition to statistical errors,
we estimate the systematic error coming
from uncertainties in the choice of fitting 
range~\cite{QCDPAX93,UKQCD62-60}.
Varying $t_0$ from $t_{\chi^2}$ up to $t_{min} +4$,
we estimate the upper (lower) bound for the systematic error 
by the difference between the maximum (minimum) value
and the central value obtained from 
the fit with $t_0 = t_{min}$.
We take $t_0$ only up to $t_{min}+4$, 
because, when $t_0$ is larger than this value, 
data in the fitting range become too noisy. 
(For the $\Delta$ baryon at $\beta=6.0$, 
we vary $t_0$ up to $t_{min}+3$ 
because a fit with $t_0=t_{min}+4=20$ does not converge.)

In this way we estimate
the errors in ground state masses 
due to statistical fluctuations
as well as those due to the possibly remaining contamination
from excited states 
which cannot be properly taken into account by the standard 
least mean square fit with a fixed fitting range. 
Note that the data are consistent with the implicit assumption 
that the ground state dominates for $t \geq t_{min}$ when 
we take into account these systematic errors. 
Consistency of this assumption is also checked 
by a two-mass fit discussed in Sec.~\ref{sec:Excited}.

\subsection{Pion masses}
\begin{figure}[htbp]
\begin{center}
\leavevmode
\epsfysize=21.0cm
  \epsfbox{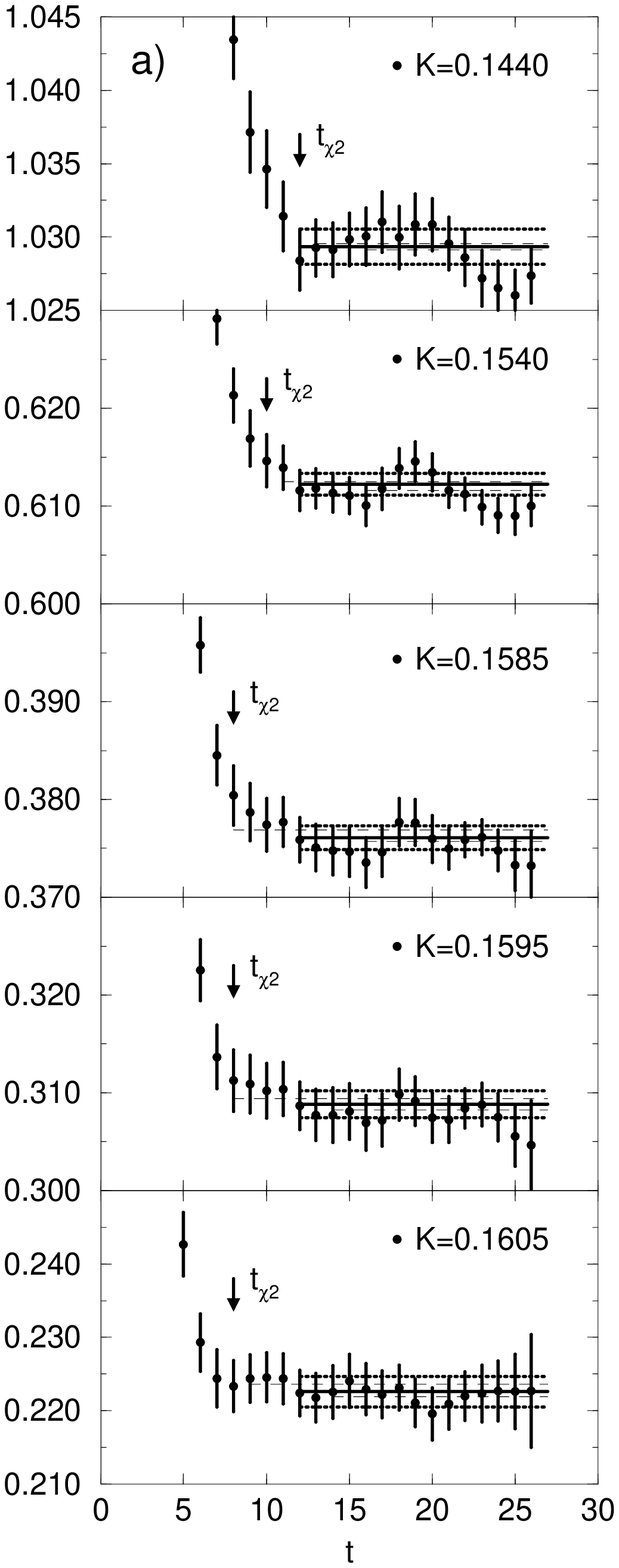}
\leavevmode
\epsfysize=21.0cm
  \epsfbox{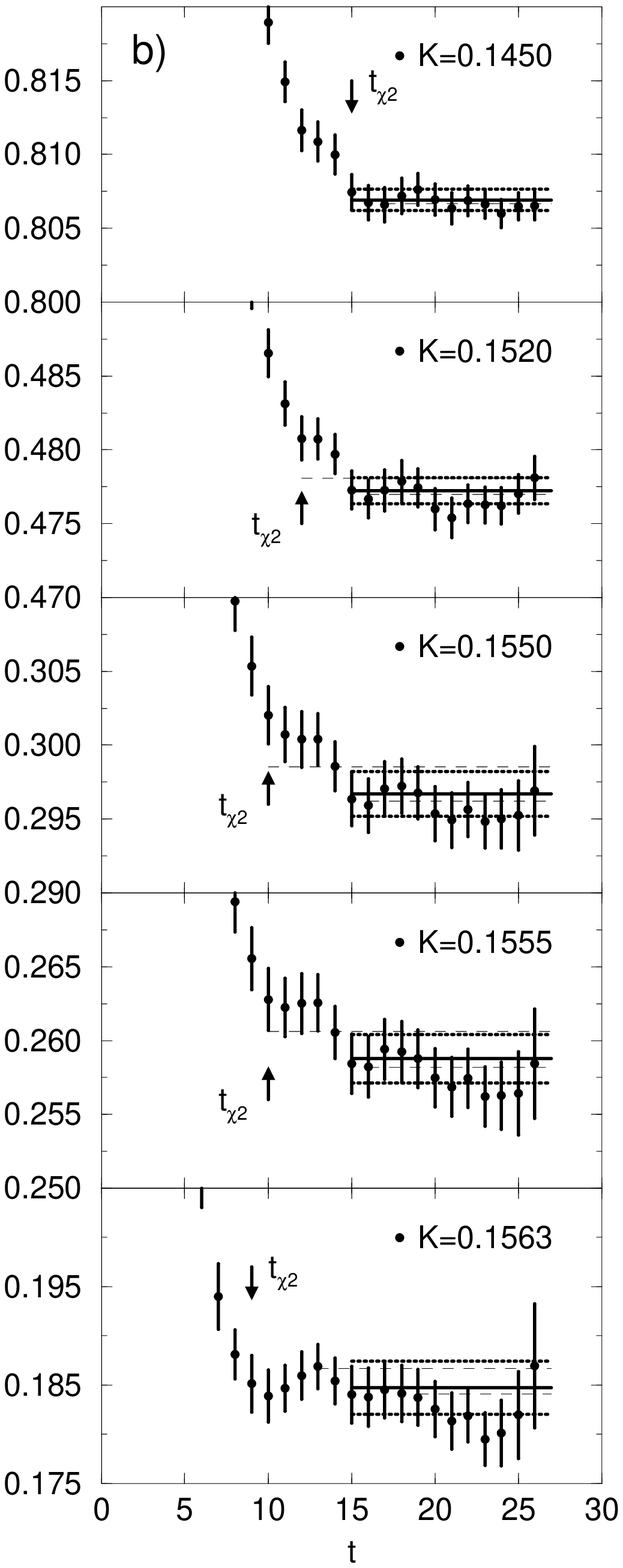}
\end{center}
\vskip -5mm
\caption{
Effective masses for the pion:
a) $\beta=5.85$, b) $\beta=6.0$.
The result of one-mass fit is reproduced by the solid line,
dotted lines and dashed lines for the fitted mass,
its statistical error and systematic upper/lower bounds,
respectively.}
\label{fig:B585-B600-Pi}
\end{figure}

We show $m_{\it eff}$ at $\beta=5.85$ and $\beta=6.0$
in Fig.\ \ref{fig:B585-B600-Pi}.
The pion effective mass has structure 
with the scale of the standard deviation
even for $t \ge t_{\chi^2}$:
In some cases $m_{\it eff}(t)$ exhibits
two-plateau structure or slow monotonic decrease.
However, the magnitude of the fluctuation for the pion is
much smaller than the other cases.
The resulting systematic error is comparable 
to the statistical uncertainty.
The results of the fits are given in table \ref{tab:mpi}.

\subsection{Rho meson masses}
Fitting to the $\rho$ meson propagator is 
more problematic than the pion propagator.
Because of this, we will discuss it at some length 
and compare the results with previous works.

\begin{figure}[htbp]
\begin{center}
\leavevmode
\epsfysize=21.0cm
  \epsfbox{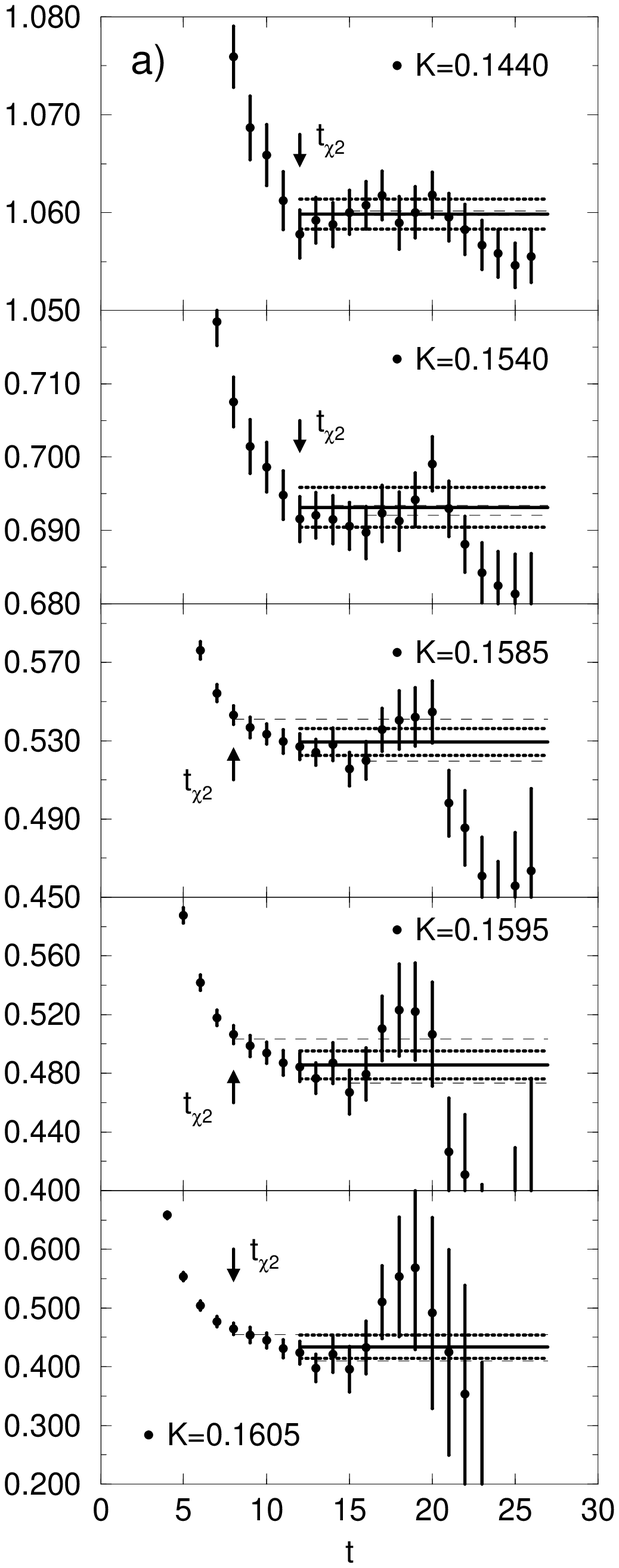}
\leavevmode
\epsfysize=21.0cm
  \epsfbox{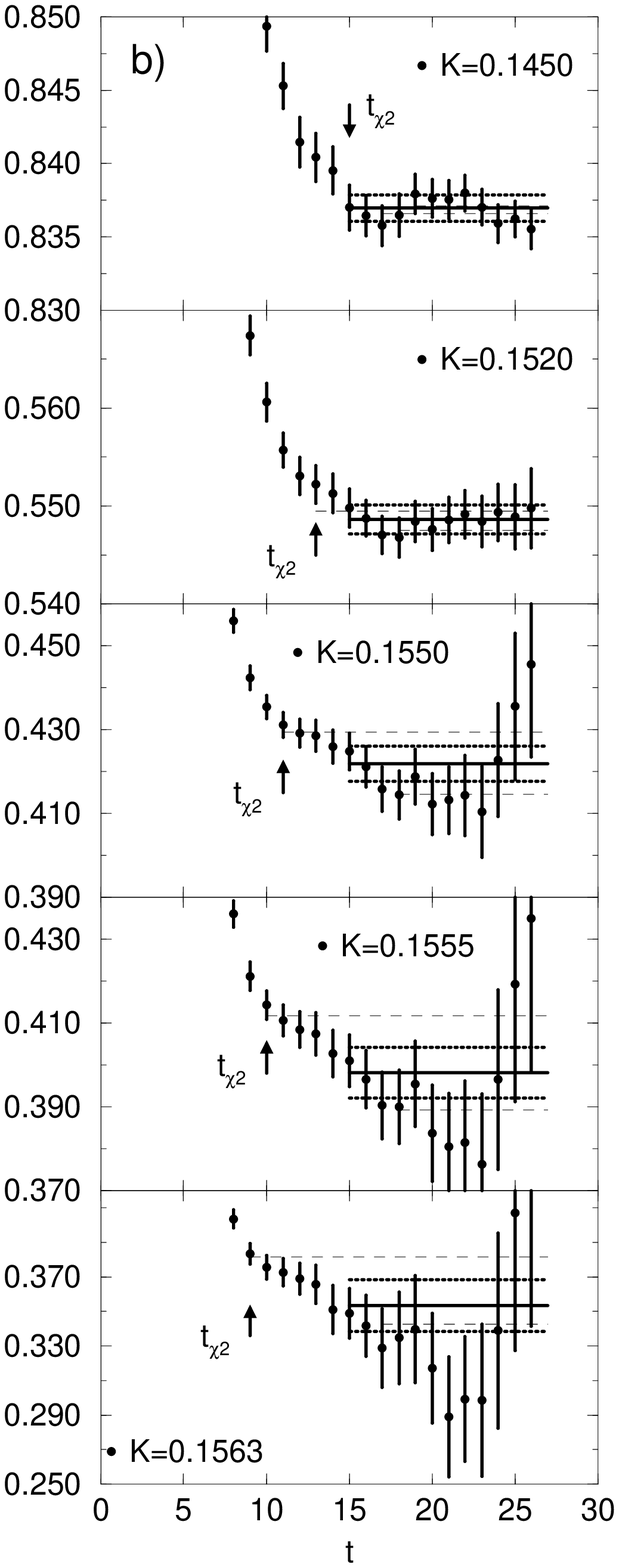}
\end{center}
\vskip -5mm
\caption{
The same as Fig.\ \protect\ref{fig:B585-B600-Pi}
for the $\rho$ meson: a) $\beta=5.85$, b) $\beta=6.0$.}
\label{fig:B585-B600-Rho}
\end{figure}

The $\rho$ meson effective mass at $\beta=5.85$ 
shown in Fig.\ \ref{fig:B585-B600-Rho}-a exhibits 
a plateau for $t \ge t_{\chi^2}=12$ 
for the smallest two $K$'s,
while it exhibits peculiar behavior
at large $t$ for the largest three $K$'s:
$m_{\it eff}(t)$ for $t=17$ --- 20 is larger than 
that for $t=12$ --- 16
and it drops abruptly at $t=21$.
We regard this behavior as due to statistical fluctuations.
We find that fits to a range $t=12$ --- $t_{max}$ are 
stable for $t_{max}=$ 14 --- 27. 
Therefore we choose $t_{max}=T/2$ even for these cases. 
The results of the fits are summarized 
in table \ref{tab:mrho}.
The systematic error upper bound is 1 --- 2 times larger 
than the statistical error for the largest three $K$'s.

Fig.\ \ref{fig:B585-B600-Rho}-b shows 
the effective mass at $\beta=6.0$.
Except for the smallest $K$, 
$m_{\it eff}(t)$ is decreasing at $t \sim t_{\chi^2}$.
Rate of the decrease becomes slow at $t\sim 12$ 
to exhibit a plateau for two or three time slices.
The value of $m_{\it eff}$ decreases further 
up to $t \sim 17$ to attain another plateau.
The plateaus are not long enough to determine unambiguously
the time slice where the contribution of excited states
can be ignored.
It should be emphasized again that 
$\chi^2/df$ are almost identical both for
the fits with $t_{min}=12$ and $t_{min}=17$:
1.35 and 1.16 for $K=0.1550$, 1.20 and 1.13 for $K=0.1555$ and
0.77 and 0.76 for $K=0.1563$, respectively.
See also Fig.\ \ref{fig:Chi2-rho}.
Therefore the value of $\chi^2$ 
does not give a guide to determine $t_{min}$.
The point $t_{min}=15$ is located between 
the two pseudo-plateaus at $t\sim 12$ and $t\sim 17$.
In table \ref{tab:mrho} are summarized the results for 
the fits with $t_{min}=15$ together with the systematic error.
Reflecting the slow monotonic decrease of effective masses,
the ratio of the systematic error to the statistical error 
is relatively large:
the systematic error amounts to about twice 
the statistical error for the largest three $K$'s.

We notice a very intriguing fact that $m_{fit}$ 
by the correlated fits to a range from $t=t_0$ to $T/2$
has a strong correlation with $m_{\it eff}$ at $t=t_0$.
A typical example is seen in Fig.\ \ref{fig:Chi2-rho}.
This holds for the other particles also.
This means that the result of the fit 
to a range $t_0$ --- $T/2$ is mainly 
determined from data at $t \sim t_0$.

In our previous work~\cite{QCDPAX92},
we analyzed the same set of $\rho$ meson propagators 
with uncorrelated fits.
Paying attention to the monotonic decrease of effective masses,
we made two different fits to estimate the systematic error 
coming from uncertainties in the choice of fitting range.
One is a fit to a range $t\sim$ 9 --- 11 at $\beta=5.85$ 
($t\sim$ 12 --- 15 at $\beta=6.0$).
We called the fit ``pre-plateau fit''.
Another is a fit to a range $t\sim$ 11 --- $t_{max}$
at $\beta=5.85$ ($t\sim$ 15 --- $t_{max}$ at $\beta=6.0$),
which we called ``plateau fit''.
The latter fitting ranges correspond approximately
to those we adopt in this work.
Because $m_{\it eff}$ is decreasing,
the $\rho$ masses obtained from the correlated fits are 
systematically larger than those from the uncorrelated fits,
due to the fact give in the preceding paragraph.
The mass value obtained in this work
is between that from the uncorrelated plateau fit 
and that from the pre-plateau fit.

In table \ref{tab:mK15} we reproduce the results 
for the $\rho$ meson masses at $\beta=6.0$ 
for $K=0.155$ and 0.1563 together with those 
by the Ape collaboration~\cite{APE18,APE24}
and the LANL group~\cite{LANL}.
In 1991, the Ape collaboration reported the result
obtained on a $24^3 \times 32$ lattice 
with a multi-origin $7^3$ cubic source~\cite{APE24}.
Then we made simulations for the same spatial size
with larger temporal extent~\cite{QCDPAX92},
$24^3 \times 54$, using the point source.
For $K=0.155$, the values of $m_{\it eff}$ at $t \sim 10$ are 
in close agreement with Ape's.
Consequently the result 0.4280(33) 
obtained from the pre-plateau fit ($t =$ 12 --- 15) 
agreed with the Ape result 0.429(3) 
within one standard deviation.
However, the result 0.4169(48) from the plateau fit 
($t =$ 15 --- 27) was smaller by approximately 
twice the statistical error.
We regarded the latter more reliable.
At that time there was a report that 
the mass value appears to depend on 
the type of the source adopted~\cite{LANLsmear}.
Therefore, in order to clarify 
whether the origin of the discrepancy 
between our result and the Ape result 
is due to the different type of source,
we made calculation at $K=0.155$ 
for 400 configurations~\cite{QCDPAX93}
using the point source, the wall source and the source 
adopted by the Ape collaboration.
The results obtained from correlated fits 
for the three different sources 
have agreed with each other; 
0.4201(29), 0.4228(19) and 0.4249(19)
for the point source, the wall source and the multi-origin source,
respectively.
The recent result reported by the LANL group 
0.422(3)~\cite{LANL} is consistent with these numbers.
It is probable that the slightly large value
by the Ape collaboration is due to small temporal extent.
The Ape collaboration has also made simulations 
using both the point source and the multi-cube
source~\cite{APE18}
with larger temporal size and smaller spatial size:
$18^3 \times 64$.
Their results 0.430(10) and 0.428(8) are 
consistent with other results within relatively large errors, 
although the central values are slightly higher than 
the results by other groups.
The slightly large central values may be 
due to the small spatial size.
For $K=0.1563$, the results obtained from the correlated fit 
in this work is consistent with those by the Ape collaboration
and the LANL group,
albeit with large errors in the results.

\subsection{Baryon masses}\label{sec:baryon}
\begin{figure}[htbp]
\begin{center}
\leavevmode
\epsfysize=21.0cm
  \epsfbox{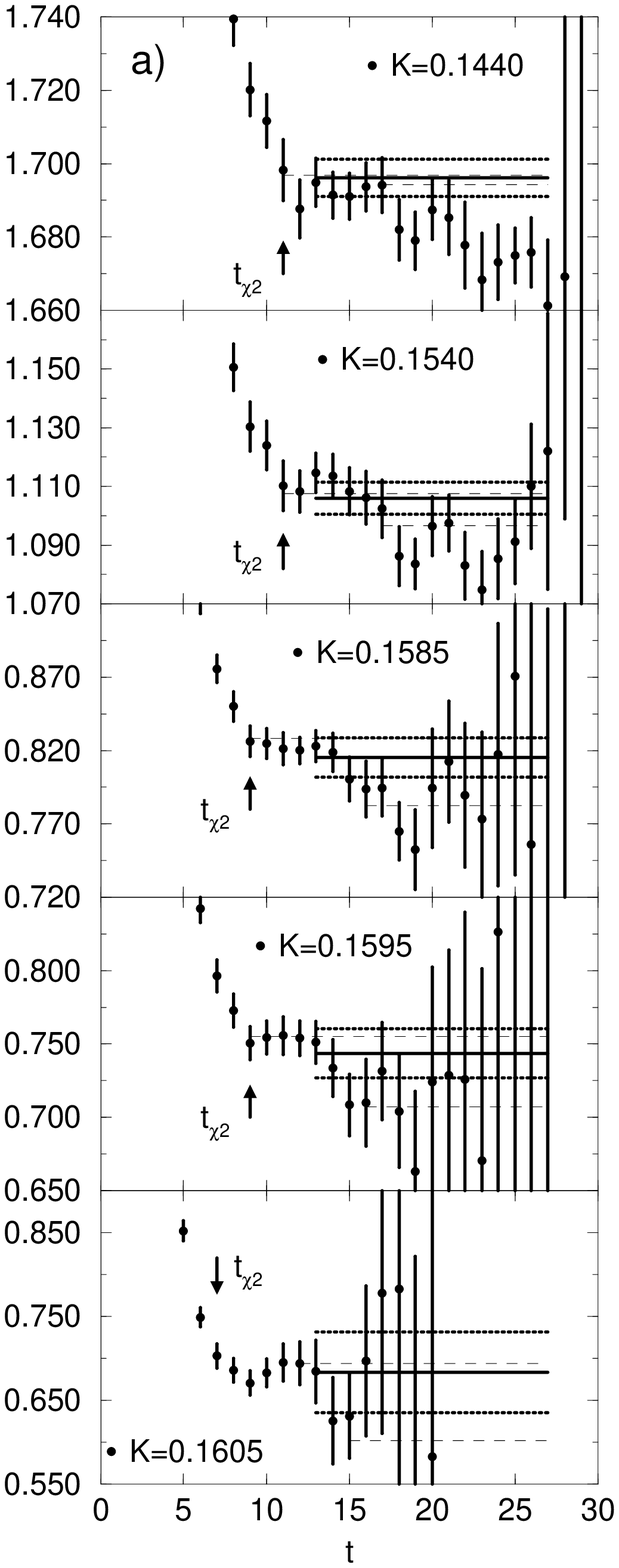}
\leavevmode
\epsfysize=21.0cm
  \epsfbox{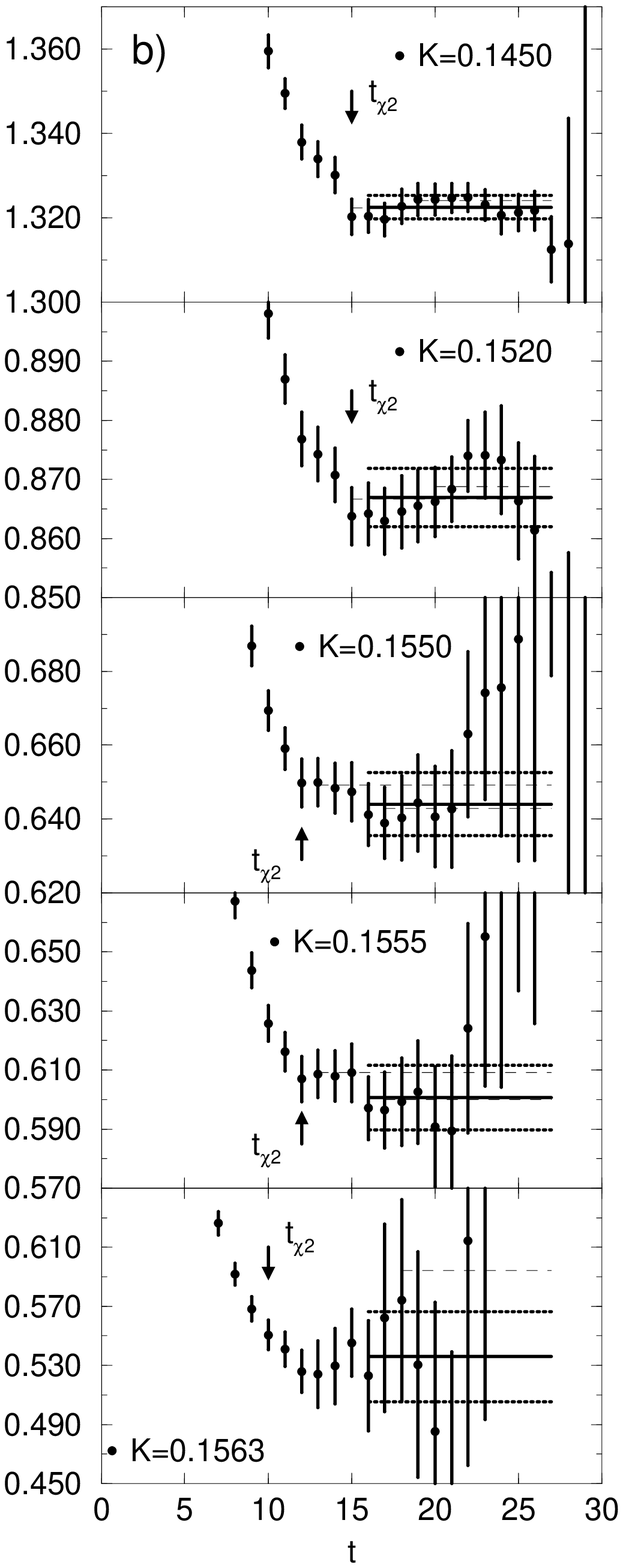}
\end{center}
\vskip -5mm
\caption{
The same as Fig.\ \protect\ref{fig:B585-B600-Pi}
for the nucleon: a) $\beta=5.85$, b) $\beta=6.0$.}
\label{fig:B585-B600-Pro}
\end{figure}

Fig.\ \ref{fig:B585-B600-Pro} shows effective masses
for the nucleon at $\beta=5.85$ and $\beta=6.0$.
Decrease of $m_{\it eff}$ at $t \sim t_{\chi^2}$
is not conspicuous compared with the case of the $\rho$ meson.
However, we see two-plateau structure for the cases of
$K=0.1585$ and 0.1595 at $\beta=5.85$
and $K=0.155$ (see also Fig.\ \ref{fig:Chi2-pro})
and 0.1555 at $\beta=6.0$. 
The choice $t_{min}=13\ (16)$ for $\beta=5.85\ (6.0)$ 
corresponds to that we select the first (last) plateau as correct 
for the case where two plateaus are observed.
Table \ref{tab:mN} summarizes the results of the fits.

\begin{figure}[htbp]
\begin{center}
\leavevmode
\epsfysize=21.0cm
  \epsfbox{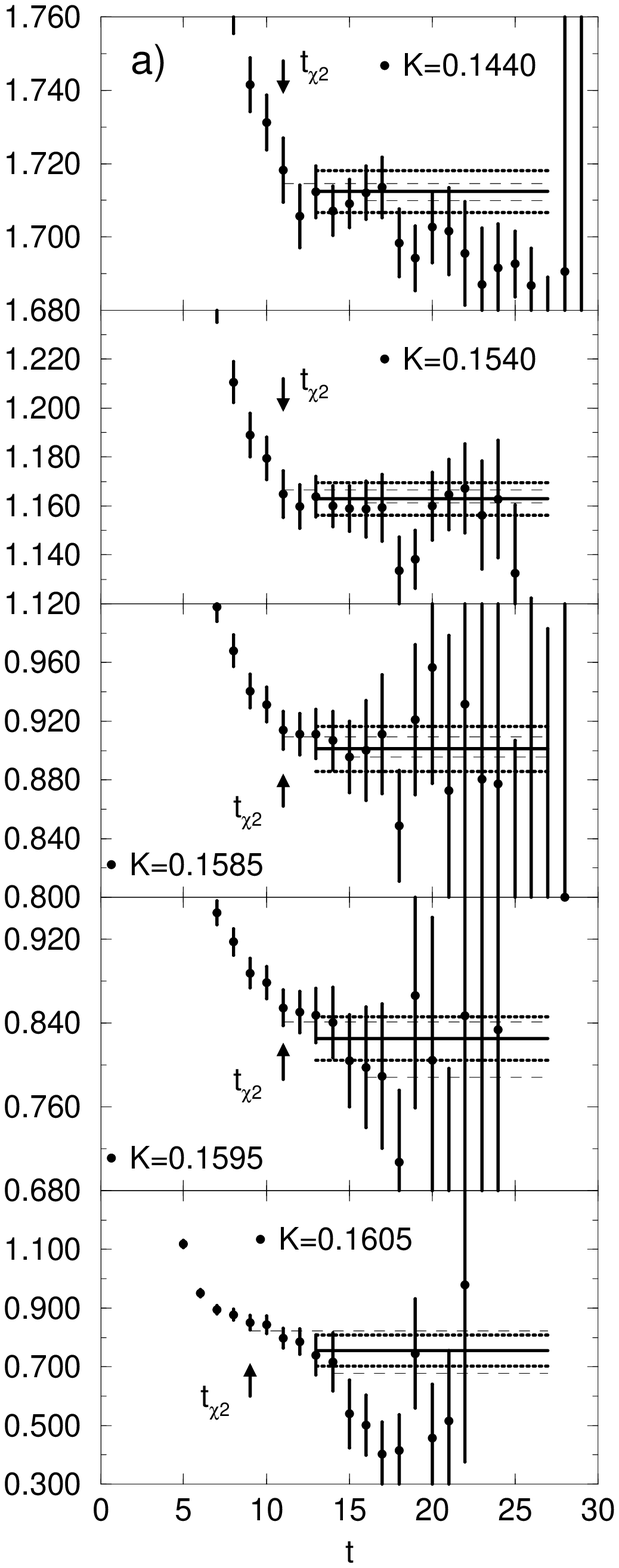}
\leavevmode
\epsfysize=21.0cm
  \epsfbox{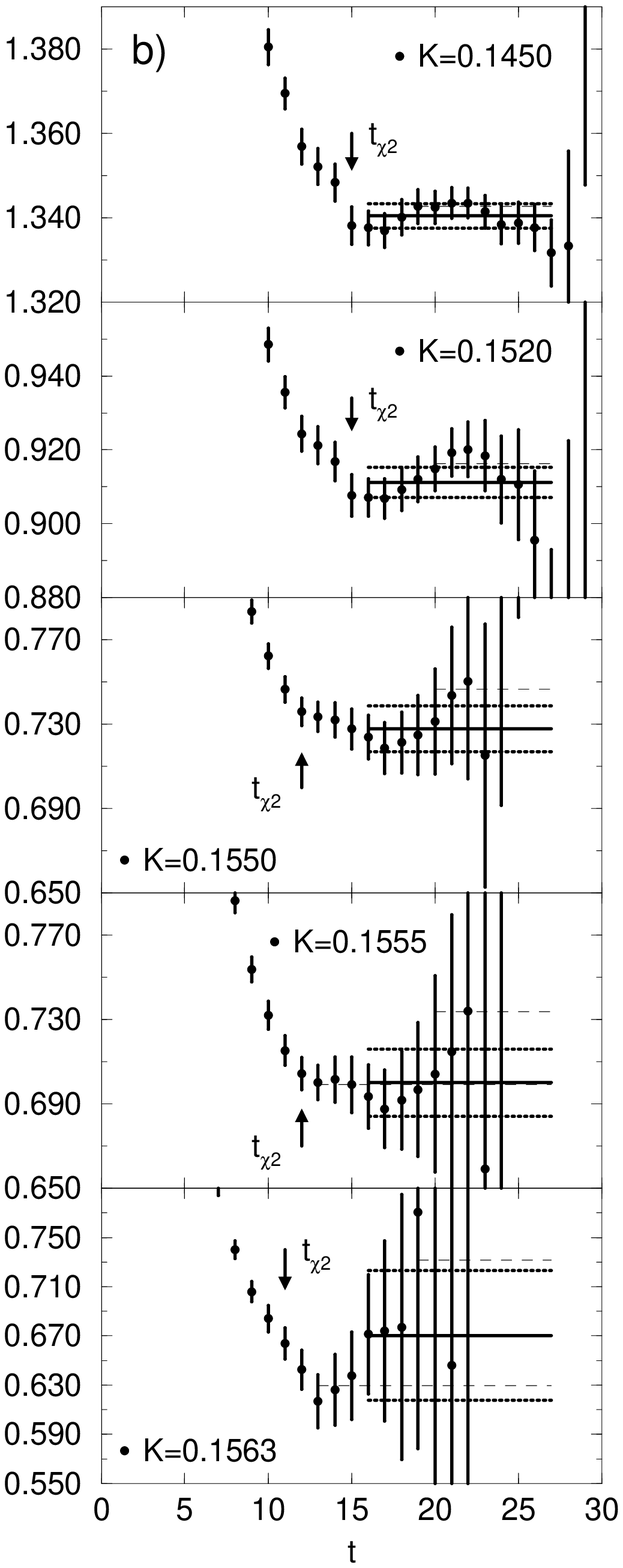}
\end{center}
\vskip -5mm
\caption{
The same as Fig.\ \protect\ref{fig:B585-B600-Pi}
for the $\Delta$ baryon: a) $\beta=5.85$, b) $\beta=6.0$.}
\label{fig:B585-B600-Del}
\end{figure}

For $\Delta$, monotonic decrease of effective masses 
at $t \sim t_{\chi^2}$ or two-plateau structure is seen
for $K=0.1595$ and 0.1605 at $\beta=5.85$ and 
for $K=0.1550$ and 0.1563 at $\beta=6.0$.
Effective mass plots are shown in 
Fig.\ \ref{fig:B585-B600-Del}.
The results of the fits are summarized in table \ref{tab:mD}.

In table \ref{tab:mK15}, the baryon masses at $\beta=6.0$
for $K=0.155$ and 0.1563 together with those 
by the Ape collaboration and the LANL group are reproduced.
The nucleon masses reported by the three groups agree within
the statistical uncertainties.
The $\Delta$ masses for $K=0.155$ are slightly scattered:
Our result is higher than the LANL result by 
two standard deviations.
However note that the values of the $\Delta$ mass
obtained on 400 configurations \cite{QCDPAX93}
(0.7054(95), 0.7008(57) and 0.7128(191) 
for the point source, the wall source and the multi-origin source,
respectively)
are in good agreement with the LANL result.
Therefore we think that the difference between the LANL result
and our present result is due to statistical errors.

\subsection{Finite lattice effects}
The linear extension of the lattice in the spatial directions
is 2.45 (2.03) fm at $\beta$ = 5.85 (6.0), 
when we use $a^{-1} = 1.93$ (2.33) GeV
determined from $m_\rho$
(see Sec.~\ref{sec:Chiral}).
These values are much larger than twice the electromagnetic radius 
of the nucleon, $2\times$ 0.82 fm.
We also note that our results on the lattice 
with spatial volume $24^3$
agree well with those on a lattice with $32^3$~\cite{LANL},
as discussed above.
Therefore we do not take into account in this work
finite lattice effects which are supposed to be small.

\subsection{Mass ratios}
\begin{figure}[htbp]
\begin{center}
\leavevmode
\epsfysize=12cm
  \epsfbox{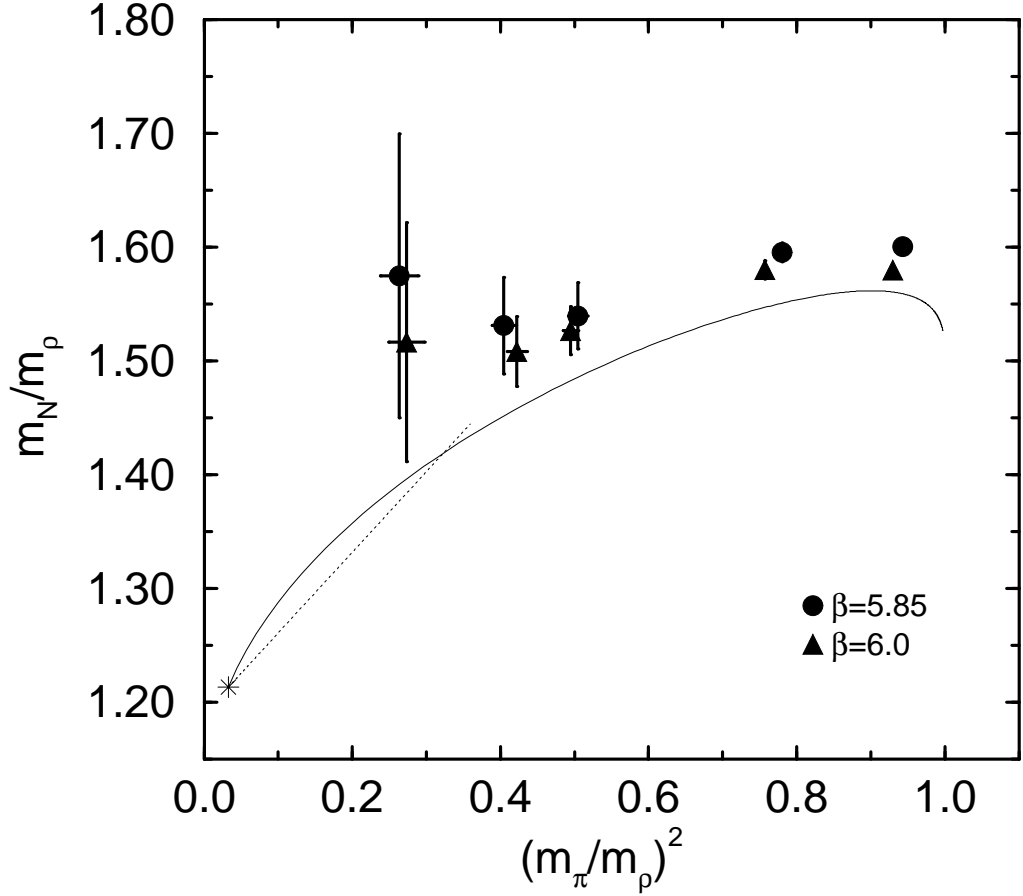}
\end{center}
\vskip -5mm
\caption{
Nucleon to $\rho$ mass ratio versus 
pion to $\rho$ mass ratio squared.
The errors shown are statistical only.
The solid curve is obtained from phenomenological
mass formulae\protect\cite{ONO}.
The dotted line is obtained by assuming that
$m_N/m_{\rho}$ and $(m_{\pi}/m_{\rho})^2$ are 
linear in the quark mass.
The experimental value is marked with a star.}
\label{fig:Mass-Rat}
\end{figure}

The mass ratio $m_{N}/m_{\rho}$ is plotted versus 
$(m_{\pi}/m_{\rho})^2$ in Fig.\ \ref{fig:Mass-Rat}.
The values of the mass ratio
are given in table \ref{tab:ed}.
The value of $m_{N}/m_{\rho}$ at $\beta=6.0$ is 
systematically smaller than that at $\beta=5.85$,
although the results at two $\beta$'s agree 
within the statistical uncertainty 
except for the case of the heaviest quark
(($m_{\pi}/m_{\rho})^2 \sim 0.94$).

\section{Excited State Masses}\label{sec:Excited}
\begin{figure}[htbp]
\begin{center}
\leavevmode
\epsfysize=18cm
  \epsfbox{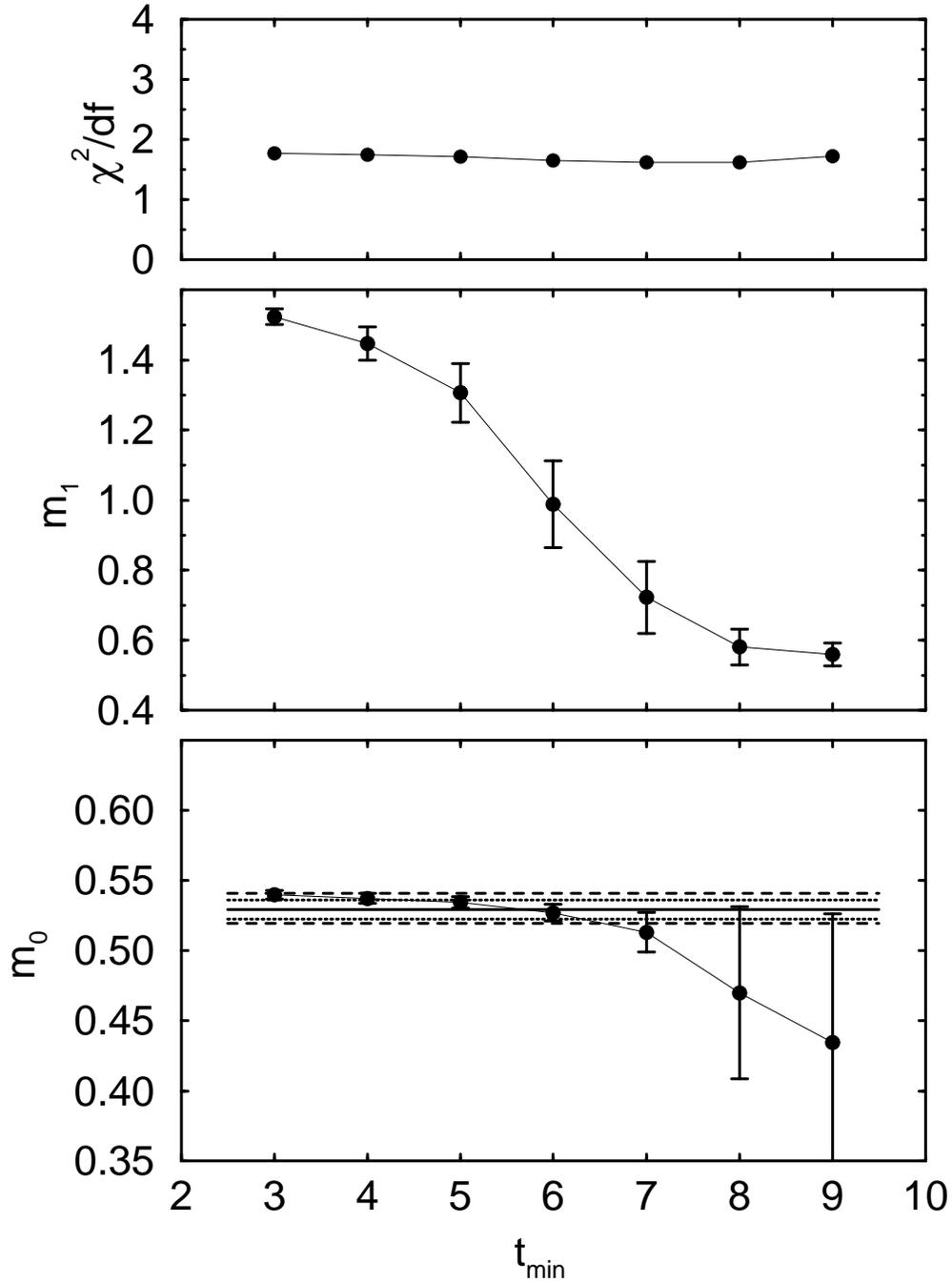}
\end{center}
\vskip -5mm
\caption{
Masses of the ground state and the excited state 
for the $\rho$ meson at $\beta=5.85$, $K=0.1585$
together with the value of $\chi^2/df$ of the two-mass fits
versus $t_{min}$.
The error bars are statistical uncertainties estimated by
the least mean square fit.
The result of the one-mass fit is reproduced by the solid line,
dotted lines and dashed lines for the fitted mass,
its statistical error and systematic upper/lower bounds,
respectively.
Note the difference in the scale of the plots 
for $m_0$ and $m_1$.}
\label{fig:B585-rho-ex}
\end{figure}

\begin{figure}[htbp]
\begin{center}
\leavevmode
\epsfysize=18cm
  \epsfbox{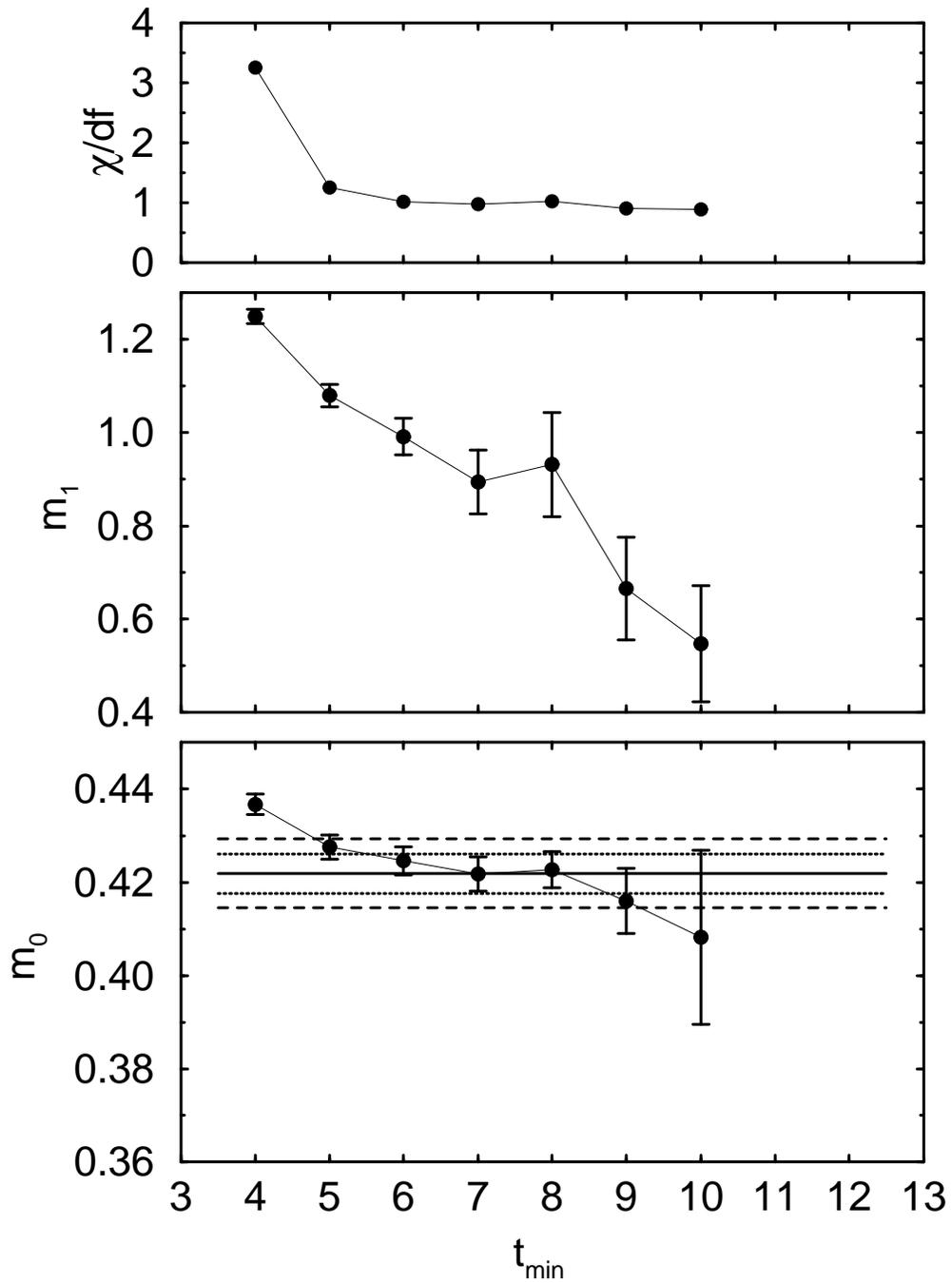}
\end{center}
\vskip -5mm
\caption{
The same as Fig.\ \protect\ref{fig:B585-rho-ex}
for the $\rho$ meson at $\beta=6.0$, $K=0.155$.}
\label{fig:B600-rho-ex}
\end{figure}

\begin{figure}[htbp]
\begin{center}
\leavevmode
\epsfysize=18cm
  \epsfbox{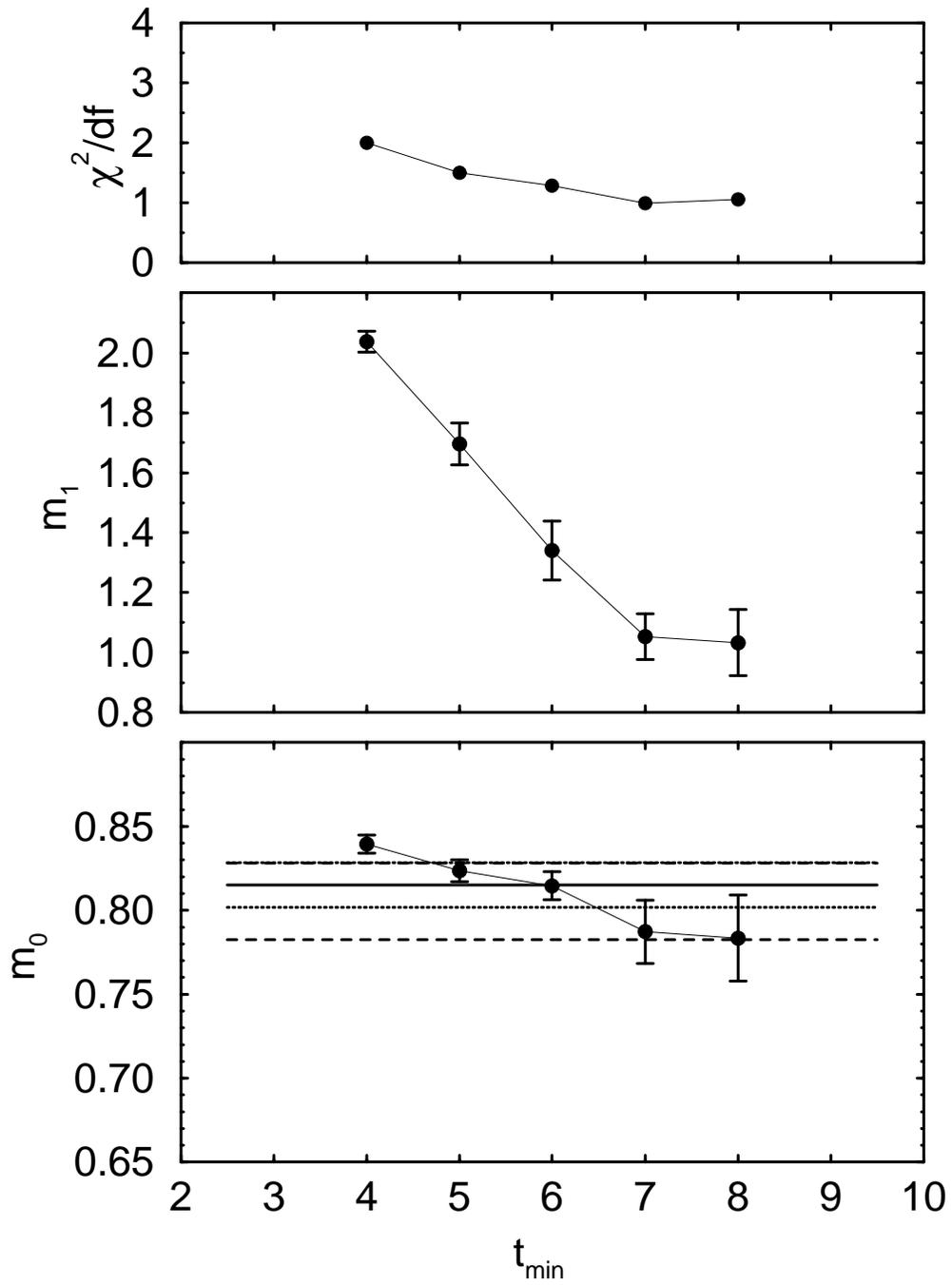}
\end{center}
\vskip -5mm
\caption{
The same as Fig.\ \protect\ref{fig:B585-rho-ex}
for the nucleon at $\beta=5.85$, $K=0.1585$.}
\label{fig:B585-pro-ex}
\end{figure}

\begin{figure}[htbp]
\begin{center}
\leavevmode
\epsfysize=18cm
  \epsfbox{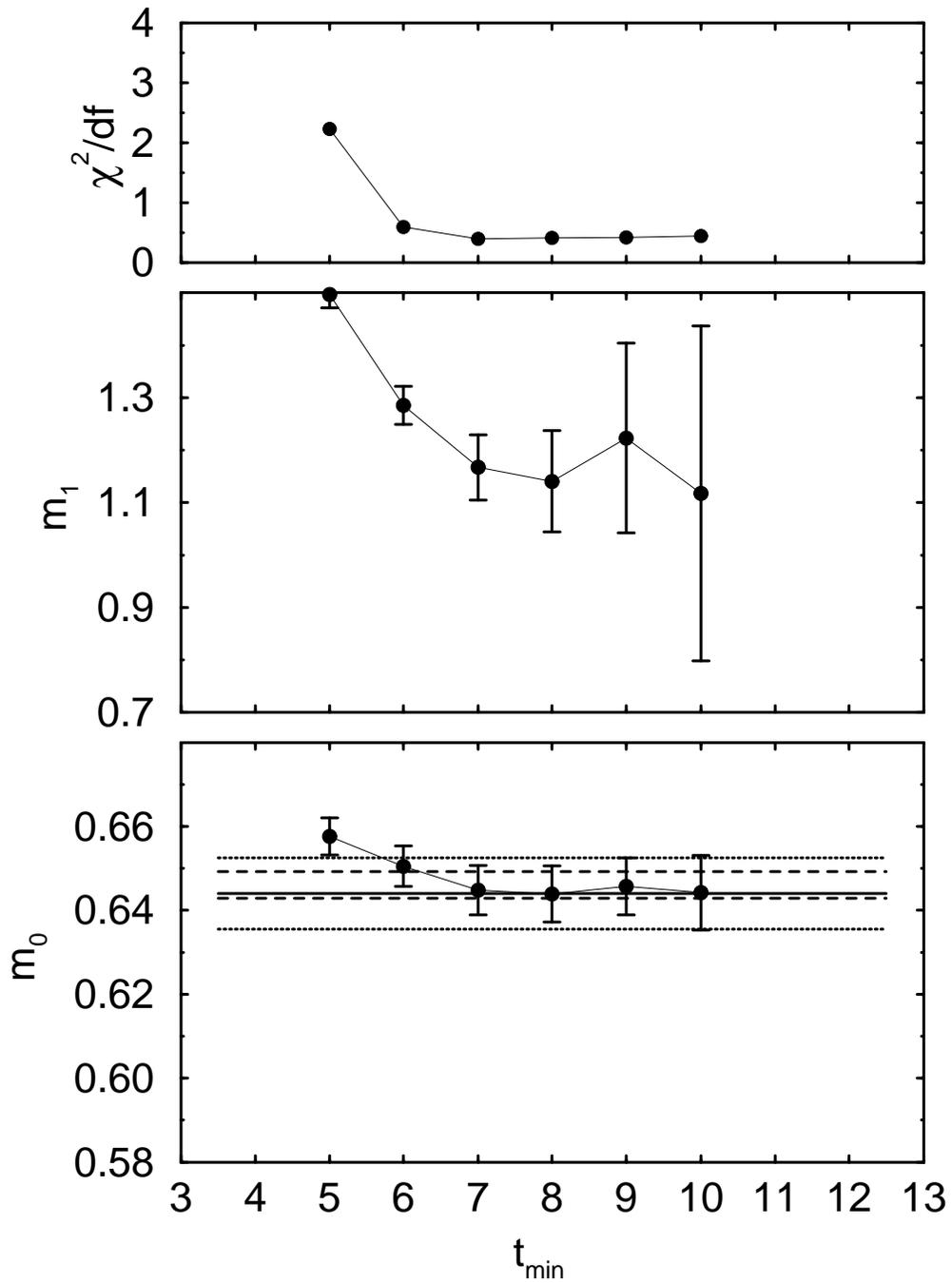}
\end{center}
\vskip -5mm
\caption{
The same as Fig.\ \protect\ref{fig:B585-rho-ex}
for the nucleon at $\beta=6.0$, $K=0.155$.}
\label{fig:B600-pro-ex}
\end{figure}

In addition to the masses of ground states,
we study the masses of first excited states
for the $\rho$ meson and the nucleon.
To this end, we perform two-mass fits 
to the corresponding propagators varying $t_{min}$.
Our results for the $\rho$ meson are shown 
in Fig.\ \ref{fig:B585-rho-ex} for $\beta=5.85$, $K=0.1585$, 
and in Fig.\ \ref{fig:B600-rho-ex} for $\beta=6.0$, $K=0.155$. 
The results for the nucleon are given 
in Figs.\ \ref{fig:B585-pro-ex} and \ref{fig:B600-pro-ex}
for $\beta=5.85$ and 6.0, respectively. 
We find the following:

1) $\chi^2/df$ is stable and small ($\sim 1$ --- 2) 
for $t_{min} \ge 4\ (5)$ in the case of the $\rho$ meson
and for $t_{min} \ge 5\ (6)$ 
in the case of the nucleon at $\beta=5.85\ (6.0)$, respectively.

2) 
When $\chi^2/df$ is small, 
the ground state masses $m_0$ from the two-mass fit are
consistent with those from the one-mass fit within the errors, 
although the errors for $m_0$ from the two-mass fit 
become extremely large at large $t_{min}$. 

3) Although $\chi^2/df$ is stable, 
the mass of the first excited state $m_1$ is 
in general quite unstable.
For example, for the $\rho$ meson at $\beta=5.85$, 
the value of $m_1$ decreases from 1.5 for $t_{min}=3$ 
to 0.6 for $t_{min}=9$ (cf.\ Fig.\ \ref{fig:B585-rho-ex}). 
Similar behavior is also seen in the results 
for the $\rho$ meson at $\beta=6.0$ 
(Fig.\ \ref{fig:B600-rho-ex}) 
and the nucleon at $\beta=5.85$ 
(Fig.\ \ref{fig:B585-pro-ex}).
The case of the nucleon at $\beta=6.0$ is exceptional:
$m_1$ is relatively stable 
(Fig.\ \ref{fig:B600-pro-ex}).

Under these circumstances, we select two $t_{min}$'s which 
give $m_0$ consistent with the result of the one-mass fit, 
under the condition that the errors are small.
We then investigate whether the results 
for the excited state mass are consistent 
with the corresponding experimental values.

\begin{figure}[t]
\begin{center}
\leavevmode
\epsfysize=9cm
  \epsfbox{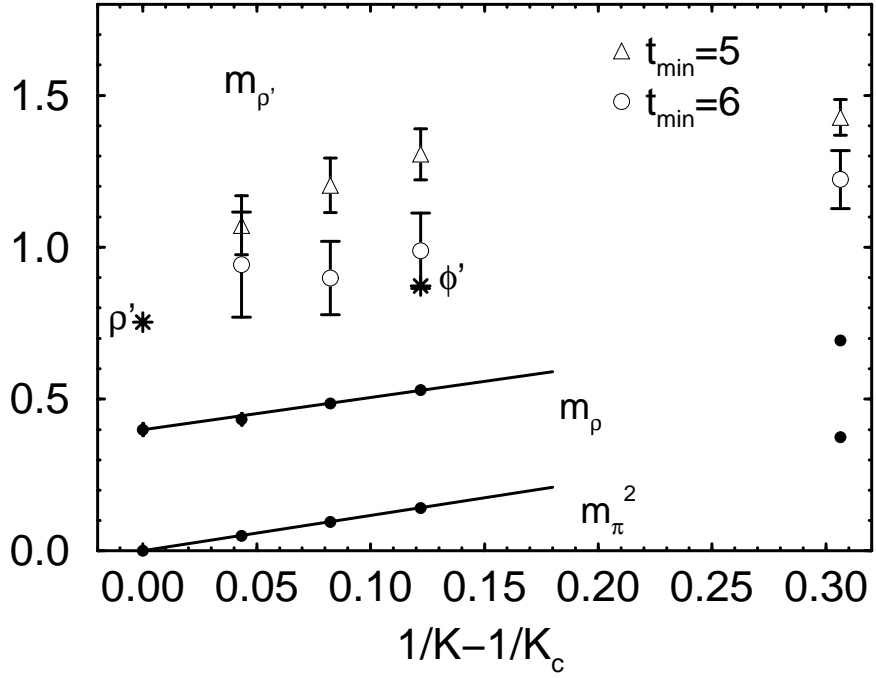}
\end{center}
\vskip -5mm
\caption{
Mass of the excited state of the $\rho$ meson 
(denoted by $\rho'$) at $\beta=5.85$ versus $1/K-1/K_c$.
The corresponding experimental values are  marked with stars.
The data for $m_{\rho}$ and $m_{\pi}^2$ are 
taken from the results of one-mass fits.
With the scale of the plot, the results for $m_{\rho}$ from 
two-mass fits are indistinguishable from the one-mass fit results.}
\label{fig:B585-rho-1K}
\end{figure}

\begin{figure}[b]
\begin{center}
\leavevmode
\epsfysize=9cm
  \epsfbox{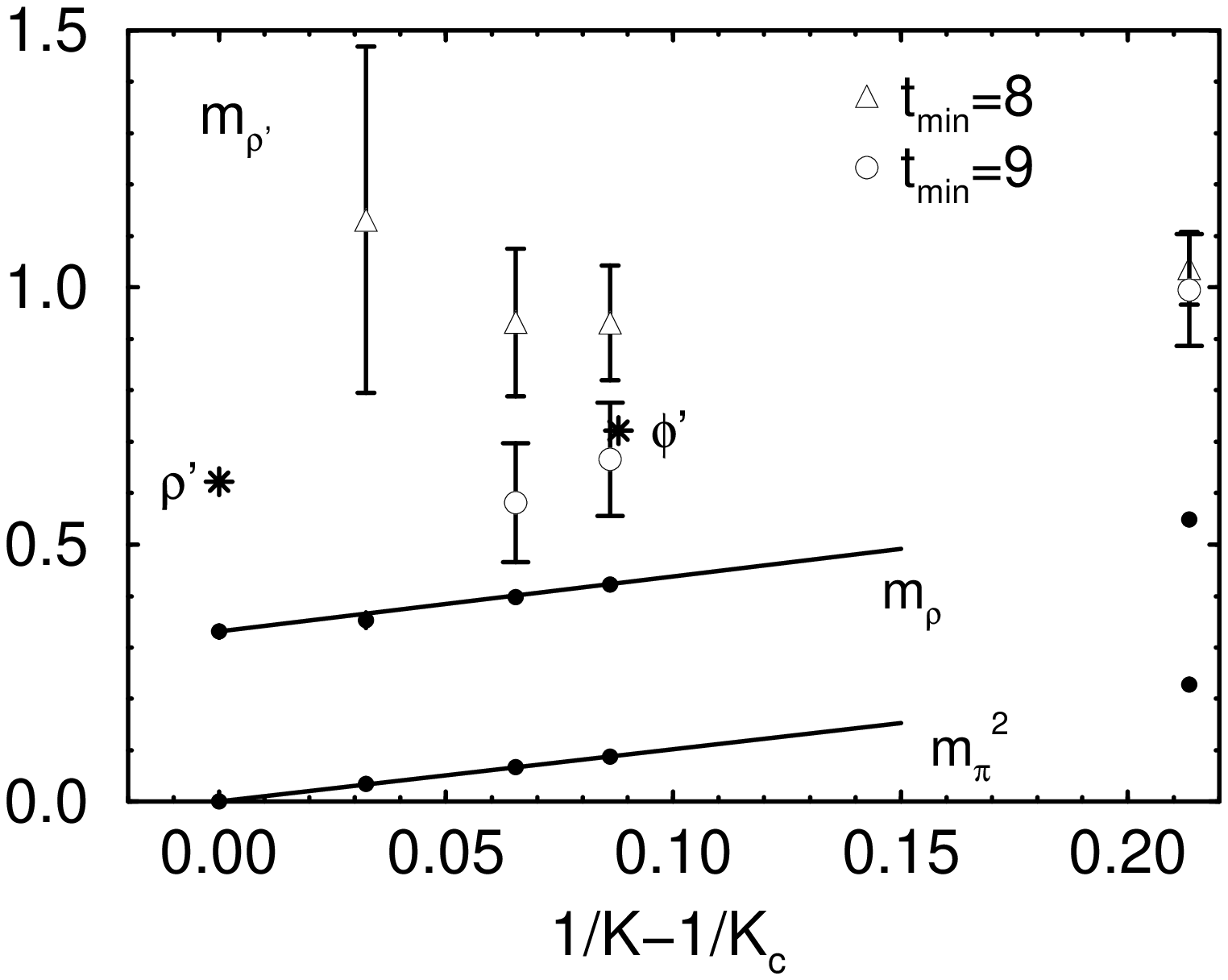}
\end{center}
\vskip -5mm
\caption{
The same as Fig.\ \protect\ref{fig:B585-rho-1K}
for $\beta=6.0$.}
\label{fig:B600-rho-1K}
\end{figure}

In Figs.\ \ref{fig:B585-rho-1K} and \ref{fig:B600-rho-1K}
are shown the first excited state masses 
of the $\rho$ meson obtained from the fit 
with $t_{min}=5$ and 6 (8 and 9) versus $1/K$ 
at $\beta=5.85\ (6.0)$, respectively.
(A two-mass fit with $t_{min}=9$ 
for the largest $K$ at $\beta=6.0$ does not converge.
Therefore the corresponding data is missing in the figure.) 
We give in the figures the experimental values 
for the masses of $\rho(1450)$ and $\phi(1680)$
which are the first excited states of the vector mesons.
The mass of $\phi(1680)$ is plotted at the third largest $K$,
because this value of $K$ corresponds to the 
strange quark mass
as mentioned in Sec.~\ref{sec:Strange}.
Apparently the results for the excited state mass depend strongly 
on the value of $t_{min}$.
For quarks lighter than the strange quark, 
the excited state mass obtained with
smaller $t_{min}$ is much larger than experiment,
while that with larger one is consistent with experiment
within large statistical errors.
Therefore, although the value of $m_1$ is unstable,
there exist two-mass fits to the $\rho$ propagators
which give both the ground state mass consistent 
with the one-mass fit and 
the first excited state mass consistent with experiment.

\begin{figure}[t]
\begin{center}
\leavevmode
\epsfysize=9cm
  \epsfbox{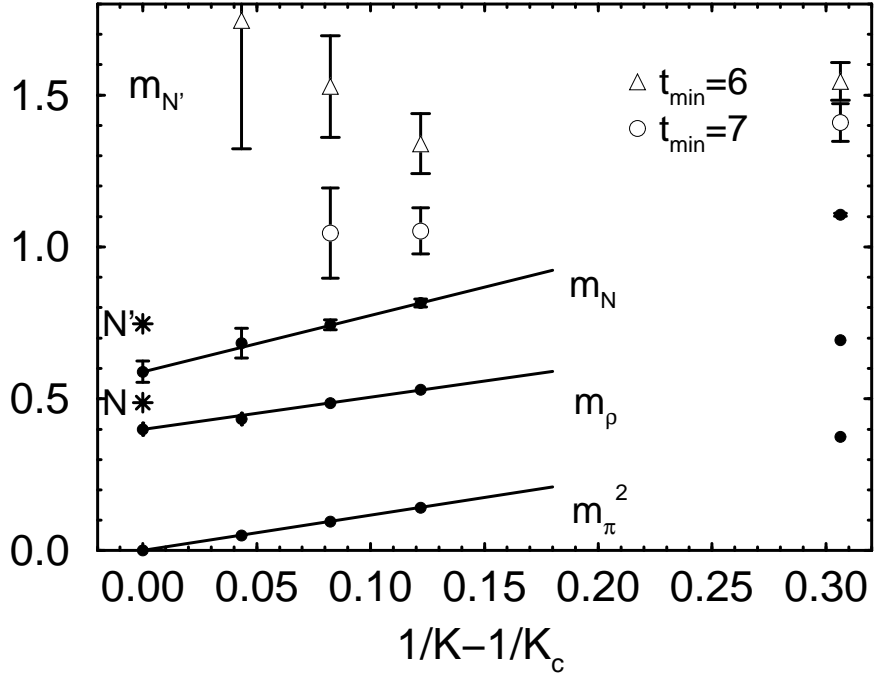}
\end{center}
\vskip -5mm
\caption{
Mass of excited state of the nucleon
(denoted by $N'$) at $\beta=5.85$ versus $1/K-1/K_c$.
The experimental values for masses of the nucleon ($N$)
and its excited state ($N'$) are marked with stars.
The data for $m_N$, $m_{\rho}$ and $m_{\pi}^2$ are
taken from the results of one-mass fits.
With the scale of the plot, the results for $m_N$ from 
two-mass fits are indistinguishable from the one-mass fit results.}
\label{fig:B585-pro-1K}
\end{figure}

Fig.\ \ref{fig:B585-pro-1K} shows 
the masses of excited state of the nucleon 
at $\beta=5.85$ versus $1/K$.
The excited state masses obtained from the fit with $t_{min}=7$ 
are much smaller than those with $t_{min}=6$.
(A two-mass fit with $t_{min}=7$ 
for the largest $K$ does not converge.)
We expect that the mass difference 
between the ground state and the first excited state 
depends only weakly on the quark mass,
because the mass difference for the spin $1/2$ baryon
satisfies this property.
The mass difference for the nucleon is 
$m_{N(1440)}-m_{N(940)}= 500$ MeV.  
The figure shows that
the excited state masses with $t_{min}=7$
lie approximately 500 MeV higher than the ground state masses. 
Therefore 
there exist two-mass fits
whose results do not contradict with experiment
also for the nucleon at $\beta=5.85$.

\begin{figure}[b]
\begin{center}
\leavevmode
\epsfysize=9cm
  \epsfbox{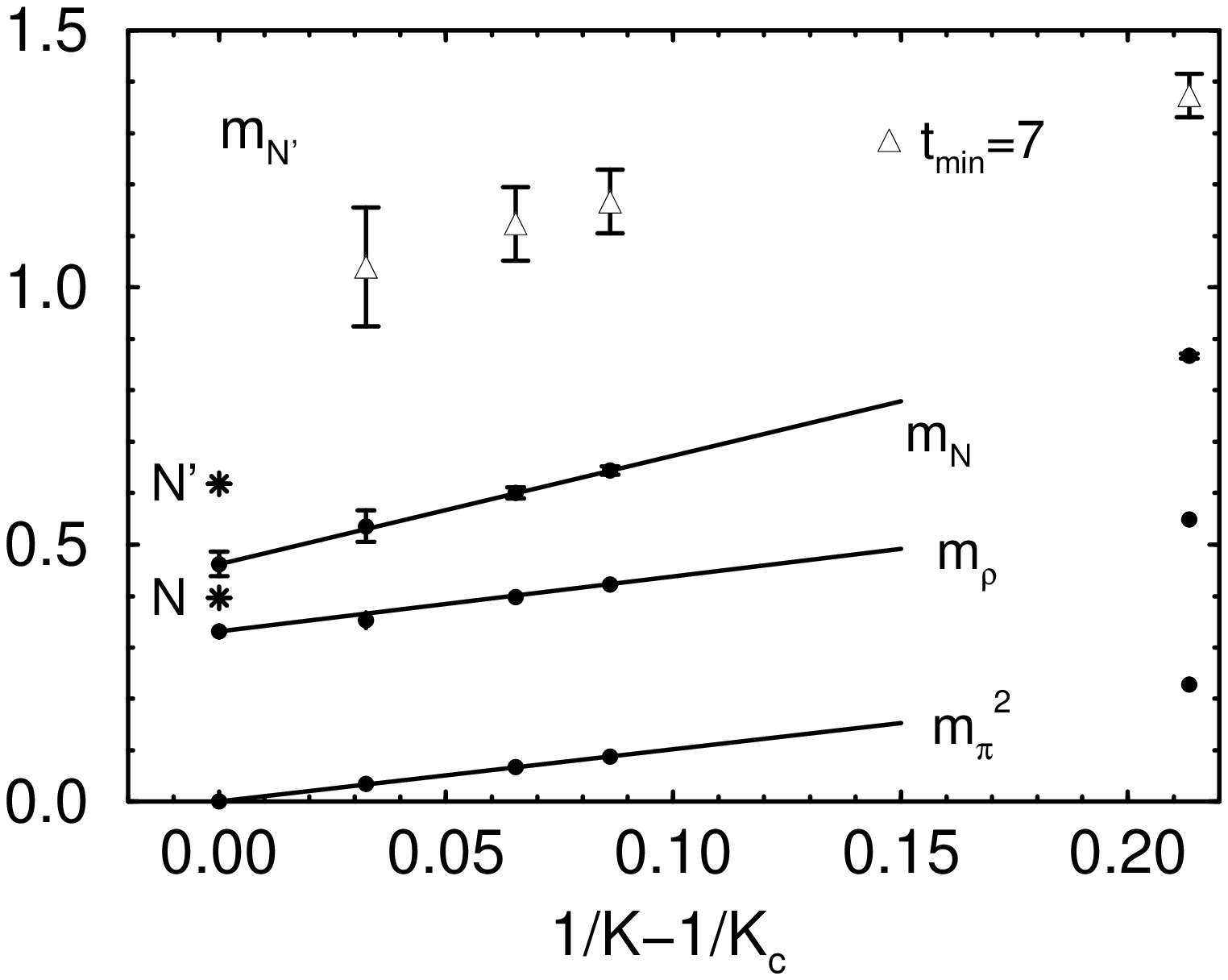}
\end{center}
\vskip -5mm
\caption{
The same as Fig.\ \protect\ref{fig:B585-pro-1K}
for $\beta=6.0$.}
\label{fig:B600-pro-1K}
\end{figure}

In Fig.\ \ref{fig:B600-pro-1K} we show the excited state masses
of the nucleon at $\beta=6.0$ with $t_{min}=7$.
The masses of the first excited state lie much more than
500 MeV above the ground state masses.
As mentioned before, 
two-mass fits for the nucleon at $\beta=6.0$ are stable 
and therefore the values of the excited state mass
do not change much even if we take other $t_{min}$.
When we recall that there exists a fit which gives a reasonable 
excited state mass at $\beta=5.85$,
this situation is puzzling.
One possible origin for the heavy excited state mass at
$\beta=6.0$ is a finite size effect,
because the physical volume is smaller at $\beta=6.0$.
There remains a possibility that when we simulate 
on a larger lattice,
a two-mass fit with larger $t_{min}$ gives a value consistent with
the nucleon excited state mass.

There are several published data for the mass of excited states
\cite{QCDPAX93,APE24,APE63,UKQCDEX,UKQCDJ}.
In table \ref{tab:ex}, we reproduce the results for 
the ratio of the excited state
mass to the ground state mass selecting the quark mass
corresponding approximately to the strange quark mass.
For the $\rho$ meson, except our results in this work
with $t_{min}=6$ (9) at $\beta=5.85$ (6.0) and the result
for the wall source in ref.~\cite{QCDPAX93},
the reported ratios are considerably larger than 
the corresponding experimental value 
$m_{\phi(1680)}/m_{\phi(1020)} = 1.65$.
For the nucleon, the mass ratios reported by the Ape collaboration
and the UKQCD collaboration
are considerably larger than our result.
One possible origin of the differences is due to the
choice of fitting range.
Because the two-mass fit is very unstable,
we certainly have to employ a more efficient way to extract 
reliable values for the excited state masses.

\section{Masses of Hadrons with 
Physical Light Quarks}\label{sec:Chiral}

\subsection{Extrapolation procedure}
Extrapolation of hadron masses to the chiral limit 
is done with the correlation being taken into account,
among the masses at different values of hopping parameter.
First we consider a least mean square fit to minimize
\begin{equation}
\chi^2 = \sum_{t,t',K,K'}
\{ G(t,K)-G_0(t,K) \} \, C^{-1}(t,K;t',K') \, 
\{G(t',K')-G_0(t',K')\}, \label{LMF}
\end{equation}
where $G_0(t,K) = A(K) e^{-m(K)t}$ is the fitting function
to hadron propagator $G(t,K)$ and $C^{-1}$ is 
the inverse of the full correlation matrix $C(t,K;t';K')$.
A linear approximation to the fitting function around the minimum
of $\chi^2$ gives the relation
between the error matrix $\Sigma$
for fit parameters and correlation matrix $C(t,K;t',K')$
for propagators:
\begin{equation}
\Sigma = ( D^T C^{-1} D )^{-1}, \label{Sigma}
\end{equation}
where $D$ is the Jacobian defined by
\begin{equation}
D_{t,K;A(K'),m(K')} = 
[{\partial G_0(t,K)} / {\partial A(K')}, 
{\partial G_0(t,K)} / {\partial m(K')} ]. \label{Jacob}
\end{equation} 
($D$ is diagonal with respect to $K$.)
The full least mean square fit to minimize 
$\chi^2$ in eq.~\ref{LMF} is different 
from the set of least mean square fits for each $K$
to minimize $\chi^2$'s in eq.~\ref{lms}:
The masses and amplitudes obtained by the two methods
are in general different.
We take those obtained from the fits to each propagator
for evaluation of the Jacobian.\footnote{
We have checked that the error matrix thus obtained is 
very close to that obtained using the Jacobian 
at the absolute minimum of eq.~\protect\ref{LMF}.
Consequently the difference in the extrapolated values 
obtained using two error matrices is 
at most 5\% of their statistical uncertainties.}

For extrapolation, we minimize $\chi^2$ given by
\begin{equation}
\chi^2 = \sum_K \{ m(K)-f(K) \} \, \Sigma^{-1}(K,K') \, 
\{m(K')-f(K') \},
\end{equation}
where the correlation matrix $\Sigma(K,K')$ 
is the sub-matrix among the masses
of the full error matrix $\Sigma$ and
$f(K)$ is the fitting function. (For the pion,
$m(K)$ is replaced by $m^2(K)$ with appropriate replacement of
$\Sigma^{-1}(K,K')$.)

\subsection{Linear extrapolation to the chiral limit}
\begin{figure}[t]
\begin{center}
\leavevmode
\epsfysize=9cm
  \epsfbox{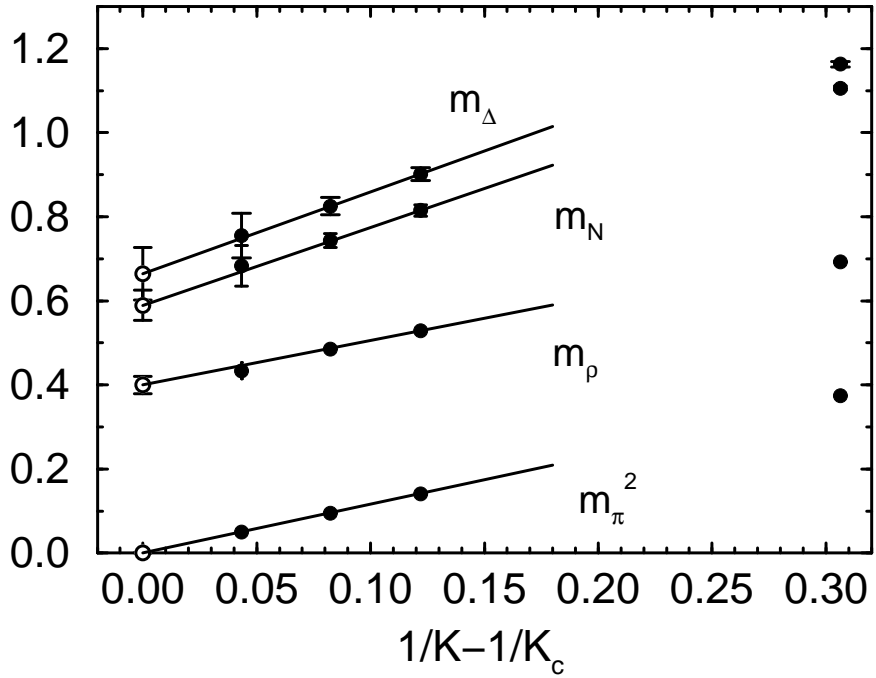}
\end{center}
\vskip -5mm
\caption{
Linear extrapolations of hadron masses at $\beta=5.85$
to the chiral limit.
The open circles at zero quark mass are extrapolated values.
The errors shown are statistical only, and do not include
the systematic errors discussed in the text.}
\label{fig:B585-1K}
\end{figure}

\begin{figure}[b]
\begin{center}
\leavevmode
\epsfysize=9cm
  \epsfbox{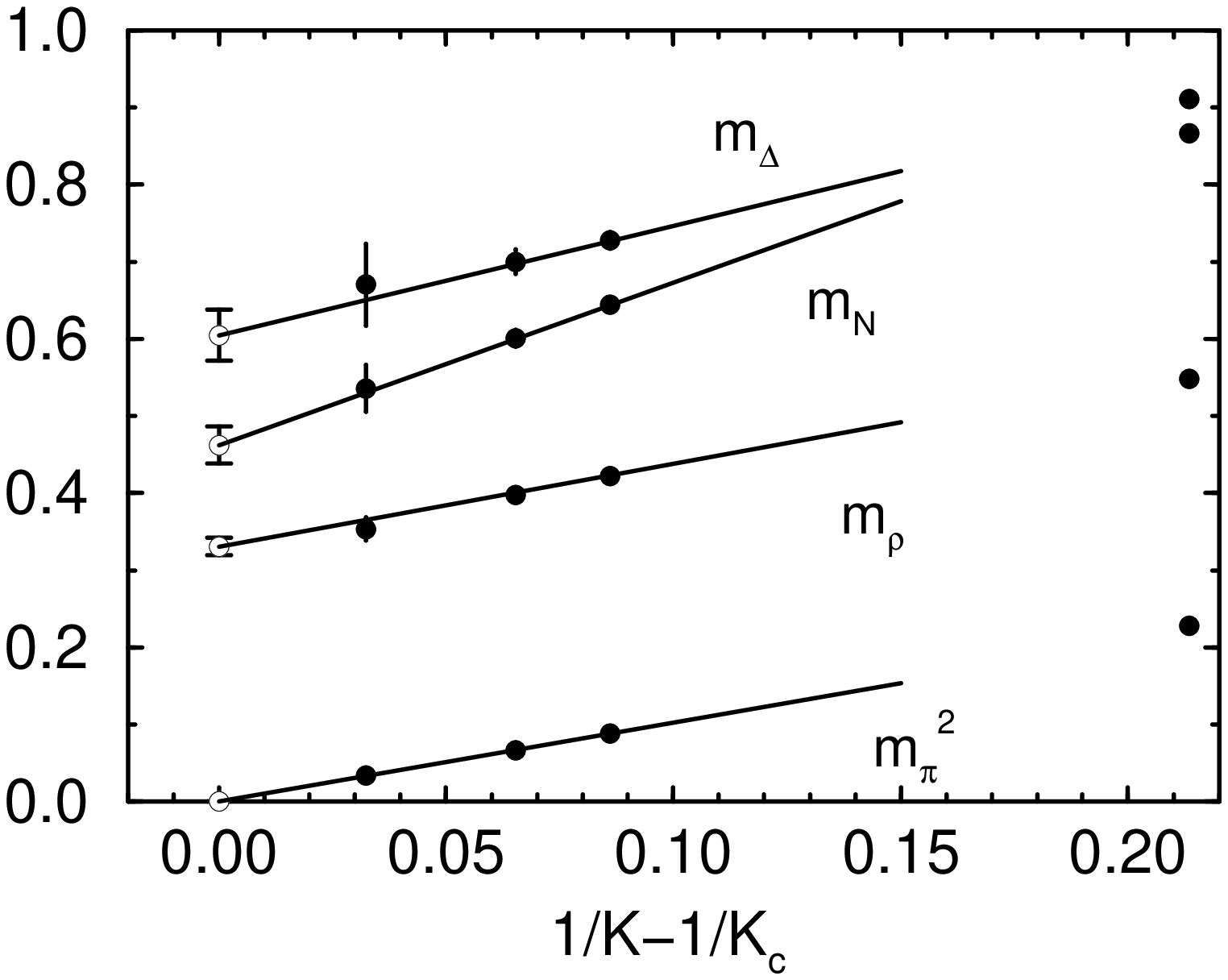}
\end{center}
\vskip -5mm
\caption{
The same as Fig.\  \protect\ref{fig:B585-1K}
for $\beta=6.0$.}
\label{fig:B600-1K}
\end{figure}

We fit the data of the mass squared for the pion
and the mass for the other hadrons
at the largest three $K$'s
to a linear function of $1/K$; $f(K) = a_0 + a_1/K$.
We find that quality of the linear fit is good
in the sense that $\chi^2/df < 2$ ($df = 1$ in this case)
and therefore we do not study in this work
the effects of possible chiral logarithms~\cite{CLOG1,CLOG2}.
We summarize the fit parameters together with 
$\chi^2/df$ in table \ref{tab:lf}.
The linear extrapolations of hadron masses at $\beta=5.85$ and 6.0
are shown in Figs.\ \ref{fig:B585-1K} 
and \ref{fig:B600-1K}, respectively.

In table \ref{tab:Kc} we summarize the results 
for the critical hopping parameter $K_c$
and the masses at $K_c$  
together with the errors estimated by the least mean square fit
and those by the jack-knife method.
We find that the error estimated by the jack-knife method 
is larger than that by the least mean square fit 
except for $K_c$ at $\beta=6.0$.
We take the error obtained by the jack-knife method 
as our estimate of the statistical uncertainty,
unless otherwise stated.

\subsection{Systematic error analyses}
We first estimate the systematic error 
on the masses in the chiral limit
coming from uncertainties in the choice of fitting range 
for extracting the ground state mass at each $K$.
To this end, we repeat linear extrapolations of the masses 
obtained from the fits to a range $t_0$ --- $T/2$,
varying $t_0$ (common to all $K$'s) from
${\rm max}_K \{t_{\chi^2}(K)\}$ to $t_{min}+4$.
We find that quality of the linear fits depends
on the choice of $t_0$:
$\chi^2/df$ are considerably large for some choices of $t_0$.
We adopt the condition $\chi^2/df<2$
for the linear fit to be accepted.
We take the difference between the fitted mass value and
the maximum/minimum mass value under the condition $\chi^2/df < 2$
as our estimate of the systematic upper/lower error.
We call the systematic error thus obtained 
the fit-range systematic error.

Data at the fourth largest $K$ slightly deviates 
from the linear fit.
In order to estimate the systematic error 
which comes from the choice of fitting function,
we make a quadratic fit
($f(K) = a_0 + a_1/K + a_2/K^2$)
to the largest four $K$'s, 
varying $t_0$ in the range used for the estimate 
of the fit-range systematic error.
We estimate the systematic error
by the difference between the maximum/minimum value
with $\chi^2/df<2$
and that of the linear fits.
We call the systematic error thus obtained 
the fit-func systematic error.

\subsection{Pion mass extrapolation and $K_c$}
Pion masses squared are fitted to 
a linear function of $1/K$ to obtain
the critical hopping parameter.
The value of $\chi^2/df$ is 0.56 (1.1) 
for the fit ($t_{min}=12$ (15)) at $\beta=5.85$ (6.0).
The fit-range systematic errors are estimated from the fits with 
$t_0=8$ --- 16 at $\beta=5.85$ and 10 --- 19 at $\beta=6.0$.  
All the fits give $\chi^2/df < 2$.
The upper (lower) bound comes from the fit with $t_0=11$ (14)
with $\chi^2/df$ of 0.36 (0.04) for $\beta=5.85$ and 
from the fit with $t_0=12$ (19) with $\chi^2/df$ of
0.44 (0.96) for $\beta=6.0$.

For data at $\beta=5.85$, no quadratic fits with $t_0=8$ --- 16
give $\chi^2/df <2$.
On the other hand, quadratic fits to data at $\beta=6.0$ 
with $t_0=13$ --- 19 give $\chi^2/df<2$.
Because $m_{\pi}^2$ is a concave function of $1/K$
when the data at the fourth largest $K$ is included, 
$K_c$ obtained from the quadratic fit 
is larger than that from the linear fit.

The values of $K_c$'s together with the fit-range systematic error
and the fit-func systematic error are given by
\begin{tabbing}
xxxxxxxxxxx \= xxxxxxxxxxxxx \= xxxxxxxxxxxx \= 
xxxxxxxxxx \= xxxxxxxxxx \= xxxxxxxxx \= \kill
       \>              \>  stat.   \> 
\ \ \ \ sys.(fit-range) \> \> sys.(fit-func.) \\
\ \ \ $\beta=5.85$ \> $K_c=$ 0.161624 \> $\pm$0.000033 
\> +0.000001 \> $-$0.000025 \> \\             
\ \ \ $\beta=6.00$ \> $K_c=$ 0.157096 \> $\pm$0.000028 
\> +0.000033 \> $-$0.000009 \> +0.000109 \\
\end{tabbing}
\vspace{-12pt}
The fit-range systematic error is comparable 
to the statistical uncertainty.

The result for $K_c$ at $\beta=6.0$ 
agrees well with that in ref.~\cite{APE24}.
Although it is slightly smaller than the LANL result 
0.15714(1)~\cite{LANL},
we conclude that
our result is
consistent with theirs within the sum of the statistical error and 
the fit-range systematic error.

In this work, we do not distinguish 
the physical point 
where $m_{\pi}/m_{\rho}$ takes its experimental value
from the critical point 
where the pion mass vanishes, because
we find that physical quantities at the two points
differ only at most 30\% of their statistical errors.

\subsection{Rho meson mass extrapolation and lattice spacing}
A linear fit to the $\rho$ meson masses 
(with $t_{min} = 12$ (15))
at the largest three $K$'s
gives $\chi^2/df$ of 1.8 (1.2) for $\beta=5.85$ (6.0).
Therefore the linear fit is acceptable.

\begin{figure}[htbp]
\begin{center}
\leavevmode
\epsfysize=20cm
  \epsfbox{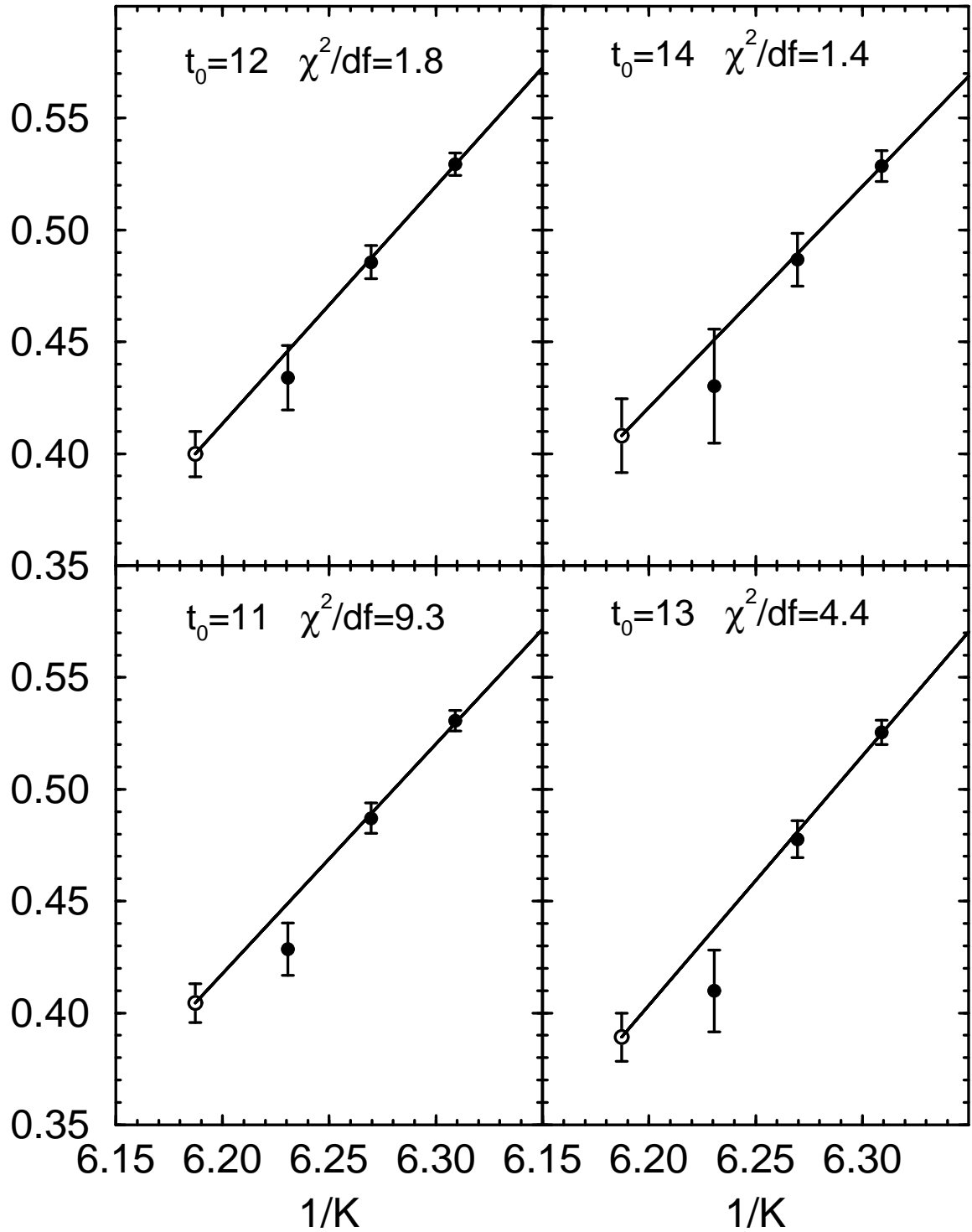}
\end{center}
\vskip -5mm
\caption{
$\rho$ meson masses at $\beta=5.85$ obtained from 
the fit with various $t_0$ together with 
linear extrapolations of these data.
The open circles are extrapolated values.
The errors shown are those estimated by the least mean square fits.}
\label{fig:B585-Rho-tK}
\end{figure}

\begin{figure}[htbp]
\begin{center}
\leavevmode
\epsfysize=20cm
  \epsfbox{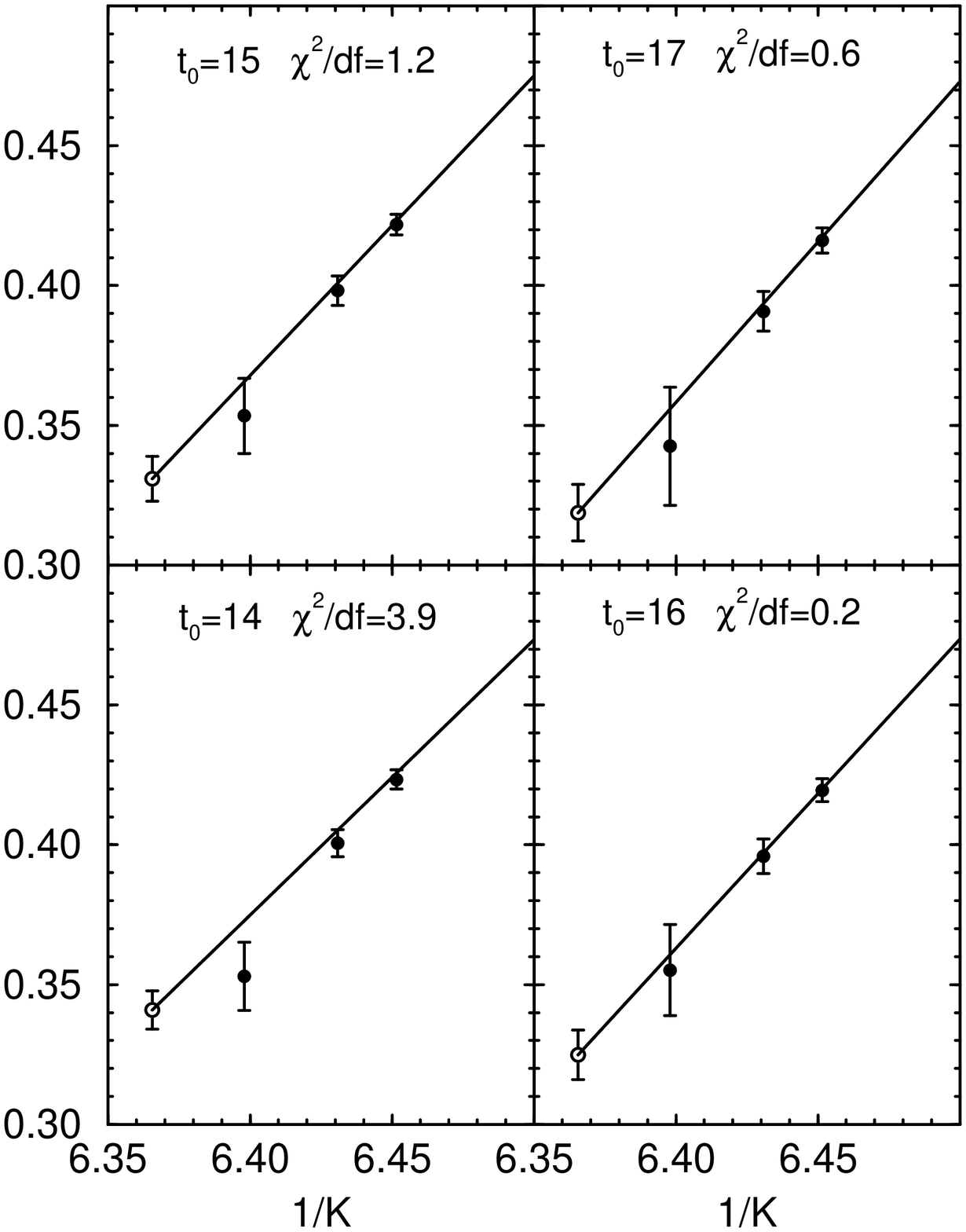}
\end{center}
\vskip -5mm
\caption{
The same as Fig.\ \protect\ref{fig:B585-Rho-tK}
but for $\beta=6.0$.}
\label{fig:B600-Rho-tK}
\end{figure}

However, we find that quality of the linear fit
strongly depends on the choice of fitting range.
See Figs.\ \ref{fig:B585-Rho-tK} and \ref{fig:B600-Rho-tK}.
In table \ref{tab:rho-1},
we summarize $\chi^2/df$, $m_{\rho}(K_c)$ and
the inverse lattice spacing defined by 
$a^{-1} = 0.77 {\rm GeV}/m_{\rho}(K_c)$ versus $t_0$.

We also make a quadratic fit to the data 
at the largest four $K$'s
to estimate the systematic error 
due to the choice of fitting function.
Table \ref{tab:rho-2} summarizes the results 
of the quadratic fits versus $t_0$.

The method to estimate the systematic error 
is the same as that adopted for the pion. 
Our final results for $m_{\rho}$ are 
\begin{tabbing}
xxxxxxxxxxx \= xxxxxxxxxxxxxxx \= xxxxxxxxx 
\= xxxxxxx \= xxxxxxxxx \= xxxxxx \= xxxxxxx \=\kill
                   \>                         \> stat. 
\> sys.(fit-range) \> \> sys.(fit-func.) \> \\
\ \ \ $\beta=5.85$ \> $m_{\rho}(K_c)$ = 0.400 \> $\pm$0.021 
\> +0.008 \> $-$0.027 \> +0.0 \> $-$0.013  \\
\ \ \ $\beta=6.00$ \> $m_{\rho}(K_c)$ = 0.331 \> $\pm$0.011 
\> +0.018 \> $-$0.020 \> +0.0 \> $-$0.008  \\
\end{tabbing}
\vspace{-12pt}
The value of $m_{\rho}(K_c)$ at $\beta=6.0$ agrees well with 
the Ape result 0.3332(75) and the LANL result 0.3328(106).
The values of $m_{\rho}(K_c)$'s are translated 
to the lattice spacing as
\begin{tabbing}
xxxxxxxxxxx \= xxxxxxxxxxx \= xxxxxxxxxx 
\= xxxxxxx \= xxxxxxx \= xxxxxx \= xxxxxxx \=\kill
                   \>                \> stat. 
\> sys.(fit-range) \> \> sys.(fit-func.) \> \\
\ \ \ $\beta=5.85$ \> $a^{-1}=$ 1.93 \> $\pm$0.10 
\> +0.14 \> $-$0.04 \> +0.08 \> $-$0.0 \ \ \ \ GeV \\
\ \ \ $\beta=6.00$ \> $a^{-1}=$ 2.33 \> $\pm$0.08 
\> +0.15 \> $-$0.12 \> +0.06 \> $-$0.0 \ \ \ \ GeV\\
\end{tabbing}
Although the statistical error on $a^{-1}$ is several percent,
we notice that the systematic error is much larger.
Summing up both the statistical and systematic errors,
we find that $a^{-1}$ can be as large as 2.25 GeV (2.62 GeV)
at $\beta=5.85\ (6.0)$ and as small as 1.79 GeV (2.13 GeV).

\begin{figure}[htbp]
\begin{center}
\leavevmode
\epsfysize=11cm
  \epsfbox{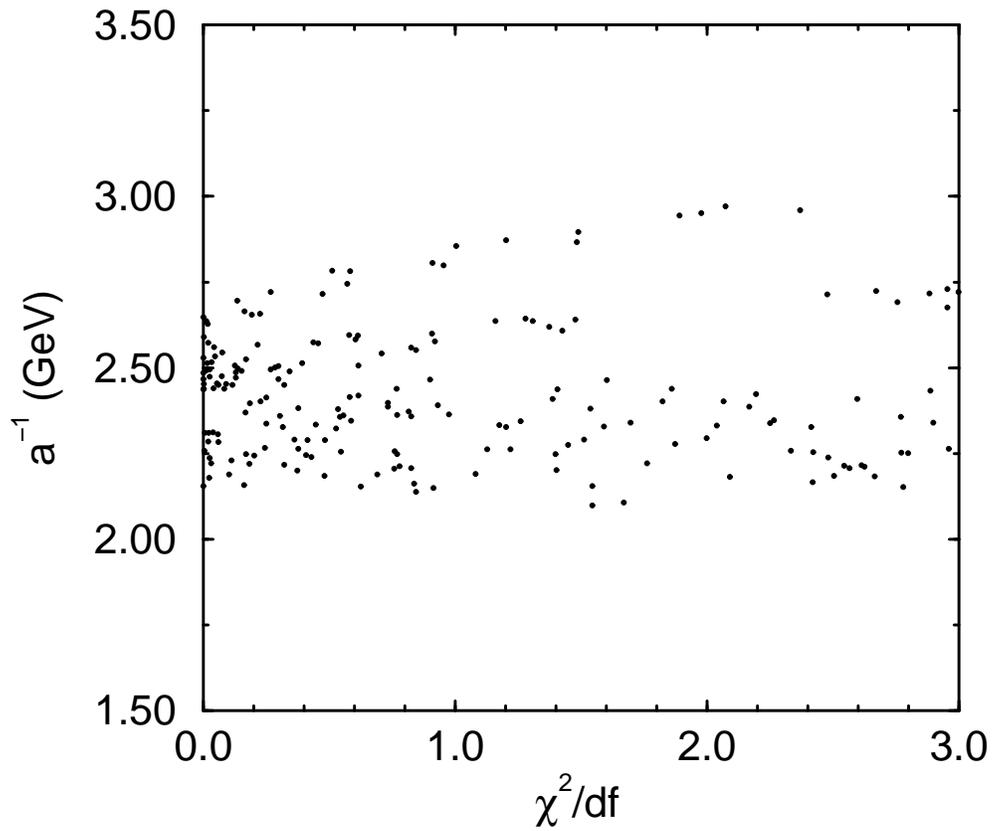}
\end{center}
\vskip -5mm
\caption{
$a^{-1}$ determined from linear fits to all possible combinations
of the $\rho$ masses obtained by 
varying $t_0$ from $t_{\chi^2}$ to 18.}
\label{fig:Ainv}
\end{figure}

In analyses of the systematic errors above,
we have taken $t_0$ common to all $K$'s.
However, it is not necessary to restrict ourselves
to take a common value of $t_0$,
because the time slice at which the contribution of 
excited states becomes negligible 
can depend on the quark mass.
We make linear fits to all possible combinations 
of the $\rho$ masses at the largest three $K$'s,
varying $t_0$ separately for each $K$ from $t_{\chi^2}$ to 18. 
Fig.\ \ref{fig:Ainv} shows $a^{-1}$ 
at $\beta=6.0$ versus $\chi^2/df$.
We see that there are linear fits with small $\chi^2/df$ 
which give both large and small $a^{-1}$.
The value of $a^{-1}$ scatters approximately 
from 2.15 GeV to 2.65 GeV.
This upper value as well as the lower value are consistent with 
those obtained above with the systematic errors included.

We estimate the value of $J$ defined by 
$m_V {{d m_V} \over {d m_{PS}^2}}$ \cite{UKQCDJ}
from the linear fits discussed above:

\begin{tabbing}
xxxxxxxxxxx \= xxxxxxxxxxx \= xxxxxxxxxx 
\= xxxxxxx \= xxxxxxx \=\kill
                   \>                \> stat. 
\> sys.(fit-range) \> \> \\
\ \ \ $\beta=5.85$ \> $J=$ 0.420  \> $\pm$0.049
\> +0.028 \> $-$0.024 \\
\ \ \ $\beta=6.00$ \> $J=$ 0.395 \> $\pm$0.026
\> +0.026 \> $-$0.026 \\
\end{tabbing}
The value of $J$ at $\beta=6.0$ is smaller than 
the experimental value 0.48(2) even when we include
the systematic errors.

\subsection{Nucleon and $\Delta$ masses}
Both linear fits and quadratic fits are made 
to the masses of the nucleon and the $\Delta$ baryon
by the same method as for the $\rho$ meson.
Results of the linear fits versus the fit-range are summarized
in tables \ref{tab:N-1} and \ref{tab:D-1}.
The fit with $t_{min}=13\ (16)$ at $\beta=5.85$ (6.0),
which is adopted in this work, 
gives a small $\chi^2/df=$ 0.37 (0.05).
For the nucleon, quality of the linear fits is good for
almost all values of $t_0$
in the sense that $\chi^2/df$ are 
approximately less than 2,
except for the fit with $t_0=9$ at $\beta=5.85$.
This feature is different from that for the $\rho$ meson.
Quality of the fits to the $\Delta$ masses at $\beta=5.85$
is good for $t_0 \le 13$ including our choice $t_{min}=13$
and that at $\beta=6.0$ is good for all $t_0$ except for $t_0=13$.

Results with various errors are give by
\begin{tabbing}
xxxxxxxxxxx \= xxxxxxxxxxxxxxxx \= xxxxxxxxxx 
\= xxxxxxx \= xxxxxxx \= xxxxxx \= xxxxxxx \=\kill
                   \>                   \>  stat. 
\>  sys.(fit-range) \> \> sys(fit-func.) \> \\
\ \ \ $\beta=5.85$ \> $m_N(K_c)=$ 0.589 \>  $\pm$0.036 
\>  +0.018 \> $-$0.058 \>  +0.0 \> $-$0.018 \\
\ \ \ $\beta=6.00$ \> $m_N(K_c)=$ 0.462 \>  $\pm$0.024 
\>  +0.020 \> $-$0.009 \>  +0.0 \> $-$0.007 \\
\end{tabbing}
\begin{tabbing}
xxxxxxxxxxx \= xxxxxxxxxxxxxxxx \= xxxxxxxxxx 
\= xxxxxxx \= xxxxxxx \= xxxxxxx \= xxxxxxx \=\kill
                   \>                          \>  stat. 
\> sys.(fit-range) \> \> sys.(fit-func.) \> \\
\ \ \ $\beta=5.85$ \> $m_{\Delta}(K_c)=$ 0.664 \> $\pm$0.063 
\> +0.034 \> $-$0.0   \> +0.0 \> $-$0.031 \\
\ \ \ $\beta=6.00$ \> $m_{\Delta}(K_c)=$ 0.605 \> $\pm$0.033 
\> +0.041 \> $-$0.011 \> +0.016 \> $-$0.007 \\
\end{tabbing}
The value of the nucleon mass in the chiral limit 
at $\beta=6.0$ lies between the LANL result 0.482(13)
and the Ape result 0.432(15).
For the $\Delta$ masses, results by the three groups 
agree well with each other, albeit with large errors;
the LANL result is 0.590(30) and the Ape result 0.58(3).
The LANL results are those at the physical point
where $m_{\pi}/m_{\rho}$ takes its experimental value.

These results are translated to the masses in physical units
using the value of $a^{-1}$ obtained from $m_{\rho}$.
The systematic error on the lattice spacing 
is not taken into account for the estimate of 
the systematic error on the baryon masses.
Results read
\begin{tabbing}
xxxxxxxxxxx \= xxxxxxxxxxxxx \= xxxxxxxxxx 
\= xxxxxxx \= xxxxxxxxx \= xxxxxxx \= xxxxxxx \=\kill
                   \>              \>  stat. 
\> sys.(fit-range) \> \> sys.(fit-func.) \> \\
\ \ \ $\beta=5.85$ \> $m_N=$ 1.135 \>  $\pm$0.088 
\> +0.034 \> $-$0.112 \> +0.0 \> $-$0.034 \ \ \ \ \ GeV\\
\ \ \ $\beta=6.00$ \> $m_N=$ 1.076 \>  $\pm$0.060 
\> +0.047 \> $-$0.020 \> +0.0 \> $-$0.017 \ \ \ \ \ GeV\\
\ \ \ $\beta=5.85$ \> $m_{\Delta}=$ 1.279 \> $\pm$0.136 
\> +0.066 \> $-$0.0   \> +0.0   \> $-$0.059 \ \ \ \ \ GeV\\
\ \ \ $\beta=6.00$ \> $m_{\Delta}=$ 1.407 \> $\pm$0.086 
\> +0.096 \> $-$0.026 \> +0.038 \> $-$0.015 \ \ \ \ \ GeV\\
\end{tabbing}
The central value of the nucleon mass at $\beta=6.0$ (5.85)
is larger than its experimental value 
by about 15\% (20\%) and that of
the $\Delta$ mass by about 15\% (4\%):
The errors amount to twice the statistical errors
except for the $\Delta$ baryon at $\beta=5.85$.
The systematic errors are comparable with
the statistical errors (3 --- 13\%).
Even when the systematic errors are included,
the baryon masses at $\beta=6.0$ do not agree with experiment.
Our data are consistent 
with the GF11 data~\cite{GF11} at finite lattice spacing, 
within statistical errors. 
In order to take the continuum limit of our results,
we need data for a wider range of $\beta$ with  
statistical and systematic errors much reduced.

\subsection{Masses of strange hadrons}\label{sec:Strange}
The hopping parameters for the strange quark 
which are estimated from the experimental value of $m_K/m_\rho$
turn out to be $K_s=0.1588$ and 0.1550 
at $\beta=5.85$ and 6.0, respectively.
Note that they are identical or almost identical to the 
third largest hopping
parameter $K=0.1585$ and 0.1550 which we have chosen 
in such a way that they approximately 
correspond to the strange quark.
The masses of $\Omega^{-}$ estimated at $K=K_s$ are 
1.696(92) GeV and 1.693(57) GeV at $\beta=5.85$ and 6.0,
respectively (statistical errors only).
They are in good agreement with the experimental 
value 1.672 GeV.
The masses of the vector meson at $K=K_s$ are
998(45) MeV and 986(26) MeV 
at $\beta=5.85$ and 6.0, respectively,
which equal the $\phi$ meson mass 1019 MeV
within about one standard deviation.
As is well known, there are ambiguities 
in determination of the hopping parameter
for the strange quark.
When the hopping parameters for the strange quark mass are
alternatively determined from $m_{\phi}/m_{\rho}$,
they are equal to $0.1585$ and 0.1547.
The results for the $\Omega^{-}$ mass at these hopping parameters
are consistent with those above within one standard deviation. 

\section{Meson Decay Constants}\label{sec:Decay}
\subsection{Vector meson decay constants}
We evaluate vector meson decay constants
defined by 
\begin{equation}
\langle 0|(\bar u \gamma_i d)^{cont.}|V({\vec p}=0) \rangle 
= \epsilon_i F_V m_V, 
\end{equation}
where $\epsilon_i$ and $m_V$ are the polarization vector 
and the mass of the vector meson, respectively,
and $(\bar u \gamma_i d)^{cont.}$ is the vector current 
in the continuum limit.
The experimental value for the $\rho$ meson is 
$F_{\rho}=216(5)$ MeV.
(This $F_V$ is related to $f_V^{-1}$ by $f_V^{-1} = F_V/m_V$.)

The expectation value of the local lattice current 
$(\bar u \gamma_i d)^{latt.}$ between the vacuum and 
the vector meson
is related to the continuum one by
the relation
\begin{equation}
\langle 0|(\bar u \gamma_i d)^{cont.}|V({\vec p}=0) \rangle =
Z_K Z_V \langle 0|(\bar u \gamma_i d)^{latt.}
                 |V({\vec p}=0) \rangle. 
\end{equation}
The coefficient $Z_K$ is a scale factor 
for the difference between the continuum and 
lattice normalizations of the quark field.
The renormalization constant $Z_V$ is the ratio of 
the conserved lattice current to the local current,
which can be estimated by perturbation theory 
or numerical simulations.
We test the following three possible choices of 
$Z_K$ and $Z_V$:

\begin{enumerate}
\item 
those in naive perturbation theory: 
$Z_K= 2K$ and $Z_V = 1 - 0.174 \, g^2$~\cite{Zpert},
\item
those in tadpole improved perturbation theory:
$Z_K= (1-3K/4K_c)$~\cite{ZKtp} and 
$Z_V = 1 - 0.82 \, \alpha_{\overline{MS}}(1/a)$~\cite{LM} \ \ 
($\alpha_{\overline{MS}}(\pi/a) 
= g^2_{\overline{MS}}(\pi/a)/4\pi$ is determined by the relation 
$1/g^2_{\overline{MS}}(\pi/a) = {\rm Tr} (U_P/3)/g^2 + 0.02461$
~\cite{LM,El-Khadra}.
We then determine $\alpha_{\overline{MS}}(1/a)$ 
using the two loop renormalization group equation.),  
\item
Monte Carlo estimate of $Z_V$ = 0.51~\cite{STDB585}\ 
(0.57~\cite{ZMC}) at $\beta=5.85\ (6.0)$ with
$Z_K = 2K$. 
(Data for $Z_V$ at $\beta=5.85$~\cite{STDB585}
are given in table \ref{tab:ZV}.
Because the results for $Z_V$ are independent 
of the quark mass in the range we investigate,
we use the averaged value.)
The error on $Z_V$ is ignored in the following.
\end{enumerate}
We abbreviate the decay constants obtained 
using the above three renormalization constants
as $F_V^{PT}\ F_V^{TP}$ and $F_V^{MC}$, respectively.  

The statistical error is obtained by the jack-knife method.
The systematic error is estimated 
varying $t_0$ as in the case of mass calculation.
The range of $t_0$ is the same as that
for the $\rho$ mass.
In table \ref{tab:fV}
we summarize the results for the decay constants at each $K$.
We quote the error only for $F_V^{TP}$,
because the errors for the others can be easily obtained from
that for $F_V^{TP}$ by multiplying the ratio of $Z$-factors.

\begin{figure}[htbp]
\begin{center}
\leavevmode
\epsfysize=9cm
  \epsfbox{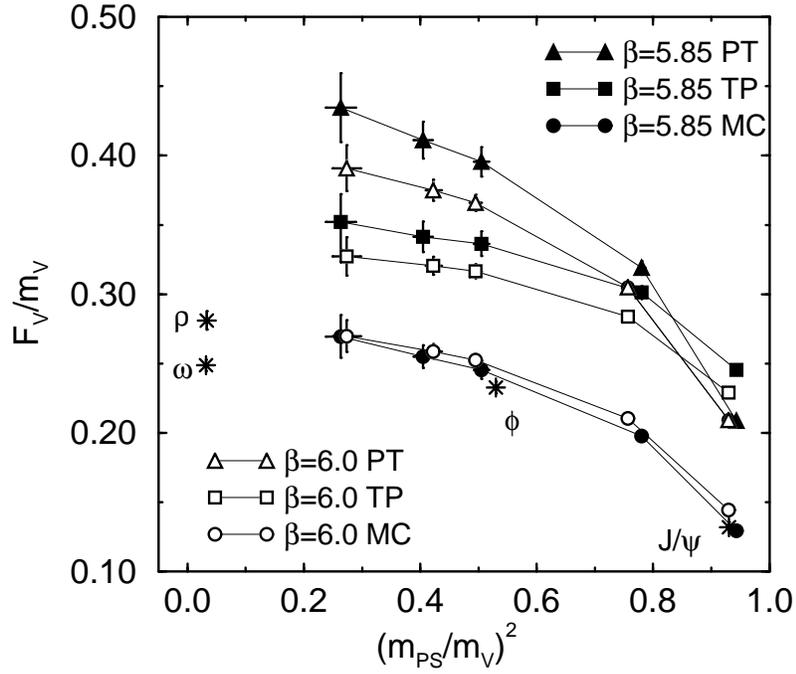}
\end{center}
\vskip -5mm
\caption{
Ratio of the vector meson decay constant 
to the vector meson mass, for the three choices of
renormalization constants discussed in the text.
The errors shown are statistical only 
and are estimated by the jack-knife method.
The corresponding experimental values for vector mesons 
are marked with stars.
The value of $m_{PS}$ for the strange quark 
is estimated by phenomenological 
mass formulae\protect\cite{ONO}
using $m_{V} = m_{\phi} = 1019$ MeV.
}
\label{fig:Fvpr}
\end{figure}

Fig.\ \ref{fig:Fvpr} shows $F_V/m_V$ versus $(m_{PS}/m_{V})^2$
together with the corresponding experimental values 
for $\rho$, $\omega$, $\phi$ and $J/\psi$. 
Note that we can compare the numerical results with the experimental
values for $\phi$ and $J/\psi$ without extrapolation.
The values with $F_V^{MC}$ at two $\beta$'s 
remarkably agree with each other.
Furthermore they agree well 
with the experimental values for $\phi$ and $J/\psi$.
This implies that scaling violation in $F_V^{MC}$ is small.
On the other hand, we find sizable scaling violation
in $F_V^{PT}$ and $F_V^{TP}$.
They are off the experimental values for 
$\phi$ and $J/\psi$ by 40 --- 100\%. 
We find that $F_V^{MC}/m_V$'s at $\beta=6.0$ agree well with 
the Ape data~\cite{APE24,APE63}.

\begin{figure}[htbp]
\begin{center}
\leavevmode
\epsfysize=12cm
  \epsfbox{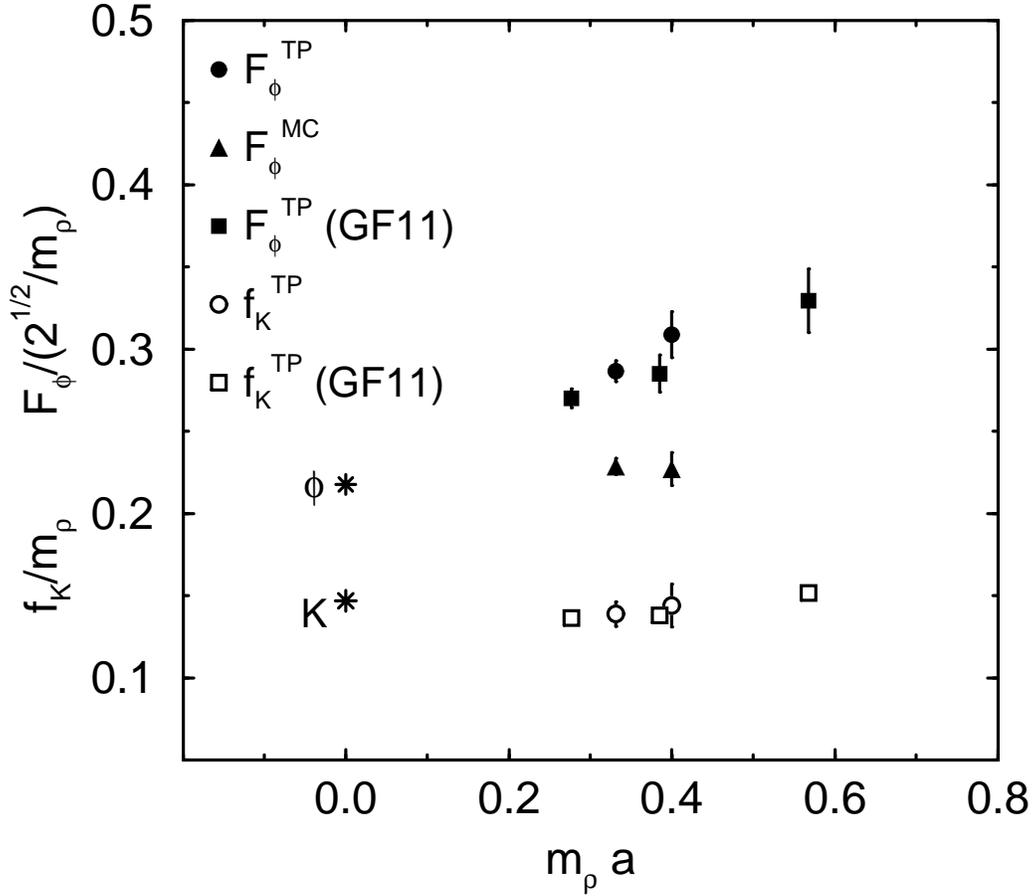}
\end{center}
\vskip -5mm
\caption{
Ratios of the $\phi$ meson decay constant and the $K$ meson
decay constant 
to the $\rho$ meson mass versus 
the $\rho$ meson mass in lattice units.
The errors in our data are statistical only.
The GF11 data are taken from ref.\ \protect\cite{GF11DE}.
The corresponding experimental values are marked with stars.}
\label{fig:Decay-s}
\end{figure}

In fig.~\ref{fig:Decay-s}
we depict the values of $F_{\phi}/m_{\rho}$ versus 
$m_{\rho} a$ together with the GF11 result\cite{GF11DE}. 
The values of the hopping parameter 
for the strange quark are given in Sec.~\ref{sec:Strange}.
Note that
the values of $F_{\phi}^{MC}/{m_{\rho}}$ 
agree with experiment
already at $m_{\rho}a$ = 0.33 --- 0.40
within 1 --- 2 standard deviations.
The values of $F_{\phi}^{TP}/{m_{\rho}}$ 
are consistent with the GF11 result, 
although the central values are
about $1\sigma$ higher than the GF11 data.
They are off the experimental value by 30 --- 40\% at these 
values of $m_{\rho}a$.
Linear extrapolation of our data 
to zero lattice spacing is consistent with experiment.

\begin{figure}[htbp]
\begin{center}
\leavevmode
\epsfysize=9cm
  \epsfbox{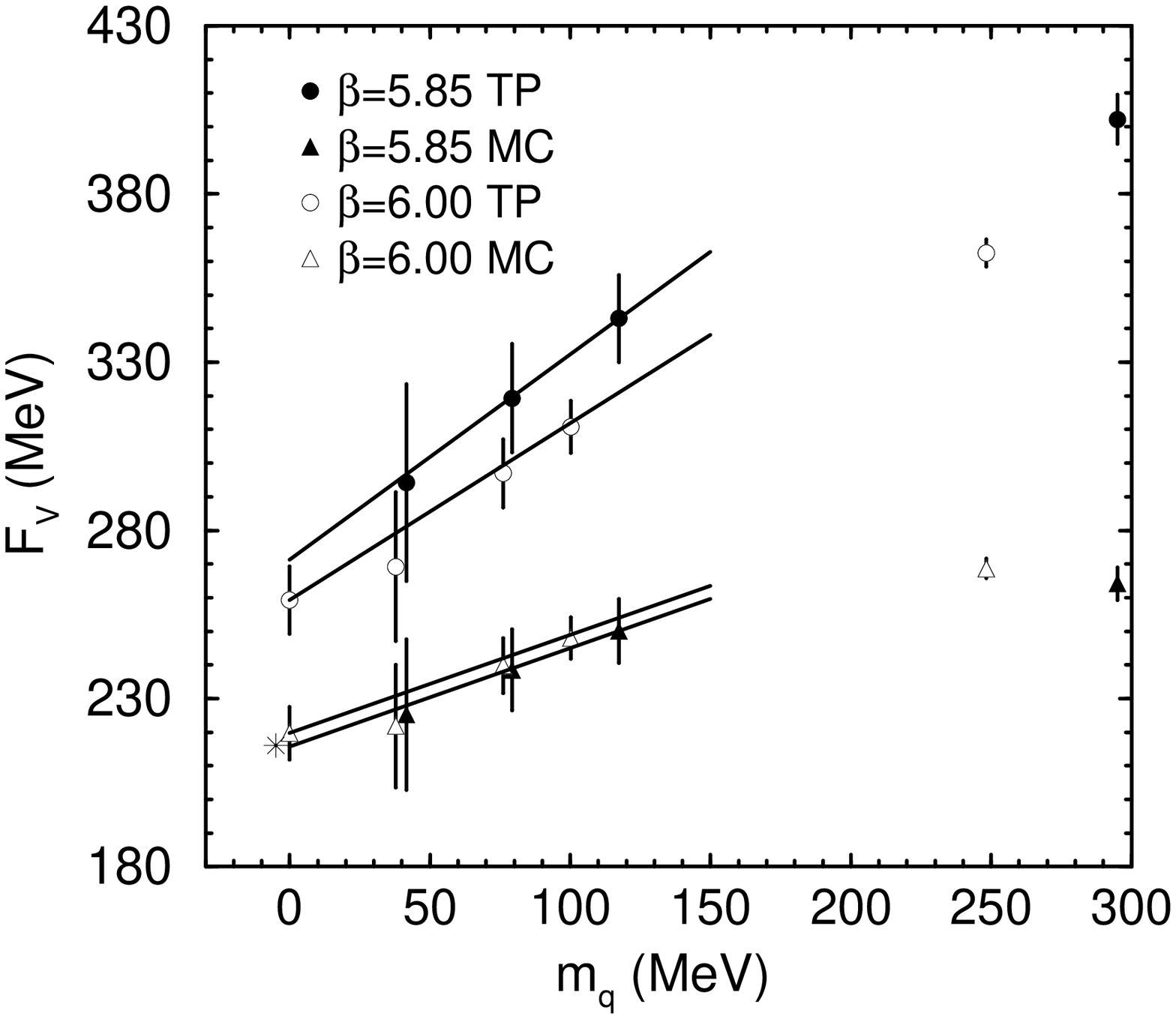}
\end{center}
\vskip -5mm
\caption{
Linear extrapolations of vector meson decay constants,
for the two choices of renormalization constants 
discussed in the text.
The open symbols at zero quark mass are extrapolated values 
for $\beta=6.0$.
The errors shown are statistical only.
The experimental value for the $\rho$ meson is marked with a star.}
\label{fig:Fv1K}
\end{figure}

The value of $F_V$ in the chiral limit 
is obtained from a linear fit in terms of $1/K$
in a similar way to that made for hadron mass extrapolation. 
We first calculate the correlation matrix 
$\Sigma(K,K')$ for $F_V(K)$ from the error matrix $\Sigma$
for the mass and amplitude (eq.~\ref{Sigma})
using the error propagation rule and then minimize $\chi^2$.
A linear fit to the data at the largest three $K$'s 
gives a reasonable $\chi^2/df$:
$\chi^2/df$ = 0.04 (0.38) for $F_V^{TP}$, 0.09 (0.44)
for $F_V^{PT}$ and $F_V^{MC}$ at $\beta=5.85$ (6.0), respectively.
Fig.\ \ref{fig:Fv1K} shows $F_V$
as functions of the quark mass
together with the fitting functions.

\begin{figure}[htbp]
\begin{center}
\leavevmode
\epsfysize=12cm
  \epsfbox{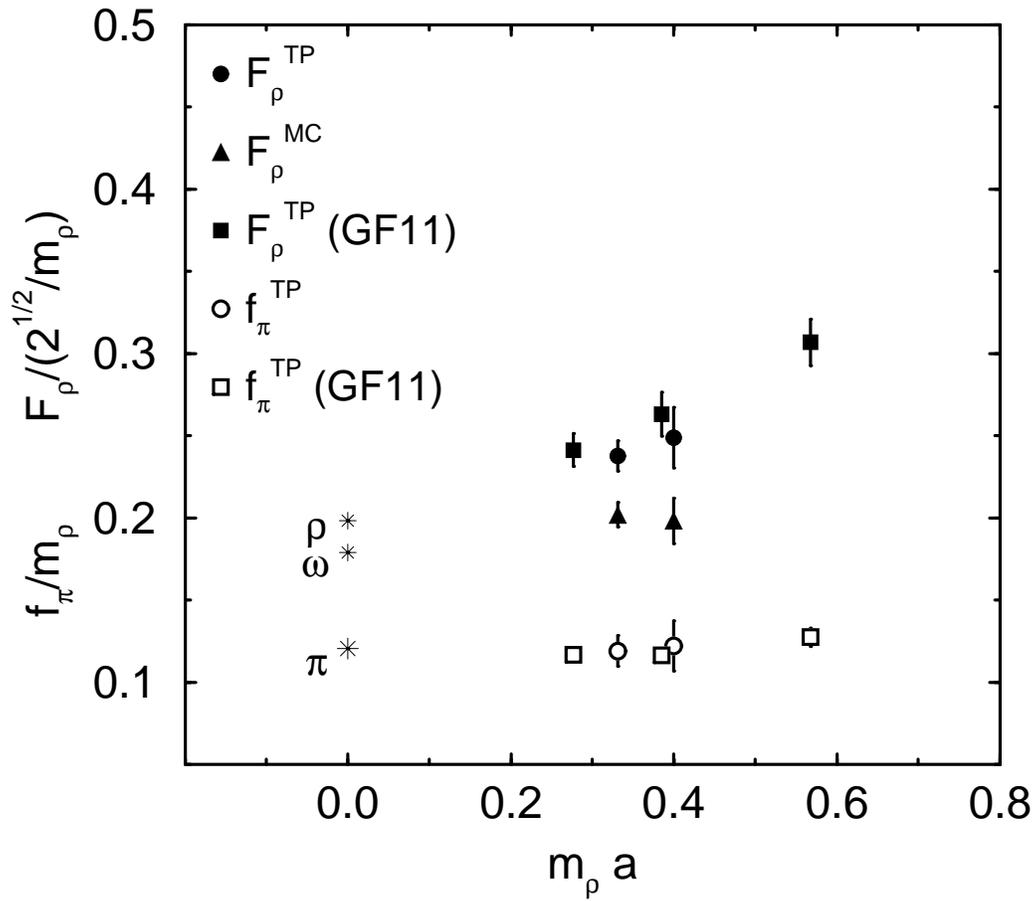}
\end{center}
\vskip -5mm
\caption{
The same as Fig.\ \protect\ref{fig:Decay-s} for 
the $\rho$ meson and the pion.}
\label{fig:Decay}
\end{figure}

The method to estimate the systematic error due 
to the choice of fitting range is similar to 
that for hadron masses at $K_c$.
The results of the linear fit for various fitting ranges 
are given in table \ref{tab:fV-1}.
Our final results for $F_{\rho}$ read
\begin{tabbing}
xxxxxxxxxxx \= xxxxxxxxxxxxx \= xxxxxxxxxxxx 
\= xxxxxxxx \= xxxxxxxx \= xxxxx \= \kill
                    \>                   \> stat. 
\> sys.(fit-range) \> \\ 
\ \ \ $\beta=$ 5.85 \> $F_{\rho}^{TP}=$ 0.141 \> $\pm$0.017 
\> +0.007 \> $-$0.035 \> \\
\ \ \               \> $F_{\rho}^{TP}=$ 271   \> $\pm$20    
\> +14    \> $-$68    \> MeV \\
\ \ \               \> $F_{\rho}^{MC}=$ 0.112 \> $\pm$0.013 
\> +0.006 \> $-$0.027 \> \\
\ \ \               \> $F_{\rho}^{MC}=$ 216   \> $\pm$15    
\> +11    \> $-$52    \> MeV \\

\ \ \ $\beta=$ 6.00 \> $F_{\rho}^{TP}=$ 0.111 \> $\pm$0.008 
\> +0.016 \> $-$0.017 \> \\ 
\ \ \               \> $F_{\rho}^{TP}=$ 259   \> $\pm$10    
\> +37    \> $-$40    \> MeV \\
\ \ \               \> $F_{\rho}^{MC}=$ 0.0944 \> $\pm$0.0064 
\> +0.010 \> $-$0.014 \> \\
\ \ \               \> $F_{\rho}^{MC}=$ 220   \>  $\pm$8      
\> +24    \> $-$33  \> MeV \\
\end{tabbing}
The values of $F_{\rho}^{PT}$ can be obtained 
from $F_{\rho}^{MC}$ 
by multiplying $Z_V^{PT}/Z_V^{MC} =$ 1.61 (1.45) 
at $\beta=5.85$ (6.0).
We show the values of $F_{\rho}/m_{\rho}$ 
in Fig.\ \ref{fig:Decay}.
It should be noted that the values of
$F_V^{MC}$ in the chiral limit at two $\beta$'s are consistent 
with the experimental value of $F_{\rho}$.
We find that our values of $F_{\rho}^{TP}/m_{\rho}$ 
are consistent with the GF11 result~\cite{GF11DE},
albeit the central values being
roughly $1 \sigma$ lower than the GF11 data;
this tendency is opposite to the case of the $\phi$ meson.
We note that linear extrapolation of our data 
for $F_{\rho}^{TP}/m_{\rho}$ to zero lattice spacing
is again consistent with experiment.

\subsection{Pseudo scalar meson decay constants}
The pseudo scalar meson decay constant is defined by 
\begin{equation}
 \langle 0|(\bar u \gamma_0 \gamma_5 d)^{cont.}
          |P({\vec p}=0) \rangle 
 = \sqrt{2} \, m_{PS} \, f_{PS}. 
\end{equation}
The experimental value is $f_{\pi}$ = 93 MeV.
We investigate three cases of renormalization constants
as in the case of $F_V$:
1) $Z_A=1-0.133 \, g^2$ in naive perturbation 
theory~\cite{Zpert} with $Z_K=2K$, 
2) $Z_A = 1 - 0.31 \, \alpha_{\overline{MS}}(1/a)$~\cite{LM}
with $Z_K=(1-3K/4K_c)$~\cite{ZKtp}
in tadpole improved perturbation theory, and
3) $Z_A = 0.69$~\cite{ZMC} at $\beta=6.0$ as a 
nonperturbative evaluation with $Z_K=2K$.
(Corresponding $Z_A$  at $\beta=5.85$ is not known.)

\begin{figure}[htbp]
\begin{center}
\leavevmode
\epsfysize=9cm
  \epsfbox{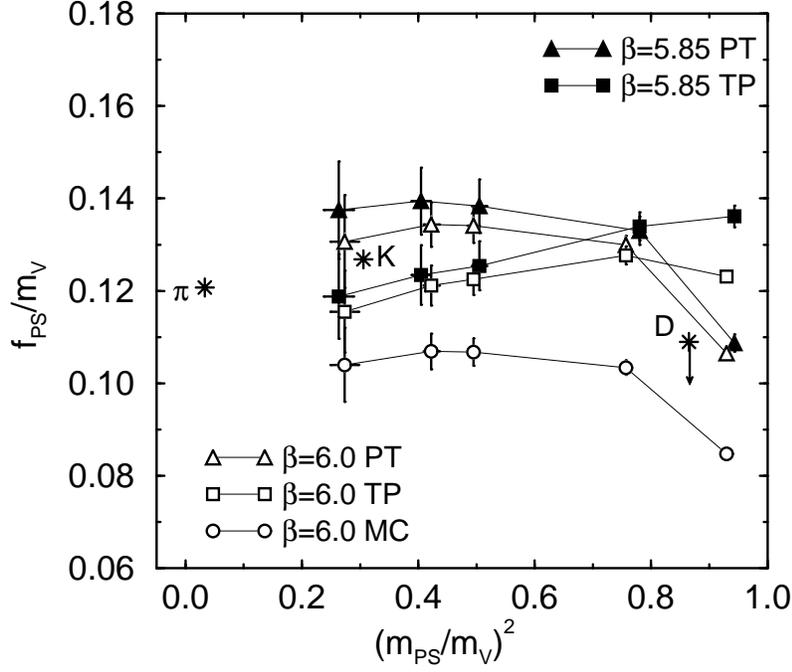}
\end{center}
\vskip -5mm
\caption{
Ratio of the pseudo-scalar meson decay constant
to the vector meson mass, for the three choices of 
renormalization constant discussed in the text.
The errors shown are statistical only
and are estimated by the jack-knife method.
The corresponding experimental values for pseudo-scalar mesons
are marked with stars.}
\label{fig:Fprhopr}
\end{figure}

We derive $f_{PS}$ from a fit to the $\tilde\pi$ propagator.
The value of $t_{min}$ is chosen to be the same as 
that for $\pi$.
The pion mass from the $\tilde\pi$ propagator is 
given in table \ref{tab:pit}.
Although the mass obtained is 1 --- 2 standard deviations 
smaller than that from the $\pi$ propagator,
they are consistent with each other if we take account of 
the systematic error. 
The decay constant at each $K$ is given in table \ref{tab:fPS}.
Our data for $f_{PS}^{PT}$ at $\beta=6.0$
and $K=0.155,0.1563$ 
are consistent with the Ape results~\cite{APE24,APE63}.
Fig.\ \ref{fig:Fprhopr} shows $f_{PS}/m_V$ 
versus $(m_{PS}/m_V)^2$ 
together with the corresponding experimental values for 
$\pi$ and $K$ and the upper bound for the $D$ meson.
Contrary to the case of the vector meson,
$f_{PS}^{MC}$ differs from the experimental value 
for the $K$ meson by a factor of about 1.2.
There is a possibility that
the lattice size $10^3 \times 20$ is not large enough to suppress
finite lattice size effects in the Monte Carlo evaluation of $Z_A$.
We think we have to calculate nonperturbatively
$Z_A$ both at $\beta=5.85$
and 6.0 on a larger lattice
in order to clarify the reason of the discrepancy.

In fig.~\ref{fig:Decay-s}
we show the values of $f_{K}^{TP}/m_{\rho}$ versus 
$m_{\rho}a$ together with
the GF11 result~\cite{GF11DE}. 
The values of $f_K$ are evaluated at the hopping parameter
given by $2/(1/K_c+1/K_s)$.
The values of $f_{K}^{TP}/{m_{\rho}}$ 
are consistent with the GF11 result, 
albeit with larger errors in our results.
Our data at finite lattice spacing
are also consistent with experiment.

\begin{figure}[htbp]
\begin{center}
\leavevmode
\epsfysize=9cm
  \epsfbox{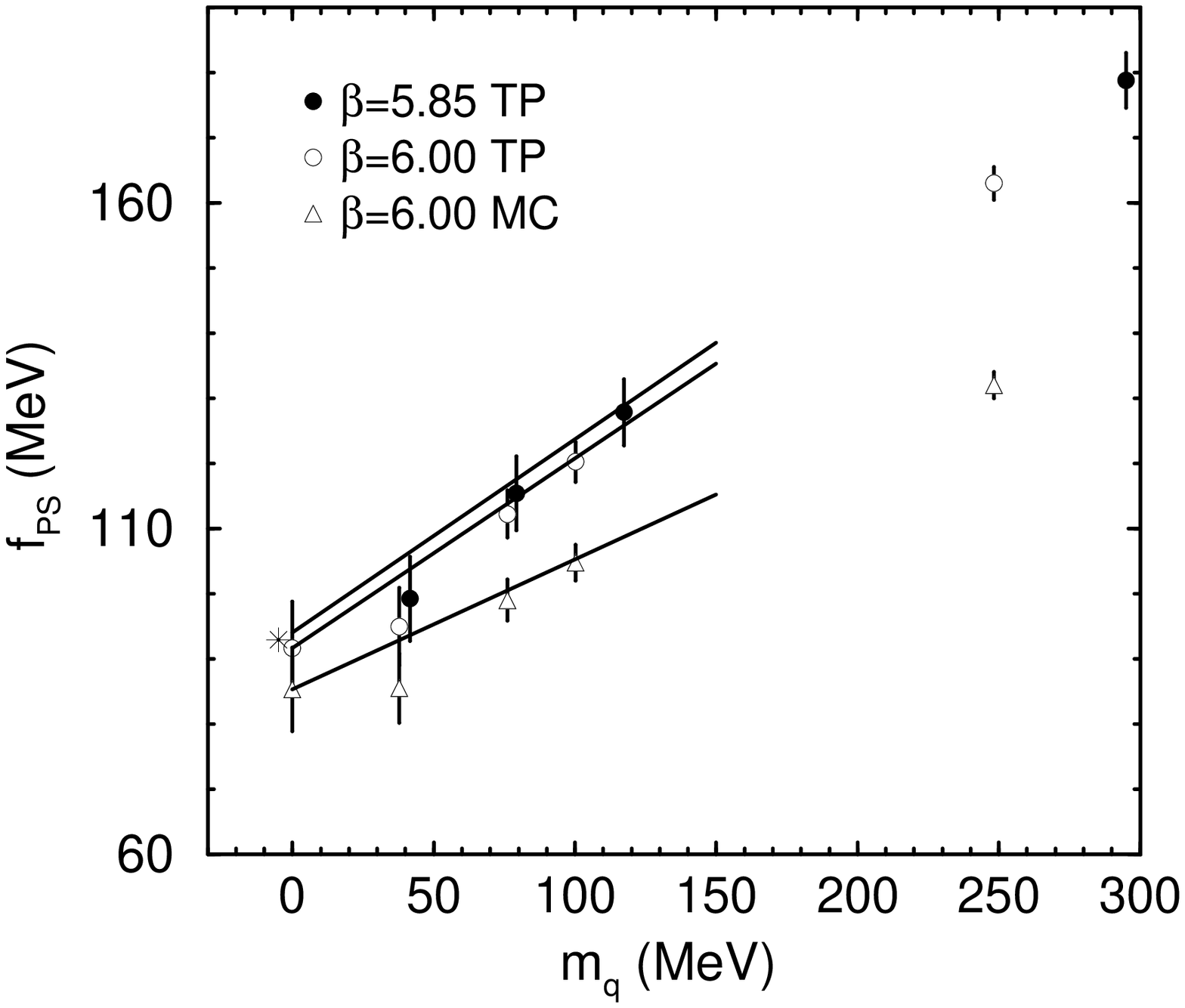}
\end{center}
\vskip -5mm
\caption{
Linear extrapolations of the pseudo-scalar 
meson decay constant,
for the two choices of renormalization constant 
discussed in the text.
The open symbols at zero quark mass are extrapolated values
for $\beta=6.0$.
The errors shown are statistical only.
The experimental value for the pion is marked with a star.}
\label{fig:Fp1K}
\end{figure}

\begin{figure}[htbp]
\begin{center}
\leavevmode
\epsfysize=14cm
  \epsfbox{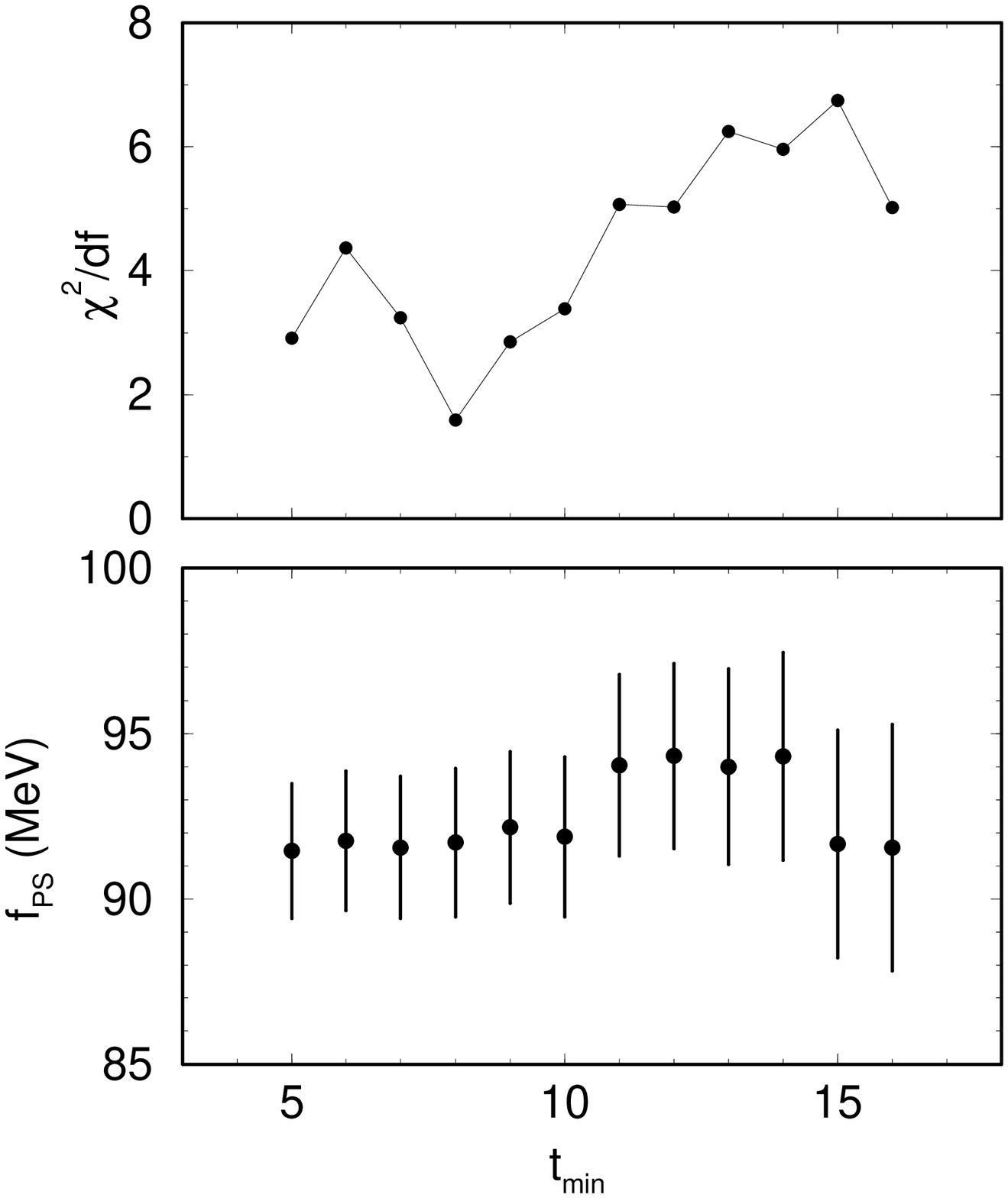}
\end{center}
\vskip -5mm
\caption{
Pseudo-scalar meson decay constant at zero quark mass
versus $t_{min}$,
together with $\chi^2/df$.
The errors are estimated by the least mean square fit.}
\label{fig:Fptmin}
\end{figure}

The extrapolation to the chiral limit is problematic.
We find that neither of the linear fit to the data 
at the three largest $K$'s
nor the quadratic fit to the data at the four largest $K$'s
gives $\chi^2/df$ small enough:
For $f_{PS}^{TP}$ at $\beta=5.85 \ (6.0)$,
$\chi^2/df =$ 9.1 (6.7) for the linear fit and
9.9 (7.4) for the quadratic fit, respectively.
Fits to $f_{PS}^{MC}$ are similar.
In Fig.\ \ref{fig:Fp1K} are shown $f_{PS}$
versus the quark mass together with the linear fits.
The data at the largest $K$ is much below the fitting lines.
Even if we change $t_{min}$, $\chi^2/df$ does not reduce much.
In Fig.\ \ref{fig:Fptmin} we show $\chi^2/df$ together with
the result for $f_{PS}^{TP}$
at $\beta=6.0$ versus $t_{min}$.
Although $\chi^2/df$ is large, 
the results of the fits are very stable.
Therefore we quote the decay constant 
obtained by the linear extrapolation
of the data with $t_{min}=12$ (15) 
at $\beta=5.85$ (6.0) as the central value of the decay constant.
We estimate the systematic errors similarly 
as in the previous cases with 
$t_0 = t_{\chi^2}$ --- 14 (16) for $\beta=5.85$ (6.0).

These analyses give
\begin{tabbing}
xxxxxxxxxxx \= xxxxxxxxxxxxx \= xxxxxxxxxxx 
\= xxxxxxxx \= xxxxxxxx \= xxxxxxx \= xxxxxxx \= xxxx \= \kill
                    \>                       \> stat.  
\> sys.(fit-range) \> \> sys.(fit-func.) \> \> \\ 
\ \ \ $\beta=$ 5.85 \> $f_{\pi}^{TP}=$ 0.0489 \> $\pm$0.0056 
\> +0.0008  \> $-$0.0017 \> +0.0 \> $-$0.0011 \> \\
\ \ \               \> $f_{\pi}^{TP}=$ 94.1   \> $\pm$11.8   
\> +1.6     \> $-$3.3 \> +0.0 \> $-$2.2 \> MeV \\
\ \ \ $\beta=$6.00  \> $f_{\pi}^{TP}=$ 0.0394 \> $\pm$0.0027 
\> +0.0011  \> $-$0.0    \> +0.0  \> $-$0.0013 \> \\
\ \ \               \> $f_{\pi}^{TP}=$ 91.7   \> $\pm$7.2    
\> +2.7     \> $-$0.0    \> +0.0  \> $-$3.0 \> MeV \\
\ \ \               \> $f_{\pi}^{MC}=$ 0.0367 \> $\pm$0.0024 
\> +0.0011  \> $-$0.0    \> +0.0  \> $-$ 0.0014 \> \\
\ \ \               \> $f_{\pi}^{MC}=$ 85.4   \> $\pm$6.4    
\> +2.5     \> $-$0.0    \> +0.0  \> $-$3.4 \> MeV \\
\end{tabbing}

The values of $f_{\pi}$ obtained with 
the tadpole improved renormalization constants 
are consistent with the experimental value 
within the statistical errors (see Fig.\ \ref{fig:Decay}).
That with the MC renormalization constant 
is also consistent with experiment
if we take account of the (small) systematic error.
However, we should take these numbers with caution,
because $\chi^2/df$ for the extrapolation 
is not small enough as mentioned above.
Note that the decay constants in the chiral limit are
consistent with the GF11 data~\cite{GF11DE},
although the errors in our results are considerably larger.

\section{Conclusions and Discussion} \label{sec:conclusions}
In analyses of numerical simulations 
toward high precision determination of light hadron masses,
one first encounters the problem of fitting range 
for hadron propagators.
We find that effective masses of hadrons in general
do not exhibit clear plateaus,
although statistics is relatively high 
(the number of configurations is 100 (200) 
at $\beta=5.85$ (6.0)).
The correlated $\chi^2$ fits do not determine unambiguously
the time slice beyond which the ground state dominates.
We also notice a very intriguing fact
that $m_{fit}$ by the correlated fits to a range from $t=t_0$
has a strong correlation with $m_{\it eff}$ at $t=t_0$.
Varying systematically the fitting range,
we estimate systematic errors in hadron masses
due to statistical fluctuations
as well as due to the contamination from excited states,
which cannot be properly taken into account 
by the standard least mean square fit 
with a fixed fitting range.
We find that
the systematic errors 
for the hadron masses with quarks
lighter than the strange quark
amount to 1 --- 2 times the statistical errors.

When the lattice scale is fixed from the $\rho$ meson mass,
the masses of the $\Omega^{-}$ baryon and the $\phi$ meson 
at two $\beta$'s
agree with experiment within about one standard deviation.
On the other hand, the central value of
the nucleon mass at $\beta=6.0$ (5.85)
is larger than its experimental value 
by about 15\% (20\%) and that of
the $\Delta$ mass by about 15\% (4\%):
Even when the systematic errors are included,
the baryon masses at $\beta=6.0$ do not agree with experiment.
In order to take the continuum limit of the nucleon mass
and the $\Delta$ mass,
we need data for a wider range of $\beta$ with  
statistical and systematic errors much reduced.
For the masses of excited states of 
the $\rho$ meson and the nucleon,
there exist two-mass fits which do not 
contradict with experiment,
except for the case of the nucleon at $\beta=6.0$.
Although this does not necessarily imply 
that the excited state masses appear consistent with experiment
because two-mass fits are very unstable,
the existence of such a fit consistent with experiment
encourages us to perform more works in this direction.

Determination of  meson decay constants is usually
accompanied by uncertainties of renormalization constants.
One can in principle employ any renormalization constant
such as that determined by naive perturbation theory or
tadpole improved perturbation theory.
We have indeed shown that when we use renormalization constants
given by tadpole improved perturbation theory,
although the decay constants for the $\phi$, $\rho$, 
$K$ and $\pi$ mesons are in general 
off experiment at finite lattice spacing,
for example, by 30 --- 40\% 
at $m_{\rho}a$ = 0.33 --- 0.40
in the case of the $F_{\phi}$,
they approach in the continuum limit 
toward values consistent with the experimental values.

It is, however, desirable to employ a renormalization constant
which gives weak $a$ dependence for the decay constants.
We have shown that when we use the renormalization constants
determined by Monte Carlo simulations,
the vector meson decay constants at two $\beta$'s 
remarkably agree with each other and reproduce 
the experimental values within the errors
for a wide range of the quark mass
with the chiral limit included.
This implies a strong advantage 
to apply renormalization constants
determined nonperturbatively.
For pseudo-scaler mesons, however, 
we find that although the decay constant $f_{PS}^{MC}$
in the chiral limit agrees with the experimental value of 
$f_{\pi}$ albeit with large errors,
it differs from the experimental value 
of $f_K$ by about 20\% at $m_{\rho} a=0.33$.
This discrepancy might be due to systematic errors in
the numerical calculation of $Z_A$.
These results imply the importance of more systematic 
nonperturbative determination
of the renormalization constants for various meson decays.

\vspace{20pt}
Numerical simulations are performed 
under the QCDPAX project which is
supported by the Grants-in-Aid of Ministry of Education,
Science and Culture (Nos.\ 62060001 and 02402003). 
Analyses of data are also supported in part by the Grants-in-Aid
of Ministry of Education, Science and Culture
(Nos.\ 07NP0401, 07640375 and 07640376).

\section*{Note Added}
After this work was completed,
three groups have reported results of high statistics studies
of the hadron spectrum~\cite{JLQCD,GERMAN,LANLNEW}
at $\beta=6.0$.
Their results are consistent with ours.

\vspace{2cm}

\newpage

\begin{table}[htb]
\begin{center}
\begin{tabular}{|c|rr|c|rr|c|}
 \hline
 \multicolumn{3}{|c}{$\beta=5.85$} & 
 \multicolumn{3}{|c|}{$\beta=6.0$} & 
 \smash{\lower2.ex\hbox{$m_{\pi}/m_{\rho}$}} \\
 \cline{1-6} 
 $K$ & \multicolumn{2}{|c|}{\#iteration} &
 $K$ & \multicolumn{2}{|c|}{\#iteration} & \\
 \hline
 0.1440 &   80 $\pm$ &   3 & 0.1450  &   90 $\pm$ &  3 & 0.97 \\
 0.1540 &  160 $\pm$ &  10 & 0.1520  &  160 $\pm$ & 10 & 0.87 \\
 0.1585 &  420 $\pm$ &  45 & 0.1550  &  380 $\pm$ & 40 & 0.70 \\
 0.1595 &  610 $\pm$ &  75 & 0.1555  &  430 $\pm$ & 45 & 0.64 \\
 0.1605 & 1850 $\pm$ & 410 & 0.1563  & 1110 $\pm$ &170 & 0.52 \\
 \hline
\end{tabular}
\end{center}
\caption{
Hopping parameters and average number 
of iterations used to solve quark propagators.
Approximate values for $m_{\pi}/m_{\rho}$ 
are also given.
Table \protect\ref{tab:ed} 
contains precise values for $m_{\pi}/m_{\rho}$.}
\label{tab:K}
\end{table}

\begin{table}[htb]
\begin{center}
\begin{tabular}{|c|rr|rr|rr|rr|}
 \hline
 \multicolumn{9}{|c|}{$\beta=5.85$} \\
 \hline
 \multicolumn{1}{|c}{$K$} & \multicolumn{2}{|c}{$\pi$} & 
 \multicolumn{2}{|c}{$\rho$} & \multicolumn{2}{|c}{$N$} &
 \multicolumn{2}{|c|}{$\Delta$} \\
 \hline
 \ &  $t_{\chi^2}$ & $\chi^2/df$ & 
      $t_{\chi^2}$ & $\chi^2/df$ & 
      $t_{\chi^2}$ & $\chi^2/df$ & 
      $t_{\chi^2}$ & $\chi^2/df$   \\[2mm] 
 0.1440 & 12 & 0.98 & 12 & 1.32 & 11 & 1.33 & 11 & 1.49 \\
 0.1540 & 10 & 0.90 & 12 & 1.01 & 11 & 1.61 & 11 & 1.61 \\
 0.1585 &  8 & 0.72 &  8 & 2.04 &  9 & 1.36 & 11 & 1.18 \\
 0.1595 &  8 & 0.45 &  8 & 1.73 &  9 & 1.13 & 11 & 1.07 \\
 0.1605 &  8 & 0.46 &  8 & 1.20 &  7 & 1.56 &  9 & 1.30 \\
 \hline
 \multicolumn{9}{|c|}{$\beta=6.0$} \\
 \hline
 \multicolumn{1}{|c}{$K$} & \multicolumn{2}{|c}{$\pi$} & 
 \multicolumn{2}{|c}{$\rho$} & \multicolumn{2}{|c}{$N$} &
 \multicolumn{2}{|c|}{$\Delta$} \\
 \hline
 \ &  $t_{\chi^2}$ & $\chi^2/df$ &
      $t_{\chi^2}$ & $\chi^2/df$ & 
      $t_{\chi^2}$ & $\chi^2/df$ &
      $t_{\chi^2}$ & $\chi^2/df$   \\[2mm] 
 0.1450 & 15 & 0.55 & 15 & 1.02 & 15 & 0.36 & 15 & 0.57 \\
 0.1520 & 12 & 1.26 & 13 & 0.71 & 15 & 0.38 & 15 & 0.56 \\
 0.1550 & 10 & 1.39 & 11 & 1.42 & 12 & 0.41 & 12 & 0.94 \\
 0.1555 & 10 & 1.35 & 10 & 1.32 & 12 & 0.64 & 12 & 1.22 \\
 0.1563 &  9 & 1.54 &  9 & 0.95 & 10 & 1.21 & 11 & 1.11 \\
 \hline
 \end{tabular}
\end{center}
\caption{
$t_{\chi^2}$ and $\chi^2/df$ at $t_{\chi^2}$.
See the text for details.}
\label{tab:chi2}
\end{table}

\begin{table}[htb]
\begin{center}
 \begin{tabular}{|clr|clr|}
 \hline
 \multicolumn{3}{|c|}{$\beta=5.85$} & 
 \multicolumn{3}{c|}{$\beta=6.0$} \\
 \hline
 $K$ & \multicolumn{1}{c}{$m_{\pi}$} & $\chi^2/df$ &
 $K$ & \multicolumn{1}{c}{$m_{\pi}$} & $\chi^2/df$ \\
  \hline
  0.1440 & 1.0293(12)$^{+2}_{-2}$  & 13.7/14 & 
  0.1450 & 0.8069(7)$^{+0}_{-2}$   &  6.1/11 \\
  0.1540 & 0.6122(11)$^{+3}_{-6}$  & 10.9/14 & 
  0.1520 & 0.4772(9)$^{+9}_{-2}$   & 11.5/11 \\
  0.1585 & 0.3761(12)$^{+8}_{-4}$  &  7.4/14 & 
  0.1550 & 0.2967(15)$^{+18}_{-5}$ & 14.1/11 \\
  0.1595 & 0.3088(14)$^{+6}_{-6}$  &  5.8/14 & 
  0.1555 & 0.2588(16)$^{+18}_{-6}$ & 17.0/11 \\
  0.1605 & 0.2226(21)$^{+10}_{-7}$ &  6.0/14 &
  0.1563 & 0.1847(27)$^{+20}_{-6}$ & 20.9/11 \\
  \hline
  \end{tabular}
\end{center}
\caption{
Pion masses in lattice units.
In parentheses are errors estimated by the jack-knife method.
Errors given in the form $\mbox{}^{+upper}_{-lower}$
are for the fitting range dependent upper/lower bound.}
\label{tab:mpi}
\end{table}

\begin{table}[htb]
\begin{center}
 \begin{tabular}{|clr|clr|}
 \hline
 \multicolumn{3}{|c|}{$\beta=5.85$} & 
 \multicolumn{3}{c|}{$\beta=6.0$} \\
 \hline
 $K$ & \multicolumn{1}{c}{$m_{\rho}$} & $\chi^2/df$ &
 $K$ & \multicolumn{1}{c}{$m_{\rho}$} & $\chi^2/df$ \\
 \hline
 0.1440 & 1.0598(15)$^{+4}_{-1}$      &  18.5/14 &
 0.1450 & 0.8370(9)$^{+2}_{-4}$       &  11.2/11 \\
 0.1540 & 0.6931(27)$^{+3}_{-11}$     &  14.2/14 &
 0.1520 & 0.5486(15)$^{+9}_{-11}$     &   6.6/11 \\
 0.1585 & 0.5294(69)$^{+115}_{-100}$  &  28.9/14 &
 0.1550 & 0.4218(42)$^{+75}_{-73}$    &  14.3/11 \\
 0.1595 & 0.4856(96)$^{+176}_{-123}$  &  23.2/14 &
 0.1555 & 0.3982(61)$^{+135}_{-90}$   &  12.4/11 \\
 0.1605 & 0.434(20)$^{+21}_{-24}$     &  14.6/14 &
 0.1563 & 0.353(15)$^{+28}_{-11}$     &   7.6/11 \\
  \hline
  \end{tabular}
\end{center}
\caption{
The same as table \protect\ref{tab:mpi} 
for the $\rho$ meson.}
\label{tab:mrho}
\end{table}

\begin{table}[htpb]
\begin{center}
 \begin{tabular}{|lllll|}
 \hline
 \multicolumn{5}{|c|}{$K$=0.155} \\
 \hline
 \ & \multicolumn{1}{c}{$\pi$}
   & \multicolumn{1}{c}{$\rho$}
   & \multicolumn{1}{c}{$N$}
   & \multicolumn{1}{c|}{$\Delta$} \\
This work $24^3\times 54$               & 
0.2967(15) & 0.4218(42) & 0.6440(85) & 0.728(11) \\     
Ape $24^3\times 32$ \cite{APE24}    &
0.298(2) & 0.429(3) & 0.647(6) & 0.745(15) \\     
Ape $18^3\times 64$ \cite{APE18}  smear        &
0.297(2) &  0.430(10)&           &             \\        
\hspace{3cm} local     &
0.297(2) &  0.428(8) &            &             \\        
LANL $32^3\times 64$ \cite{LANL}  &
0.297(1) &  0.422(3) & 0.641(4) & 0.706(8)  \\  
  \hline
  \multicolumn{5}{|c|}{$K$=0.1563} \\
\hline
 \ & \multicolumn{1}{c}{$\pi$}
   & \multicolumn{1}{c}{$\rho$}
   & \multicolumn{1}{c}{$N$}
   & \multicolumn{1}{c|}{$\Delta$} \\
This work $24^3\times 54$               &
0.1847(27) & 0.353(15) & 0.536(30) & 0.670(53) \\
Ape $24^3\times 32$ \cite{APE24}    &
0.184(3) & 0.377(8) & 0.522(14) & 0.636(45) \\
LANL $32^3\times 64$ \cite{LANL}    &
0.185(1)  & 0.363(9) & 0.540(12) & 0.631(27) \\
\hline
 \end{tabular}
\end{center}
\caption{
Comparison of hadron masses in lattice units
at $\beta=6.0$, $K=0.155$ and 0.1563.}
\label{tab:mK15}
\end{table}

\newpage

\begin{table}[htb]
\begin{center}
 \begin{tabular}{|clr|clr|}
 \hline
 \multicolumn{3}{|c|}{$\beta=5.85$} & 
 \multicolumn{3}{c|}{$\beta=6.0$} \\
 \hline
 $K$ & \multicolumn{1}{c}{$m_{N}$} & $\chi^2/df$ &
 $K$ & \multicolumn{1}{c}{$m_{N}$} & $\chi^2/df$ \\
 \hline
  0.1440 & 1.6961(50)$^{+7}_{-18}$    & 15.8/13 & 
  0.1450 & 1.3225(28)$^{+15}_{-2}$    &  3.8/10 \\
  0.1540 & 1.1060(55)$^{+15}_{-94}$   & 22.3/13 & 
  0.1520 & 0.8669(49)$^{+19}_{-3}$    &  4.2/10 \\
  0.1585 & 0.815(13)$^{+13}_{-33}$    & 17.5/13 & 
  0.1550 & 0.6440(85)$^{+53}_{-12}$   &  3.8/10 \\
  0.1595 & 0.744(17)$^{+12}_{-36}$    & 18.1/13 & 
  0.1555 & 0.6007(109)$^{+84}_{-7}$   &  6.2/10 \\
  0.1605 & 0.683(48)$^{+10}_{-82}$    & 23.4/13 & 
  0.1563 & 0.536(30)$^{+58}_{-0}$     & 15.7/10 \\
 \hline
 \end{tabular}
\end{center}
\caption{
Nucleon masses in lattice units.
In parentheses are errors estimated by the jack-knife method.
Errors given in the form $\mbox{}^{+upper}_{-lower}$
are for the fitting range dependent upper/lower bound.}
\label{tab:mN}
\end{table}

\begin{table}[htb]
\begin{center}
 \begin{tabular}{|clr|clr|}
 \hline
 \multicolumn{3}{|c|}{$\beta=5.85$} & 
 \multicolumn{3}{c|}{$\beta=6.0$} \\
 \hline
 $K$ & \multicolumn{1}{c}{$m_{\Delta}$} & $\chi^2/df$ &
 $K$ & \multicolumn{1}{c}{$m_{\Delta}$} & $\chi^2/df$ \\
 \hline
  0.1440 & 1.7124(57)$^{+21}_{-25}$   & 17.7/13 & 
  0.1450 & 1.3404(29)$^{+22}_{-1}$    &  6.2/10 \\
  0.1540 & 1.1629(67)$^{+37}_{-15}$   & 20.3/13 & 
  0.1520 & 0.9112(41)$^{+51}_{-0}$    &  6.0/10 \\
  0.1585 & 0.9011(153)$^{+83}_{-57}$  & 16.5/13 & 
  0.1550 & 0.7278(109)$^{+188}_{-0}$  & 12.1/10 \\
  0.1595 & 0.825(21)$^{+16}_{-37}$    & 15.1/13 & 
  0.1555 & 0.7001(159)$^{+336}_{-10}$ & 15.3/10 \\
  0.1605 & 0.755(53)$^{+67}_{-78}$    & 19.4/13 & 
  0.1563 & 0.670(53)$^{+61}_{-41}$    &  9.0/10 \\
 \hline
 \end{tabular}
\end{center}
\caption{
The same as table \protect\ref{tab:mN} 
for the $\Delta$ baryon.}
\label{tab:mD}
\end{table}

\begin{table}[htb]
\begin{center}
  \begin{tabular}{|lll|lll|}
  \hline
  \multicolumn{3}{|c|}{$\beta=5.85$} & 
  \multicolumn{3}{c|}{$\beta=6.0$} \\
  \hline
  $K$ & $m_{\pi}/m_{\rho}$ & $m_N/m_{\rho}$ & 
  $K$ & $m_{\pi}/m_{\rho}$ & $m_N/m_{\rho}$ \\ 
  \hline
0.1440& 0.9712(8) & 1.6004(45) & 0.1450 & 0.9641(5) & 1.5801(25) \\
0.1540& 0.8833(32) & 1.5956(82) & 0.1520 & 0.8699(21) & 1.5802(79) \\
0.1585& 0.7104(90) & 1.540(29)  & 0.1550 & 0.7033(69) & 1.527(21)  \\
0.1595& 0.636(12)  & 1.531(42)  & 0.1555 & 0.650(10)  & 1.509(31)  \\
0.1605& 0.513(25)  & 1.57(12)   & 0.1563 & 0.523(23)  & 1.52(10)   \\
  \hline
  \end{tabular}
\end{center}
\caption{
Mass ratios $m_{\pi}/m_{\rho}$ and $m_N/m_{\rho}$.
The errors quoted are statistical only and are estimated by 
the jack-knife method.}
\label{tab:ed}
\end{table}

\begin{table}[htb]
\begin{center}
  \begin{tabular}{|ll|ll|ll|}
  \hline
   & & \multicolumn{2}{|c|}{$\rho$ meson} &
       \multicolumn{2}{c|}{nucleon} \\
  \hline 
          & $\beta$ & comment & ratio &
                      comment & ratio \\
  \hline
   This work                  & 5.85  &
$t_{min}$=5 & 2.47(16) & $t_{min}=6$ & 1.64(12) \\
                              &       &
$t_{min}$=6 & 1.87(24) & $t_{min}=7$ & 1.29(10) \\       
                              & 6.0   &
$t_{min}$=8 & 2.21(27) & $t_{min}=7$ & 1.81(10) \\    
                              &       &
$t_{min}$=9 & 1.58(26) &      &          \\   
   Ape \cite{APE24}       & 6.0   &
         & 2.13(21) &         & 2.13(4) \\   
   UKQCD \cite{UKQCDEX}   & 6.2   &
         & 2.53(16) &         & 2.01(16) \\   
   APE \cite{APE63}       & 6.3   &
         & 1.93(10) &         & 1.93(12) \\
   UKQCD \cite{UKQCDJ}    & 6.2 Clover  &
         & 2.23(14) &         &          \\
   QCDPAX \cite{QCDPAX93} & 6.0   &  point      &
           1.99(15) & point   & 1.55(20) \\
                              &       &  wall       &
           1.70(26) &  wall   & 1.47(21) \\  
          experimental value  &       &             &
           1.65     &         &          \\
  \hline
  \end{tabular}
\end{center}
\caption{
Ratios of the excited state mass to the ground state mass.
We have taken the quark mass corresponding approximately to
the strange quark mass.}
\label{tab:ex}
\end{table}

\begin{table}[htb]
\begin{center}
  \begin{tabular}{|l|lll|lll|}
  \hline
  \ & \multicolumn{3}{|c|}{$\beta=5.85$} & 
      \multicolumn{3}{c|}{$\beta=6.0$} \\ 
  \hline
  \ &  $a_0$ & $a_1$ & $\chi^2/df$ &   
       $a_0$ & $a_1$ & $\chi^2/df$ \\
  \hline
$m_{\pi}^2$  &  $-$7.18(4) & 1.16(1) & 0.56 &
              $-$6.51(6)   & 1.02(1) & 1.06 \\
$m_{\rho}$   &  $-$6.16(37) & 1.06(6) & 1.76 &
              $-$6.50(39)   & 1.07(6) & 1.20 \\
$m_N$        & $-$10.87(51) & 1.85(8) & 0.37 &
              $-$12.97(79)  & 2.11(12) & 0.05 \\
$m_{\Delta}$ & $-$11.37(80) & 1.95(13) & 0.05 &
              $-$8.4(1.3)   & 1.42(21) & 0.29 \\
  \hline
  \end{tabular}
\end{center}
\caption{
Fit parameters of the linear fits to the masses 
at the largest three $K$'s.
Errors on $a_0$ and $a_1$ are those from least mean square fits.}
\label{tab:lf}
\end{table}

\begin{table}[htb]
\begin{center}
  \begin{tabular}{|lllll|}
  \hline
  \multicolumn{5}{|c|}{$\beta=5.85$} \\
  \hline
  \ & value & err-lms & err-jack & jack/lms \\
  $K_c$             & 0.161624 & 0.000027 & 0.000033 & 1.2 \\
  $m_{\rho}(K_c)$   & 0.400    & 0.010    & 0.021    & 2.1 \\
  $m_N(K_c)$        & 0.589    & 0.014    & 0.036    & 2.6 \\
  $m_{\Delta}(K_c)$ & 0.664    & 0.022    & 0.063    & 2.9 \\
  \hline
  \multicolumn{5}{|c|}{$\beta=6.0$} \\
  \hline
  \ & value & err-lms & err-jack & jack/lms \\
  $K_c$             & 0.157096 & 0.000038 & 0.000028 & 0.7 \\
  $m_{\rho}(K_c)$   & 0.3309   & 0.0080   & 0.0114   & 1.4 \\
  $m_N(K_c)$        & 0.462    & 0.015    & 0.024    & 1.6 \\
  $m_{\Delta}(K_c)$ & 0.605    & 0.025    & 0.033    & 1.3 \\
  \hline
  \end{tabular}
\end{center}
\caption{
Values of $K_c$ and masses extrapolated to $K_c$ 
determined from the linear fits to the data 
at the largest three $K$'s.
Errors obtained by least mean square fits (err-lms) and those
by the jack-knife method (err-jack) together with their ratios
(jack/lms) are also given.}
\label{tab:Kc}
\end{table}

\begin{table}[htb]
\begin{center}
  \begin{tabular}{|rrrr|rrrr|}
  \hline
  \multicolumn{4}{|c|}{$\beta=5.85$} &
  \multicolumn{4}{c|}{$\beta=6.0$} \\
  \hline
  $t_0$ & $m_{\rho}$ & $a^{-1}$ & $\chi^2/df$ &
  $t_0$ & $m_{\rho}$ & $a^{-1}$ & $\chi^2/df$ \\
  \hline
   8 & 0.4359 & 1.766 &  36.90 & 11 & 0.3525 & 2.184 &  2.51 \\
   9 & 0.4213 & 1.828 &  11.17 & 12 & 0.3476 & 2.215 &  2.54 \\
  10 & 0.4196 & 1.835 &  14.80 & 13 & 0.3425 & 2.248 &  0.77 \\
  11 & 0.4045 & 1.904 &   9.30 & 14 & 0.3409 & 2.259 &  3.94 \\
  12 & 0.3998 & 1.926 &   1.76 & 15 & 0.3309 & 2.327 &  1.20 \\
  13 & 0.3892 & 1.978 &   4.39 & 16 & 0.3248 & 2.370 &  0.17 \\
  14 & 0.4081 & 1.887 &   1.36 & 17 & 0.3188 & 2.416 &  0.58 \\
  15 & 0.3794 & 2.030 &   0.43 & 18 & 0.3112 & 2.474 &  0.03 \\
  16 & 0.3728 & 2.066 &   0.11 & 19 & 0.3206 & 2.401 &  0.00 \\
  \hline
  \end{tabular}
\end{center}
\caption{
Results of the linear fits to the $\rho$ meson
masses  versus $t_0$.
The inverse lattice spacing is defined by
$a^{-1} = 0.77 {\rm GeV}/m_{\rho}(K_c)$.}
\label{tab:rho-1}
\end{table}

\begin{table}[htb]
\begin{center}
  \begin{tabular}{|rrrr|rrrr|}
  \hline
  \multicolumn{4}{|c|}{$\beta=5.85$} &
  \multicolumn{4}{c|}{$\beta=6.0$} \\
  \hline
  $t_0$ & $m_{\rho}$ & $a^{-1}$ & $\chi^2/df$ &
  $t_0$ & $m_{\rho}$ & $a^{-1}$ & $\chi^2/df$ \\
  \hline
  12 & 0.3881 & 1.984 & 1.22 &  13 & 0.3413 & 2.256 & 0.79 \\ 
  13 & 0.3767 & 2.044 & 4.02 &  14 & 0.3393 & 2.269 & 4.33 \\
  14 & 0.3997 & 1.927 & 1.20 &  15 & 0.3253 & 2.367 & 0.91 \\
  15 & 0.3641 & 2.115 & 0.21 &  16 & 0.3194 & 2.411 & 0.07 \\
  16 & 0.3593 & 2.143 & 0.30 &  17 & 0.3116 & 2.471 & 0.41 \\
     &        &       &      &  18 & 0.3029 & 2.542 & 0.00 \\
     &        &       &      &  19 & 0.3146 & 2.448 & 0.01 \\
  \hline
  \end{tabular}
\end{center}
\caption{
Results of the quadratic fits 
to the $\rho$ meson masses versus $t_0$.
The inverse lattice spacing is defined by
$a^{-1} = 0.77 {\rm GeV}/m_{\rho}(K_c)$.}
\label{tab:rho-2}
\end{table}

\begin{table}[htb]
\begin{center}
  \begin{tabular}{|rrr|rrr|}
  \hline
  \multicolumn{3}{|c|}{$\beta=5.85$} &
  \multicolumn{3}{c|}{$\beta=6.0$} \\
  \hline
  $t_0$ & $m_{N}$ & $\chi^2/df$ & 
  $t_0$ & $m_{N}$ & $\chi^2/df$ \\
  \hline
   9 & 0.6085 & 3.20 &  12 & 0.4828 & 0.03 \\
  10 & 0.6071 & 0.37 &  13 & 0.4802 & 0.24 \\
  11 & 0.6039 & 0.83 &  14 & 0.4758 & 0.79 \\
  12 & 0.5946 & 2.15 &  15 & 0.4759 & 1.88 \\
  13 & 0.5893 & 0.37 &  16 & 0.4623 & 0.05 \\
  14 & 0.5680 & 2.28 &  17 & 0.4538 & 2.00 \\
  15 & 0.5501 & 1.59 &  18 & 0.4553 & 2.28 \\
  16 & 0.5312 & 0.49 &  19 & 0.4731 & 0.10 \\
  17 & 0.5630 & 2.55 &  20 & 0.4559 & 0.10 \\
  \hline
  \end{tabular}
\end{center}
\caption{
Results of the linear fits to the nucleon 
masses versus $t_0$.}
\label{tab:N-1}
\end{table}

\begin{table}[htb]
\begin{center}
  \begin{tabular}{|rrr|rrr|}
  \hline
  \multicolumn{3}{|c|}{$\beta=5.85$} &
  \multicolumn{3}{c|}{$\beta=6.0$} \\
  \hline
  $t_0$ & $m_{\Delta}$ & $\chi^2/df$ & 
  $t_0$ & $m_{\Delta}$ & $\chi^2/df$ \\
  \hline
  11 & 0.6982 &  0.03 &  12 & 0.6055 &  0.01 \\
  12 & 0.6928 &  0.01 &  13 & 0.6059 &  4.10 \\
  13 & 0.6640 &  0.05 &  14 & 0.6164 &  1.78 \\
  14 & 0.6899 &  4.02 &  15 & 0.6279 &  0.50 \\
  15 & 0.5375 & 34.96 &  16 & 0.6048 &  0.29 \\
  16 & 0.4757 & 23.96 &  17 & 0.6206 &  0.12 \\
     &        &       &  18 & 0.6462 &  0.17 \\
     &        &       &  19 & 0.5935 &  0.34 \\
  \hline
  \end{tabular}
\end{center}
\caption{
Results of the linear fits to the $\Delta$ 
masses versus $t_0$.}
\label{tab:D-1}
\end{table}

\begin{table}[htb]
\begin{center}
  \begin{tabular}{|cl|}
  \hline
  $K$ & \multicolumn{1}{c|}{$Z_V$} \\
  \hline
   0.1440  &  0.5121(9) \\
   0.1540  &  0.5164(10) \\
   0.1585  &  0.5126(36) \\ 
   0.1595  &  0.5112(48) \\
   0.1605  &  0.5101(76) \\
   \hline
   Average &  0.5125(30) \\
  \hline
  \end{tabular}
\end{center}
\caption{
Renormalization constants $Z_V$ 
for the local lattice current at $\beta=5.85$ 
obtained in a previous work \protect\cite{STDB585}.}
\label{tab:ZV}
\end{table}

\begin{table}[htb]
\begin{center}
  \begin{tabular}{|rrlr|rrlr|}
  \hline
  \multicolumn{4}{|c|}{$\beta=5.85$} &
  \multicolumn{4}{c|}{$\beta=6.0$}   \\
  \hline
  $K$ & $F_V^{PT}$ &  
  \multicolumn{1}{c}{$F_V^{TP}$} & $F_V^{MC}$ &
  $K$ & $F_V^{PT}$ &  
  \multicolumn{1}{c}{$F_V^{TP}$} & $F_V^{MC}$ \\
  \hline
 0.1440 & 0.2214 & 0.2600(33)$^{+17}_{-0}$     & 0.1374 &
 0.1450 & 0.1753 & 0.1919(18)$^{+5}_{-13}$     & 0.1210 \\
 0.1540 & 0.2211 & 0.2089(38)$^{+5}_{-17}$     & 0.1372 &
 0.1520 & 0.1673 & 0.1558(17)$^{+14}_{-19}$    & 0.1155 \\
 0.1585 & 0.2094 & 0.1781(68)$^{+126}_{-130}$  & 0.1299 &
 0.1550 & 0.1544 & 0.1336(33)$^{+65}_{-78}$    & 0.1065 \\
 0.1595 & 0.1996 & 0.1658(84)$^{+162}_{-142}$  & 0.1239 &
 0.1555 & 0.1493 & 0.1276(43)$^{+106}_{-88}$   & 0.1031 \\
 0.1605 & 0.1885 & 0.1528(152)$^{+153}_{-201}$ & 0.1170 &
 0.1563 & 0.1382 & 0.1157(95)$^{+181}_{-87}$   & 0.0953 \\
  \hline
  \end{tabular}
\end{center}
\caption{
$\rho$ meson decay constants in lattice units.
In parentheses are errors estimated by the jack-knife method.
Errors given in the form $\mbox{}^{+upper}_{-lower}$
are for the fitting range dependent upper/lower bound.}
\label{tab:fV}
\end{table}

\begin{table}[htb]
\begin{center}
  \begin{tabular}{|r|rrr|rrr|r|rrr|rrr|}
  \hline
  \multicolumn{7}{|c|}{$\beta=5.85$} &
  \multicolumn{7}{c|}{$\beta=6.0$} \\
  \hline
  & \multicolumn{3}{|c|}{$F_V^{TP}$} &
    \multicolumn{3}{c|}{$F_V^{MC}$}  &
  & \multicolumn{3}{|c|}{$F_V^{TP}$} &
    \multicolumn{3}{c|}{$F_V^{MC}$}  \\
  \hline
  $t_0$ & {\tenrm Latt.} & {\tenrm Phys.} & {\small $\chi^2/df$} &
          {\tenrm Latt.} & {\tenrm Phys.} & {\small $\chi^2/df$} & 
  $t_0$ & {\tenrm Latt.} & {\tenrm Phys.} & {\small $\chi^2/df$} &
          {\tenrm Latt.} & {\tenrm Phys.} & {\small $\chi^2/df$} \\
   8 & 0.169 & 325 & 53.7  & 0.133 & 257 & 55.7 & 
  11 & 0.127 & 297 &  1.9  & 0.107 & 250 &  2.0 \\
   9 & 0.157 & 303 &  6.0  & 0.125 & 240 &  7.0 &
  12 & 0.124 & 289 &  1.7  & 0.105 & 244 &  1.9 \\
  10 & 0.156 & 301 & 11.1  & 0.124 & 239 & 11.9 &
  13 & 0.120 & 280 &  0.1  & 0.102 & 237 &  0.2 \\
  11 & 0.145 & 280 &  5.8  & 0.115 & 222 &  6.3 &
  14 & 0.120 & 278 &  3.4  & 0.101 & 235 &  3.6 \\
  12 & 0.141 & 271 &  0.0  & 0.112 & 216 &  0.1 &
  15 & 0.111 & 259 &  0.4  & 0.094 & 220 &  0.4 \\
  13 & 0.130 & 251 &  2.3  & 0.104 & 201 &  2.6 &
  16 & 0.107 & 248 &  0.0  & 0.091 & 211 &  0.0 \\
  14 & 0.148 & 286 &  0.3  & 0.118 & 226 &  0.3 &
  17 & 0.101 & 234 &  0.1  & 0.085 & 199 &  0.1 \\
  15 & 0.116 & 224 &  0.0  & 0.094 & 180 &  0.0 &
  18 & 0.094 & 220 &  0.1  & 0.080 & 187 &  0.1 \\
  16 & 0.105 & 203 &  1.1  & 0.085 & 164 &  1.1 &
  19 & 0.104 & 243 &  0.1  & 0.088 & 206 &  0.1 \\
  \hline
  \end{tabular}
\end{center}
\caption{
Results of the linear fits to the $\rho$ 
meson decay constants versus $t_0$, 
in lattice units and in physical units (MeV).}
\label{tab:fV-1}
\end{table}

\begin{table}[htb]
\begin{center}
  \begin{tabular}{|c|lr|rcr|c|lr|rcr|}
  \hline
  \multicolumn{6}{|c}{$\beta=5.85$} &
  \multicolumn{6}{|c|}{$\beta=6.0$} \\
  \hline
  & \multicolumn{2}{|c|}{$t_0=t_{min}=12$} &
    \multicolumn{3}{|c|}{$t_0=t_{\chi^2}$} &
  & \multicolumn{2}{|c|}{$t_0=t_{min}=15$} &
    \multicolumn{3}{|c|}{$t_0=t_{\chi^2}$} \\
  \hline
  $K$ & \multicolumn{1}{c}{$m_{\tilde\pi}$} & 
        {\small $\chi^2/df$} &
{\small $t_{\chi^2}$} & $m_{\tilde\pi}$ & {\small $\chi^2/df$} &
  $K$ & \multicolumn{1}{c}{$m_{\tilde\pi}$} &
        {\small $\chi^2/df$} &
{\small $t_{\chi^2}$} & $m_{\tilde\pi}$ & {\small $\chi^2/df$} \\
 0.1440 & 1.0299(14) & 1.32 &  8 & 1.0304 & 1.68 &
 0.1450 & 0.8059(9) & 0.49 & 12 & 0.8068 & 1.06 \\
 0.1540 & 0.6106(21) & 0.96 &  7 & 0.6117 & 1.18 &
 0.1520 & 0.4747(14) & 0.35 &  7 & 0.4767 & 0.91 \\
 0.1585 & 0.3753(34) & 0.97 &  5 & 0.3774 & 1.41 &
 0.1550 & 0.2937(24) & 0.88 &  6 & 0.2967 & 0.89 \\
 0.1595 & 0.3070(42) & 1.00 &  4 & 0.3097 & 1.45 &
 0.1555 & 0.2559(30) & 0.85 &  6 & 0.2593 & 0.83 \\
 0.1605 & 0.2127(64) & 1.02 &  4 & 0.2175 & 1.31 &
 0.1563 & 0.1804(66) & 0.68 &  5 & 0.1897 & 0.75 \\
  \hline
  \end{tabular}
\end{center}
\caption{
Pion masses determined from $\tilde\pi$ propagators.}
\label{tab:pit}
\end{table}

\begin{table}[htb]
\begin{center}
  \begin{tabular}{|rrr|rrrr|}
  \hline
  \multicolumn{3}{|c|}{$\beta=5.85$} &
  \multicolumn{4}{c|}{$\beta=6.0$}   \\
  \hline
  $K$ & $f_{PS}^{PT}$ &  
  \multicolumn{1}{c|}{$f_{PS}^{TP}$} & 
  $K$ & $f_{PS}^{PT}$ &  
  \multicolumn{1}{c}{$f_{PS}^{TP}$} & $f_{PS}^{MC}$ \\
  \hline
 0.1440 & 0.1152 & 0.1443(25)$^{+12}_{-5}$  &
 0.1450 & 0.0892 & 0.1030(10)$^{+12}_{-0}$  & 0.0710 \\ 	
 0.1540 & 0.0922 & 0.0929(22)$^{+10}_{-1}$  &
 0.1520 & 0.0713 & 0.0701(11)$^{+18}_{-0}$  & 0.0567 \\
 0.1585 & 0.0732 & 0.0664(26)$^{+12}_{-14}$ &
 0.1550 & 0.0566 & 0.0517(13)$^{+16}_{-10}$ & 0.0450 \\
 0.1595 & 0.0677 & 0.0600(30)$^{+19}_{-15}$ &
 0.1555 & 0.0535 & 0.0482(15)$^{+15}_{-15}$ & 0.0426 \\
 0.1605 & 0.0597 & 0.0515(33)$^{+32}_{-9}$  &
 0.1563 & 0.0462 & 0.0408(26)$^{+30}_{-23}$ & 0.0368 \\
  \hline
  \end{tabular}
\end{center}
\caption{
Pseudo-scalar meson decay constants in lattice units.
In parentheses are errors estimated by the jack-knife method.
Errors given in the form $\mbox{}^{+upper}_{-lower}$
are for the fitting range dependent upper/lower bound.}
\label{tab:fPS}
\end{table}

\clearpage

\newpage

\clearpage

\begin{thebibliography}{99}
\bibitem{STD1648B585}
Y.~Iwasaki and T.~Yoshi\'e, Phys.~Lett.\ {\bf B216} (1989) 387;
Y.~Iwasaki, Nucl.~Phys.~{\bf B} (Proc.~Suppl.) {\bf 9} (1989) 254.
\bibitem{STDB585}
T.~Yoshi\'e, Y.~Iwasaki and S.~Sakai, 
Nucl.~Phys.~{\bf B} (Proc.~Suppl.) {\bf 17} (1990) 413.
%
\bibitem{APE8889}
Ape Collaboration (P.~Bacilieri {\it et al.}),
Phys.~Lett. {\bf B214} (1988) 115;
Ape Collaboration (P.~Bacilieri {\it et al.}),
Nucl.~Phys. {\bf B317} (1989) 509;
Ape Collaboration (S.~Cabasino {\it et al.}),
Nucl.~Phys.~{\bf B} (Proc.~Suppl.) {\bf 17} (1990) 431.
%
\bibitem{QCDPAX92}
QCDPAX Collaboration (Y.~Iwasaki {\it et al.}),
Nucl.~Phys.~{\bf B} (Proc.Suppl.) {\bf 30} (1993) 397.
%
\bibitem{QCDPAX93}
QCDPAX Collaboration (Y.~Iwasaki {\it et al.}),
Nucl.~Phys.~{\bf B} (Proc.Suppl.) {\bf 34} (1994) 354.
%
\bibitem{APE18}
Ape Collaboration (C.~R.~Allton {\it et al.}),
Nucl.~Phys.~{\bf B} (Proc.Suppl.) {\bf 34} (1994) 360. 
%
\bibitem{APE24}
Ape Collaboration (S.~Cabasino {\it et al.}),
Phys.~Lett. {\bf B258} (1991) 195.
%
\bibitem{APE63}
Ape Collaboration (M.~Guagnelli {\it et al.}), 
Nucl.~Phys. {\bf B378} (1992) 616.
%
\bibitem{HEMCGC}
K.~M.~Bitar {\it et al.}, Phys.~Rev. {\bf D46} (1992) 2169.
%
\bibitem{GF11}
F~.Butler, H.~Chen, J.~Sexton, A.~Vaccarino and D.~Weingarten,
Nucl.~Phys. {\bf B430} (1994) 179.
%
\bibitem{LANL}
T.~Bhattacharya and R.~Gupta,
Nucl.~Phys.~{\bf B} (Proc. Suppl.) {\bf 42} (1995) 935.
%
\bibitem{UKQCD62-60}
UKQCD Collaboration (C.~R.~Allton {\it et al.}),
Phys.~Rev. {\bf D49} (1994) 474.
%
%
\bibitem{LANLsmear}
D.~Daniel {\it et al.}, Phys.~Rev. {\bf D46} (1992) 3130.
%
\bibitem{PAXmachine}
Y.~Iwasaki {\it et al.}, 
Computer Physics Communications {\bf 49} (1988) 449;
T.~Shirakawa {\it et al.}, {\it Proceedings of Supercomputing '89} 
495, Reno, USA, Nov.\ 13-17, 1989;
Y.~Iwasaki {\it et al.}, 
Nucl.~Phys.~{\bf B} (Proc.~Suppl.) {\bf 17} (1990) 259.
%
\bibitem{ONO}
S.~Ono, Phys.~Rev. {\bf D17} (1978) 888. 
%
\bibitem{Zpert}
G.~Martinelli and Y.~C.~Zhang, 
Phys.~Lett. {\bf B123} (1983) 433.
%
\bibitem{ZKtp}
G.~P.~Lepage, 
Nucl.~Phys.~{\bf B} (Proc.Suppl.) {\bf 26} (1992) 45;
A.~S.~Kronfeld, 
Nucl.~Phys.~{\bf B} (Proc.Suppl.) {\bf 30} (1993) 445;
P.~B.~Mackenzie, 
Nucl.~Phys.~{\bf B} (Proc.Suppl.) {\bf 30} (1993) 35.
%
\bibitem{LM}
G.~P.~Lepage and P.~B.~Mackenzie, 
Phys.~Rev. {\bf D48} (1993) 2250.
%
\bibitem{El-Khadra}
A.~X.~El-Khadra {\it et al.}, 
Phys.~Rev.~Lett. {\bf 69} (1992) 729.
%
\bibitem{ZMC}
L.~Maiani and G.~Martinelli, Phys.~Lett. {\bf B178} (1986) 265.
%
\bibitem{GF11DE}
F.~Butler, H.~Chen, J.~Sexton, A.~Vaccarino and D.~Weingarten,
Nucl.~Phys. {\bf B421} (1994) 217.
%
\bibitem{UKQCDEX}
UKQCD Collaboration (C.~R.~Allton {\it et al.}), 
Phys.~Rev. {\bf D47} (1993) 5128.
%
\bibitem{UKQCDJ}
UKQCD Collaboration (P.~Lacock and C.~Michael),
Phys.~Rev. {\bf D52} (1995) 5213.
%
\bibitem{CLOG1}
S.~Sharpe, Phys.~Rev. {\bf D41} (1990) 3233;
Phys.~Rev. {\bf D46} (1992) 3146.

\bibitem{CLOG2}
C.~Bernard and M.~Golterman,
Phys.~Rev. {\bf D46} (1992) 853.

\bibitem{JLQCD}
JLQCD Collaboration (S.~Aoki {\it et al.}),
Tsukuba preprint UTHEP-323 (hep-lat/9510013),
to appear in the Proceedings of Lattice '95.
%
\bibitem{GERMAN}
M. G\"ockeler {\it et al.}, 
DESY preprint DESY 95-128 (hep-lat/9508004).
%
%
\bibitem{LANLNEW}
T.~Bhattacharya, R.~Gupta, G.~Kilcup and S.~Sharpe,
Los Alamos preprint LA-UR-95-2354 (hep-lat/9512021);
T.~Bhattacharya and R.~Gupta, Los Alamos preprint LA-UR-95-2355
(hep-lat/9510044).

\end{thebibliography}
\end{document}